\documentclass[12pt,twoside]{book}
\usepackage{amssymb}
\usepackage{amsmath}
\usepackage[dvips]{graphicx}
\usepackage{fancyhdr}
\fancypagestyle{plain}{%
   \fancyhead{} %
    
}
\begin{document}

\setcounter{page}{1}
%\newcounter{equation}[section]
\newtheorem{t1}{Theorem}[section]
\newtheorem{d1}{Definition}[section]
\newtheorem{c1}{Corollary}[section]
\newtheorem{l1}{Lemma}[section]
\newtheorem{r1}{Remark}[section]
\newcommand{\cA}{{\cal A}}
\newcommand{\cB}{{\cal B}}
\newcommand{\cC}{{\cal C}}
\newcommand{\cD}{{\cal D}}
\newcommand{\cE}{{\cal E}}
\newcommand{\cF}{{\cal F}}
\newcommand{\cG}{{\cal G}}
\newcommand{\cH}{{\cal H}}
\newcommand{\cI}{{\cal I}}
\newcommand{\cJ}{{\cal J}}
\newcommand{\cK}{{\cal K}}
\newcommand{\cL}{{\cal L}}
\newcommand{\cM}{{\cal M}}
\newcommand{\cN}{{\cal N}}
\newcommand{\cO}{{\cal O}}
\newcommand{\cP}{{\cal P}}
\newcommand{\cQ}{{\cal Q}}
\newcommand{\cR}{{\cal R}}
\newcommand{\cS}{{\cal S}}
\newcommand{\cT}{{\cal T}}
\newcommand{\cU}{{\cal U}}
\newcommand{\cV}{{\cal V}}
\newcommand{\cX}{{\cal X}}
\newcommand{\cW}{{\cal W}}
\newcommand{\cY}{{\cal Y}}
\newcommand{\cZ}{{\cal Z}}
\def\cl{\centerline}
\def\bd{\begin{description}}
\def\be{\begin{enumerate}}
\def\ben{\begin{equation}}
\def\benn{\begin{equation*}}
\def\een{\end{equation}}
\def\eenn{\end{equation*}}
\def\benr{\begin{eqnarray}}
\def\eenr{\end{eqnarray}}
\def\benrr{\begin{eqnarray*}}
\def\eenrr{\end{eqnarray*}}
\def\ed{\end{description}}
\def\ee{\end{enumerate}}
\def\al{\alpha}
\def\b{\beta}
\def\bR{\bar\R}
\def\bc{\begin{center}}
\def\ec{\end{center}}
\def\d{\dot}
\def\D{\Delta}
\def\del{\delta}
\def\ep{\epsilon}
\def\g{\gamma}
\def\G{\Gamma}
\def\h{\hat}
\def\iny{\infty}
\def\La{\Longrightarrow}
\def\la{\lambda}
\def\m{\mu}
\def\n{\nu}
\def\noi{\noindent}
\def\Om{\Omega}
\def\om{\omega}
\def\p{\psi}
\def\pr{\prime}
\def\r{\ref}
\def\R{{\bf R}}
\def\ra{\rightarrow}
\def\s{\sum_{i=1}^n}
\def\si{\sigma}
\def\Si{\Sigma}
\def\t{\tau}
\def\th{\theta}
\def\Th{\Theta}
\def\vep{\varepsilon}
\def\vp{\varphi}
\def\pa{\partial}
\def\un{\underline}
\def\ov{\overline}
\def\fr{\frac}
\def\sq{\sqrt}
\def\WW{\begin{stack}{\circle \\ W}\end{stack}}
\def\ww{\begin{stack}{\circle \\ w}\end{stack}}
\def\st{\stackrel}
\def\Ra{\Rightarrow}
\def\R{{\mathbb R}}
\def\bi{\begin{itemize}}
\def\ei{\end{itemize}}
\def\i{\item}
\def\bt{\begin{tabular}}
\def\et{\end{tabular}}
\def\lf{\leftarrow}
\def\nn{\nonumber}
\def\va{\vartheta}
\def\wh{\widehat}
\def\vs{\vspace}
\def\Lam{\Lambda}
\def\sm{\setminus}
\def\ba{\begin{array}}
\def\ea{\end{array}}
\def\ds{\displaystyle}
\def\lan{\langle}
\def\ran{\rangle}
\baselineskip 15truept
\large
%\thispagestyle
%\pagenumbering{roman}

\frontmatter

\title{Analysis of Quantum Correlation in Successive Spin Measurements, Classical Communication of Spin $S$ Singlet States and Quantum Nonlocality of two Qubit Entangled States.}, \author{Ali Ahanj\\Department of Physics,\\University of Pune, Ganeshkhind, Pune-411007\\\\\\\\ Dissertation submitted in partial fulfillment of the requirements\\for the degree of\\Doctor of Philosophy in Physics}
\maketitle
\newpage
\textbf{
\begin{center}
CERTIFICATE
\end{center}
}

CERTIFIED that the work incorporated in the thesis \\\textbf{``Analysis of Quantum Correlation in Successive Spin Measurements, Classical Communication of Spin $S$ Singlet states and Quantum Nonlocality of two Qubit Entangled States"}\\ submitted by \textbf{Ali Ahanj} was carried out by the candidate under my guidance. Such material as has been obtained from other sources has been duly acknowledged in the thesis.\\\\\\ 
\begin{flushright}
Dr. P. S. Joag\\(supervisor)
\end{flushright}
\newpage
\textbf{
\begin{center}
DECLARATION
\end{center}
}
I declare that the thesis entitled \\\textbf{``Analysis of Quantum Correlation in Successive Spin Measurements, Classical Communication of Spin $S$ Singlet states and Quantum Nonlocality of two Qubit Entangled States"}\\ submitted by me for the degree of Doctor of Philosophy, is the record of work carried out by me during the priod time \date{Nov/2003} to \date{Aug/2007} under the guidence of \textbf{Dr. P. S. Joag} and this has not formed the basis for the award of any degree, diploma, associate ship, fellowship, title in this or any other university or other institution of higher learning.\\I further declare that the material obtained from other sources has been duly acknowledged in the thesis.\\ 
\begin{flushright}
Ali Ahanj
\end{flushright}
\newpage
\begin{center}
\textbf{List of Publications}
\end{center}
1. \textbf{Non locality without inequality for almost all two-qubit entangled states based on Cabello's non locality argument}\\S. Kunkri, S. Choudhary, A. Ahanj and P. S. Joag, Phys. Rev A \textbf{73} 022346 (2006); quant-Ph/0512025\\\\
2. \textbf{Classical simulation of two spin-$s$ singlet state correlations involving spin measurements}\\ A. Ahanj, P. S. Joag and S. Ghosh, Phys. Lett A \textbf{368} 34-37 (2007); quant-Ph 0603053\\\\
3. \textbf{Quantum correlations in successive spin measurements}\\ A. Ahanj and P. S. Joag, Submitted to International Quantum Information Journal; quant-Ph/0602005\\\\
4. \textbf{Simulation of two spin-$s$ singlet correlations for all $s$ involving spin measurements}\\ A. Ahanj, P. S. Joag and S. Ghosh, Submitted to Quantum Information and Computation ;quant-Ph/0706.2287

\newpage
\begin{center}
\large {\bf Acknowledgments}
\end{center}
\begin{sloppypar}
It gives me great pleasure to express my gratitude to my supervisor Dr. Pramod S. Joag for his constant guidance, encouragement and support. He has been always more than a guide to me and remains an ideal in my life both as a physicist and as a human being.
\end{sloppypar}
\begin{sloppypar}
	I wish to thank Dr. Sibasish Ghosh who I have collaborated in some of this research and I have learn a lot from him.
\end{sloppypar}
\begin{sloppypar}
	My sincere thanks go to Mr. Ali  Saif M. Hassan who worked a long with me on some allied aspects of Quantum Informations theory. I enjoyed our association as members of the same research group.
\end{sloppypar} 
\begin{sloppypar}
	Thank also to Samir Kunkri and Sujit K. Choudhary with whom I have collaborated in some of this research.
\end{sloppypar} 
\begin{sloppypar}
	I would like to thank the Department of Physics, University of Pune for providing necessary facility.
\end{sloppypar} 
\begin{sloppypar}
	My Parents deserve not only thanks for unconditional support during all phases of my life, but also a sort of apology of my pursuing of a career so far away from home for so many years.
\end{sloppypar}
\begin{sloppypar}
	I would like to express my gratitude to my wife Zohreh and my  children Zahra and Parsa for their love and tolerance without which I could not have done any meaningful work.
\end{sloppypar}
\begin{sloppypar}
	 Last but not the least, I wish the express my regards to Dr. Mehdi Golshani who has inspired me immensely. 
\end{sloppypar}

\tableofcontents
\mainmatter
\chapter{Introduction}
\begin{center}
\scriptsize\textsc{$\cdots$ But I can safely say that nobody understands Quantum Mechanics$\cdots$\\ Richard Feynmann}
\end{center}
Quantum Mechanics (QM) represents one of the pillars of modern physics: so far a huge amount of theoretical predictions deriving from this theory have been confirmed by very accurate experimental data. No doubts can be raised on the validity of this theory. Nevertheless, even after one century since its birth, many problems related to the interpretation of this theory persist: non-local effects of entangled states, wave function reduction and the concept of measurement in QM, the transition from a microscopic probabilistic world to a macroscopic deterministic word perfectly described by classical mechanics and so on. A possible way out from these problems would be if QM represents a statistical approximation of an unknown deterministic theory, where all observables have well defined values fixed by unknown variables, the so called Hidden Variable Theories (HVT). Therefore, the debate whether QM is a complete theory and probabilities have a non-epistemic character (i.e. nature is intrinsically probabilistic) or whether it is a statistical approximation of a deterministic theory and probabilities are due to our ignorance of some parameters (i.e. they are epistemic) dates to the beginning of the theory itself.\\
The fundamental paper where this problem clearly emerged appeared in 1935 when Einstein, Podolsky and Rosen asked this question by considering an explicit example \cite{epr}.\\ For this purpose, they introduced the concept of element of reality according to the following definition: if, without disturbing a system in any way , one can predict with certainty the value of a physical quantity, then there is an element of physical reality corresponding to this quantity. They formulated also the reasonable hypothesis ( consistent with special relativity) that every non-local action was forbidden. A theory is complete when it describes every element of reality. They concluded that either one or more of their premises was wrong or Quantum Mechanics was not a complete theory, in the sense that not every element of physical reality had a counterpart in the theory.\\ This problem led to the search of a ``complete theory" by adding hidden variables to the wave function in order to implement realism. For a long time, there was a general belief among quantum physicists that quantum mechanics can not be replaced by some complete theory (HVT) due to Von Neumann's impossibility proof ( who imposed an unwarranted constraint on HVT).  But in sixties we got two theorems due to J. S. Bell \cite{bell64} \cite{bell66} and Kochen and Specker \cite{kochenspecker}. These theorems showed that quantum mechanics can not be replaced by some classes of HVT, namely local and non-contextual HVT. The most celebrated of this kind of HVT was presented by Bohm in 1952 \cite{bohmtheory}. Bohm, just prior to developing his HV interpretation, introduced a simplified scenario involving two spin-half particles with correlated spins, rather than two particles with correlated positions and momenta as used by EPR. The EPR-Bohm scenario has the advantage of being experimentally accessible.\\
In 1964 John Bell \cite{bell64} derived an inequality ( which is a statistical result, and is called Bell's inequality BI) using locality and reality assumptions of EPR-Bohm, and showed that the singlet state of two spin-$1/2$ particles violates this inequality, and hence the contradiction with quantum mechanics.\\ 
Contemporary versions of the argument are based on the Clauser, Horne, Shimony and Holt (CHSH) inequality\cite{chsh}, rather than the original inequality used by Bell. There is a very good reason for that. While Bell's argument applied only to the singlet state, the CHSH inequality is violated by all pure entangled states \cite{gisinperes}. Early versions of CHSH inequalities involved only two observers, each one having a choice of two (mutually incompatible) experiments. The various outcomes of each experiment were lumped into two sets, arbitrarily called $+1$ and $-1$. Possible generalizations involve more than two observers, or more than two alternative experiments for each observer, or more than two distinct outcomes for each experiment. We may consider $n$-partite systems, each subject to a choice of $m$ $v$-valued measurements. This gives a total of $(mv)^n$ experimentally accessible probabilities. The  set of Bell inequalities is then the set of inequalities that bounds this region of probabilities to those accessible with a local hidden variable model. Thus for each value of $n$, $m$ and $v$ the set of local realistic theories is a polytopes bounded by a finite set of linear Bell inequalities. The CHSH inequalities apply to a situation $(n,m,v)=(2,2,2)$. Gisin \textsl{et al} \cite{gisin98} have found a family of Bell inequalities for the case with the number of measurements is arbitrary, i.e. $(n,m,v)=(2,m,2)$. Collins et al \cite{cglmp} and Kaszlikowski et al \cite{kgzmz} have produced inequalities for arbitrarily high dimensional systems, i.e. $(n,m,v)=(2,2,v)$. The most complete study of Bell inequalities is for the case $(n,m,v)=(n,2,2)$. $n$-particle generalizations of the CHSH inequality were first proposed by Mermin \cite{mermin90}, and Belinskii and Klyshko \cite{klyshko}, and have been extended by Werner and Wolf \cite{wernerwolf}, and Zukowski and Brukner \cite{zukbru} to give the complete set for two dichotomic observables per site.\\
On the theoretical side, `` violation of Bell's inequalities" had become synonymous with ``non-classical correlation", i.e., entanglement. One of the first papers in which finer distinctions were made was the construction of states with the property that they satisfy all the usual assumptions leading to the Bell inequalities, but can still not be generated by a purely classical mechanism ( are not ``separable" in modern terminology) \cite{werner89}. This example pointed out a gap between the obviously entangled states ( violating a Bell inequality) and the obviously non-entangled ones, which are merely classical correlated ( separable). In 1995 Popescu \cite{popescu95} ( and later \cite{bbpssw96}) narrowed this gap considerably by showing that after local operations and classical communication one could ``distill" entanglement, leading once again to violations, even from states not violating any Bell inequality initially. To summarize this phase: it became clear that violations of Bell inequalities, while still a good indicator for the presence of non-classical correlations by no means capture all kinds of ``entanglement".\\
In entangled states, quantum mechanics predicts strong correlations between measurements on two ( or more ) systems that have previously interacted but which are separated at the time of the measurement. However, John Bell \cite{bell64} showed that the quantum correlations exhibited by entangled states, could not be reproduced by local hidden variable models, that is, models where Alice and Bob share an infinite amount of locally created hidden variables. This proved in a sense the nonlocal character of QM. This nonlocal aspect is one of the strangest properties of quantum physics, and understanding this notion remains an important problem.\\ Bell inequalities, while usually considered relevant only to foundational studies of quantum theory, answer a fundamental information-theoretic question: what correlations can be produced between separate classical subsystems, which have interacted in the past, if no communication between the subsystems is allowed? Violation of a Bell inequality, however, does nothing to quantify what classical information processing resources are required to simulate a particular set of quantum correlations. Information is something that we are able to quantify, thus the answer to this question provides a measure of the non-locality of two particles in entangled states. The first answers were given in 1999: the maximally-entangled state of two qubits (``singlet") can be simulated in any case using eight bits of communication \cite{tapp99}, or using a different strategy that uses 2 bits on average but may require unlimited communication in the worst case \cite{gisingisin99} \cite{steiner00}. All these results were superseded in 2003, when Toner and Bacon \cite{toner03} proved that the singlet can be simulated exactly using local variables plus just one bit of communication per pair. Another resource than communication has been proposed as a tool to study non-locality: the non-local machine (NLM) invented by Popescu and Rohrlich \cite{popescurohrlich}. This machine (NLM) would violate the BI ($\rm{BI}\leq2$) up to its algebraic bound of $\rm{BI}=4$ (while it is known that QM reaches up only to $\rm{BI}=2\sqrt{2}$) without violating the no-signaling constraint. The singlet state can be simulated by local variables plus just one use of the NLM per pair \cite{cgmp}. Finally, in order to simulate the correlations of a pure non-maximally entangled state, a strictly larger amount of resources may be needed than for the simulation of the maximally entangled state \cite{brunnergisin}.\\ 
Conceptually, as well as mathematically, space and time are differently described in quantum mechanics. While time enters as an external parameter in the dynamical evolution of a system, spatial coordinates are regarded as quantum mechanical observables. Moreover, spatially separated quantum systems are associated with the tensor product structure of the Hilbert state-space of the composite system. This allows a composite quantum system to be in a state that is not separable regardless of the spatial separation of its components. We speak about \emph{entanglement in space}. On the other hand, time in quantum mechanics is normally regarded as lacking such a structure. Because of different roles time and space  play in quantum theory one could be tempted to assume that the notion of \emph{``entanglement in time"} cannot be introduced in quantum physics. In chapter III in this thesis we will investigate about entanglement in time and we will derive temporal Bell's inequalities. The notion of temporal Bell's inequalities was first introduced by Leggett and Garg \cite{leggettgarg} in a different context. They focus on one and the same physical system and analyze correlations between measurement outcomes at different times. The aim of temporal Bell inequalities, in the original spirit of Leggett and Garg \cite{leggettgarg}, was to test quantum mechanics at the macroscopic level whenever a macroscopic observable of the system is monitored.\\
Bell inequalities are statistical predictions about measurements made on two particles, typically photons or particles with spin $\frac{1}{2}$. So some people were trying to show a direct contradiction (which is not a statistical one) of quantum mechanics with local realism. Greenberger, Horne and Zeilinger (GHZ)\cite{ghz} found a way to show more immediately, without inequalities, that results of quantum mechanics are inconsistent with the assumptions of EPR. It focuses on just one event, not the statistics of many events. Their proof relies on \textit{eight } dimensional Hilbert space, unlike  the case of Bell's theorem, which is valid in \textit{four} dimensions. Heywood and Redhead \cite{heywoodredhead} have provided a direct contradiction (without inequalities) of quantum mechanics with local realism for a particular state of two spin-$1$ particles.\\
Finally, Hardy \cite{hardy92} gave a proof of non locality for two particles with spin $\frac{1}{2}$ that only requires a total of \textit{four} dimensions in Hilbert space like Bell's proof but does not require inequalities.\\
This was accomplished by considering a particular experimental setup consisting of two over-lapping Mach-Zehnder interferometers, one for positrons one for electrons, arranged so that if the electron and positron each take a particular path then they will meet and annihilate one another with probability equal to 1. This arrangement is required to produce assymetric entangled state which only exhibits non locality without any use of inequality. The argument has been generalized to two spin s particles by Clifton and Niemann \cite{clinie} and to N spin half particles by Pagonis and Clinton \cite{pagcli}.\\ Later, Hardy showed that this kind of non locality argument can be made for almost all entangled states of two spin-$\frac{1}{2}$ particles except for maximally entangled one \cite{hardy93}. This proof was again simplified by Goldstein \cite{goldstein} and extended it to the case of bipartite systems whose constituents belong to Hilbert spaces of arbitrary dimensions. Recently, Cabello has introduced a logical structure to prove Bell's theorem without inequality for three particles GHZ and W \cite{cabello02}.\\
The present-day entanglement theory has its roots in the key discoveries: quantum algorithms, quantum communication complexity,  quantum cryptography, quantum dense coding and quantum teleportation. 
These recent theoretical research  has shown that quantum devices are more powerful than their classical counterparts. Indeed, the flourishing field of quantum information theory \cite{nielsen} aims to provide an information-theoretic quantification of the power underlying quantum resources. One important feature of quantum theory lies in the statistical correlations produced by measurements on local components of a quantum system. Quantum information processing has provided a new point of view to understand quantum non locality. In particular, the framework of communication complexity has provided tools to study non locality. For example, we know that if Alice and Bob share only a set of local hidden variables (shared randomness), they cannot reproduce quantum correlations, but  if they are allowed to use some additional resources, it may be possible for them to reproduce the quantum correlations. It is precisely this amount of additional resources which we consider in this thesis; they allow us to quantify quantum non locality.\\ 
In this thesis, we study about three subjects\\ 1- Classical simulation of two spin-$s$ singlet correlations for all $s$ involving spin measurements,\\ 2- Quantum correlations in successive single spin measurements,\\3- Non locality without inequality for almost all two-qubit entangled states based on Cabello's non locality argument.\\ The chapters are arranged as follows:\\ \textbf{Chapter II}: We give a classical protocol to exactly simulate quantum correlations implied by the spin $s$ singlet state for all integer as well as half-integer spin values $s$. The class of measurements we consider here are only those corresponding to spin observables. The required amount of communication is found to be $\left\lceil \log_2(s+1)\right\rceil$ in the worst case scenario, where $\left\lceil x \right\rceil$ is the least integer greater than or equal to $x$. We also obtain another classical protocol to exactly simulate quantum correlations corresponding to the spin $s$ singlet state for the infinite sequences of spins satisfying $2s+1=P^n$ ($P$ and $n$ are positive integer number). Thus this protocol also simulates singlet state correlations for all spins.  Finally we show a classical protocol for simulation of the quantum correlation implied by non maximally entangled states of two qubits  by using two bits of communication. Our model is working in the special case that one of Alice's input or Bob's input lie on $X-Y$ plane.\\ \textbf{Chapter III}: In this chapter we consider a hidden variable theoretic description of successive measurements of non-commuting spin observables on a input spin-$s$ state. Although these spin observables are non-commuting, they act on different states and so the joint probabilities for the outputs of successive measurements are well defined. We show that,
in this scenario, hidden variable theory (HVT) leads to Bell-type
inequalities for the correlation between the outputs of successive
measurements. We account for the maximum violation of these
inequalities by quantum correlations ({\it i.e.}, the correlations
of successive measurements on a quantum state) by varying spin value
and the number of successive measurements. Our approach can be used
to obtain a measure of the deviation of Quantum Mechanics from the
theory obeying realism and time-locality in terms of the amount of
classical information needed to be transferred between successive
measurements in order to simulate the above-mentioned correlations
in successive measurements.\\ \textbf{Chapter IV}: In this chapter we deal with non locality argument proposed by Cabello, which is more general than Hardy's non locality argument, but still maximally entangled states do not respond. However, for most of the other entangled states, maximum probability of success of this argument is more than that of the Hardy's argument. So it seems that in some sense, for demonstrating the nonlocal features of most of the entangled states, Cabello's argument is a better candidate.\\ \textbf{Chapter V}: Finally our conclusion and new open problems are summarized in this chapter.
\newpage

\chapter{Classical Simulation of Quantum Correlations}

\section{Introduction}	
At the advent of quantum mechanics, some physicists were puzzled by the strange properties of quantum systems, compared to classical physics, such as randomness and non locality. Einstein, Podolsky and Rosen showed that when two parties, say Alice and Bob, share an entangled state, the outcome of a measurement on Alice's part is not determined only locally, but may also be conditioned on the outcome of a distant measurement on Bob's part. Therefore, they questioned whether ``the quantum-mechanical description of physical reality could be considered complete" \cite{epr}.\\
To resolve this paradox, later called the EPR paradox, it was argued that the apparent randomness in quantum experiments could actually come from unknown (``hidden") variables created locally along with the supposedly quantum state, and that this randomness would disappear as soon as these hidden variables were revealed. However, John Bell showed in 1964 that the quantum correlations exhibited by the EPR gedanken experiment, as re-expressed by Bohm \cite{bohm57}, could not be reproduced by so-called local hidden variable models, that is, models where Alice and Bob share an infinite amount of locally created hidden variables \cite{bell64}. This proved in a sense the nonlocal character of quantum mechanics. This nonlocal aspect is one of the strangest properties of quantum physics, and understanding this notion remains an important problem.\\
Recent theoretical research into quantum algorithms \cite{p.w.shor}, quantum communication complexity \cite{raz}, and quantum cryptography \cite{benett84} has shown that quantum devices are more powerful than their classical counterparts. Indeed, the flourishing field of quantum information theory \cite{nielsen} aims to provide an information-theoretic quantification of the power underlying quantum resources. One important feature of quantum theory lies in the statistical correlations produced by measurements on local components of a quantum system. Quantum information processing has provided a new point of view to understand quantum non locality. In particular, The framework of communication complexity has provided tools to study non locality. For example, we know that if Alice and Bob share only a set of local hidden variables (shared randomness), they cannot reproduce quantum correlations, but  if they are allowed to use some additional resources, it may be possible for them to reproduce  quantum correlations. It is precisely this amount of additional resources which we consider here; they allow us to quantify quantum non locality.\\
The most obvious resource that Alice and Bob can use in addition to shared randomness is classical communication. One may naturally ask how much information should be sent from one party (Alice)  to the other (Bob) in order to reproduce the correlations distributed by entangled pairs (simulate entanglement).\footnote{It is easy to see that the Bell (CHSH) expression can exceed 2 if Bob's output depended on that of Alice.}\\
The objective was to quantify the non-locality of EPR pairs in terms of the amount of communication necessary to simulate the correlation obtained by bipartite measurement of an EPR pair.`` The key to understanding violations of Bell's inequality is not operator algebras but information transmission \cite{maudlin92}." 
This approach increases our understanding of the relationships between classical information and quantum information. It also helps us gauge the amount of information hidden in the EPR pair itself or, in some sense, the amount of information that must be space-like transmitted, in a local hidden variable model, in order for nature to account for the Bell inequalities \cite{methot}.\\
Entanglement simulation was first introduced by  Madlin in a 1992 paper published in a philosophical journal \cite{maudlin92} and was revived independently by Brassard, Cleve and Tapp in 1999 \cite{brassard99}. 
In this scenario, Alice and Bob try and output $\alpha$ and $\beta$
respectively, through a classical protocol, with the same
probability distribution as if they shared the bipartite entangled
system and each measured his or her part of the system according to
a given random Von Neumann measurement. As we have mentioned above,  
such a protocol must involve communication between Alice and Bob,
who generally share finite or infinite number of random variables.
The amount of communication is quantified \cite{pironio03} either as
the average number of cbits $\overline {C}(P)$ over the directions
along which the spin components are measured ( average or expected
communication) or the worst case communication, which is the maximum
amount of communication $C_{w}(P)$ exchanged between Alice and Bob
in any particular execution of the protocol. The third method is
asymptotic communication i.e., the limit
$lim_{n\rightarrow\infty}\overline{C}(P^{n})$ where $P^{n}$ is the
probability distribution obtained when $n$ runs of the protocol are 
carried out in parallel i.e., when the parties receive $n$ inputs
and produce $n$ outputs in one go. Note that, naively, Alice can
just tell Bob the direction of her measurement to get an exact
classical simulation, but this corresponds to  an infinite amount of
communication. The question whether a simulation can be done with
finite amount of communication was raised independently by Maudlin
 \cite{maudlin92}, Brassard, Cleve and Tapp \cite{brassard99}, and
Steiner \cite{steiner00}.\\
Brassard, Cleve and Tapp  showed in \cite{brassard99} that \textit{eight} bits of classical communication is sufficient in the worst case for an exact simulation of entanglement in an arbitrary Von Neumann measurement. These protocols use an infinite amount of shared randomness, and indeed Massar \textit{et al.} \cite{massarcleve99} proved that the communication complexity can be bounded in worst case only if the amount the shared randomness is infinite. In 2000, inspired by Feldmann \cite{feldmann95}, Steiner, independently of Maudlin, showed that for projective measurements in the real plane, 1.48 bits were enough on average \cite{steiner00}. Cerf, Gisin and Massar \cite{cerf2000} proved that for an arbitrary projective measurement 1.19 bits of communication sufficed on average. Csirik \cite{csirik02} gave a protocol with a worst-case communication upper bound of \textit{six} bits, for arbitrary projective measurements. Finally in 2003, Toner and Bacon \cite{toner03} showed that a local hidden variable model supplemented with one bit of communication in the worst case is enough to reproduce the quantum correlations of the spin $\frac{1}{2}$ singlet state  for arbitrary projective measurements.\\
Another resource which Alice and Bob can use to reproduce the quantum correlations is postselection. Here, Alice and Bob are allowed to produce a special outcome of their measurement, noted $\bot$, meaning `` no result''.  This corresponds to the physical situation where Alice and Bob's detectors are partially inefficient and sometimes do not click. In 1999, Gisin and Gisin \cite{gisingisin}, inspired by Steiner's communication protocol, gave a protocol which simulates quantum correlations with shared randomness and with a probability 1/3 of aborting for either party.\\
Independently of the above story, S. Popescu and D. Rohrlich raised the following question: Can there be stronger correlations than the quantum mechanical ones that remain causal (i.e., that do not allow superluminal signaling) \cite{popescurohrlich}? Recall that  quantum correlations violate  Bell inequality, but do not allow any faster than light signaling. Popescu and Rohrlich answered by presenting an hypothetical machine that does not allow signaling, yet violates the CHSH inequality  \cite{chsh} more than quantum mechanics. They concluded by asking why Nature is non-local, but not maximally non-local, where the maximum would only be limited by the no-signaling constraint? Cerf, Gisin, Massar and Popescu \cite{cgmp} have shown that a third resource could be used to simulate the quantum correlations: a nonlocal box (NLM). The nonlocal box is a primitive shared between Alice and Bob, with two inputs and two outputs, where the outputs ( conditioned on the inputs) are maximally nonlocal in the sense that  they violate a Bell inequality (CHSH) maximally while remaining causal. They have shown that only one use of a nonlocal box suffices to simulate quantum correlations in the bipartite spin-$\frac{1}{2}$ case. Next N. Brunner, N. Gisin and V. Scarani \cite{brunnergisin} proved that a single classical of communication or a single use of the NLM does not provide enough non-locality to simulate non-maximally entangled states of two qubits, while, these resources are enough to simulate the singlet.\\
Nevertheless, these results address the simplest scenario, that is, simulating the correlations resulting from measurements on the singlet state ( mostly for projective measurements, with a few extensions to POVMs). There are a few results about non maximally entangled pairs, multi party states, higher dimensional states, or more general measurements. One significant result in this direction is a protocol from Massar, Bacon, Cerf and Cleve  to reproduce the correlations of arbitrary measurements on any entangled pair of $d$-dimensional states (qudits) using $\textit{O}(d~log_{2}d)$ bits of communication but no local hidden variables \cite{massarcleve99}.\\
Until now, an exact classical simulation of quantum correlations,
for {\it all} possible projective measurements, is accomplished only
for spin $s = 1/2$ singlet state, requiring 1 cbit of classical
communication \cite{toner03}. It is important to know how does the
amount of this classical communication change with the change in the
value of the spin $s$, in order to quantify the advantage offered by
quantum communication over the classical one. Further, this
communication cost quantifies, in terms of classical resources, the
variation of the nonlocal character of quantum correlations with
spin values. In the present section we give two classical protocols to simulate the measurement correlation in a singlet state of two spin-$s$ systems, for \textit{all}  values of $s$, considering only  measurement of spin observables. We show that,
using $\lceil {\rm log}_2 (s + 1) \rceil$ bits of classical
communication, one can simulate the above-mentioned measurement
correlation. We will also give a classical protocol for simulation non maximally entangled state.\\
There are two main motivations for studying the classical communication cost in quantum information processing(CCCIQIP).\\ The first motivation is to better understand the fundamental laws of quantum information-the synthesis of quantum mechanics with information theory- processing.  The second motivation is the fact that CCCIQIP can be regarded as a natural generalization of quantum communication complexity, a subject of much recent interest.\\
We will explain about spin measurements in section 2, next we describe the quantum correlations for singlet state of two spin-s systems in section 3. We briefly describe the protocol of Toner and Bacon in section 4. In sections 5 and 6, we give two classical protocols to exactly simulate  two spin-$s$ singlet state. We give a protocol for simulation of the quantum correlations implied by non maximally entangled states of  two qubits  in section 7. Finally, our conclusions are summarized in section 8.
\section{Projective Measurements for $s>\frac{1}{2}$}
Now we consider systems with  finite-dimensional state space, namely the state ${\mathbb{C}}^{d}$, $(d=2s+1)$. For $d=2$ they can be identified with a spin system of spin $\frac{1}{2}$. Let $|i\rangle$, $i=1,\ldots,n$ be an orthonormal basis and expand all  vectors in terms of these basis vectors. Then every Hermitian operator can be expressed as
\begin{equation}
\hat{A}=\sum_{i=1}^{d^{2}-1}a_{i}\Lambda_{i}+a_{0}\frac{1}{d}\hat{I},
\end{equation}
where $\hat{I}$ is the unit matrix, the $a_{i}$ are real numbers given by $a_{i}=\frac{1}{2}\rm tr(\hat{A}\Lambda_{i})$, $a_{0}=\rm tr\hat{A}$, and the $\Lambda_{i}$ form a basis of the Lie algebra of SU(d) with the property
\begin{equation}
\rm tr (\Lambda_{i}\Lambda_{j})=2\delta_{ij}~~,~~\rm tr\Lambda_{i}=0.
\end{equation}
$a_0$ and the ($d^2-1$)-dimensional vector $a_i$ uniquely specify the operator $\hat{A}$ with respect to a given basis $|i\rangle$. These vectors are elements of an Euclidean vector space $V$ of dimension ($d^2-1$). For details see \cite{macfarlane68}.  $\Lambda_{i}$ are Hermitian matrices obeying (for repeated latin indices, the summation is understood over the index range ${1,\ldots,d^{2}-1}$ , excluding $0$)
\begin{equation}
\Lambda_{i}\Lambda_{j}=\frac{2}{d}\delta_{ij}\hat{I}+(d_{ijk}+if_{ijk})\Lambda_{k},
\end{equation}
where
\begin{equation}
[\Lambda_i,\Lambda_j]=2if_{ijk} \Lambda_k ~,~ \{\Lambda_i,\Lambda_j\}=\frac{4}{d}\delta_{ij}\hat{I}+2d_{ijk} \Lambda_k .
\end{equation}
Here  $[,]$ is the commutator, $\left\{,\right\}$ is the anti commutator of matrices. The $d_{ijk}$ are totally symmetric, the $f_{ijk}$ are totally antisymmetric tensors. One then also has the vector of $d^{2}-1$ operators  
\begin{equation}
\vec{\Lambda}=\left\{\hat{u}_{12},\hat{u}_{13},\hat{u}_{23},\ldots,\hat{v}_{12}
,\hat{v}_{13},\hat{v}_{23},\ldots,\hat{w}_{1},\hat{w}_{2},\hat{w}_{3},\ldots,\hat{w}_{d-1}\right\},
\end{equation}
where
\begin{eqnarray}
\hat{u}_{jk}&=&\hat{P}_{jk}+\hat{P}_{kj},\\
\hat{v}_{jk}&=&i(\hat{P}_{jk}-\hat{P}_{kj}),\\
\hat{w}_l&=&-\sqrt{\frac{2}{l(l+1)}}(\hat{P}_{11}+\ldots+\hat{P}_{ll}-l\hat{P}_{l+1,l+1}),
\end{eqnarray}
with $1\leq j<k \leq d $,  $ 1\leq l \leq d-1$. Projection operators $\hat{P}_{kl}$ are defined by $\hat{P}_{kl}=|k\rangle \langle l |.$ In the case of the density matrices representing the statistical operator, one then has matrix elements $\rho_{jk}=\rm tr(\rho\hat{P}_{jk})$ and corresponding expectation values for $\hat{u}_{jk}$, $\hat{v}_{jk}$ and $\hat{w}_{l}$ in terms of density matrix elements.\\
 For vectors $\vec{a}$ and $ \vec{b}$ with $d^{2}-1$ components we define the following products:
\begin{eqnarray}
\vec{a}.\vec{b}&\equiv & \sum_{i}a_i b_i,\nonumber\\
(\vec{a}\ast \vec{b})_k&\equiv& \sum_{ij}d_{ijk}a_i b_j,\\
(\vec{a}\times \vec{b})_k &\equiv & \sum_{ij}f_{ijk}a_ib_j.\nonumber
\end{eqnarray}
Every Hermitian matrix is of the form 
\begin{equation}
\hat{A}=\vec{a}.\hat{\Lambda}+a_0\frac{1}{d}\hat{I}, ~~~a_i\in \mathbb{R}.
\end{equation}
The multiplication law of matrices
\begin{equation}
(\vec{a}.\vec{\Lambda})(\vec{b}.\vec{\Lambda})=\frac{2}{d}\vec{a}.\vec{b}\hat{I}+(\vec{a}\ast \vec{b}+i\vec{a}\times \vec{b})\vec{\Lambda}.
\end{equation}
So, analogous to the Pauli matrices, one can consider $(2s
+ 1)^2 - 1$ number of trace-less but trace-orthogonal Hermitian $(2s + 1) \times
(2s + 1)$ matrices ${\Lambda}_1$, ${\Lambda}_2$, $\ldots$,
${\Lambda}_{(2s + 1)^2 - 1}$ ({\it i.e.}, ${\rm Tr} {\Lambda}_i = 0$ for all $i$ but ${\rm Tr} ({\Lambda}_i {\Lambda}_j) = 0$ if $i \ne j$; see, for example, \cite{mahler95})
such that a general projective measurement on the individual
spin-$s$ system corresponds to the measurement of an observable of
the form $\hat{c}.\vec{\Lambda}$, where $\hat{c}$ is a unit vector
in $\mathbb{R}^{(2s + 1)^2 - 1}$ and $\vec{\Lambda}$ is the $((2s + 1)^2 - 1)$-tuple $({\Lambda}_1, {\Lambda}_2, \ldots, {\Lambda}_{(2s + 1)^2 - 1})$ of the above-mentioned $\Lambda$ matrices. In general, the quantum correlation $\langle{\psi}^-_s|\hat{c}.\vec{\Lambda} \otimes
\hat{d}.\vec{\Lambda}|{\psi}^-_s\rangle$ will be a bilinear function
in the components on $\hat{c}$ and $\hat{d}$. In the special
case when $({\Lambda}_i \otimes {\Lambda}_j) |{\psi}^-_s\rangle =
{\alpha} {\delta}_{ij} |{\psi}^-_s\rangle$ for all $i, j = 1, 2,
\ldots, (2s + 1)^2 - 1$, the quantum correlation will be of the
form ${\alpha} \hat{c}.\hat{d}$. 
Classical simulation of this general quantum correlation seems to be
quite hard one possible reason being the absence of Bloch sphere
structure for higher spin systems.\\ Here, we will consider only measurement of spin observables, namely the observables of the form
$\hat{a}.{\vec {S}}$ on each individual spin-$s$ system, where
$\hat{a}$ is an arbitrary unit vector in $\mathbb{R}^3$ and ${\vec{S}}
= (S_x, S_y, S_z)$. For the $(2s + 1) \times (2s + 1)$ matrix
representations of the spin observables $S_x$, $S_y$ and $S_z$,
please see  page 191 - 192 of ref. \cite{sakurai99}.

\section{Correlation of Two Spin-$S$ Singlet State}
 In the case of classical simulation of the quantum
correlation $\langle{\psi}^-_{1/2}|\hat{a}.\vec{\sigma} \otimes
\hat{b}.\vec{\sigma}|{\psi}^-_{1/2}\rangle$ of the two-qubit singlet
state $|{\psi}^-_{1/2}\rangle$, Alice considers measurement of
traceless observable $\hat{a}.\vec{\sigma}$ and Bob considers that
of the traceless observable $\hat{b}.\vec{\sigma}$ . These are spin
observables. Consider a pair of spin $s$ particles, prepared in a singlet state. We shall need the explicit form of the vector $|{\psi}^-_s\rangle_{AB}$ that represents two spin $s$ particles with total angular momentum zero. For each particle, we have 
\begin{eqnarray}
\vec{S}_{A}.\hat{z}|m\rangle_A&=&m|m\rangle_A\\
\vec{S}_{B}.\hat{z}|m\rangle_B&=&m|m\rangle_B\nonumber
\end{eqnarray}
where $m=s,s-1,\ldots,-s$ and $\hbar=1$ for simplicity. In order to satisfy $(\vec{S}_A.\hat{z}+ \vec{S}_B.\hat{z}) |{\psi}^-_s\rangle_{AB}=0$, the singlet state must have the form $|{\psi}^-_s\rangle_{AB}=\sum_m c_m |m\rangle_A \otimes |- m\rangle_B.$  Therefore, the data that can be obtained separately by each observer are given by identical density matrices $\rho$, which are diagonal, with elements $\left|c_m\right|^2$ summing up to $1$. Moreover, all the probabilities $\left|c_m\right|^2$ are equal. The reason simply is that a zero angular momentum state is spherically symmetric, and the $z$-axis has no special status. Any other polar axis would yield the same diagonal density matrix $\rho$, with the same elements $\left|c_m\right|^2$. As the choice of another axis is a mere rotation of the coordinates, represented in quantum mechanics by a unitary transformation, $\rho\rightarrow U\rho U^{\dagger}$, it follows that $\rho$ commutes with all the rotation matrices $U$. Now, for any given $s$, these matrices are irreducible. Therefore $\rho$ is a multiple of the unit matrix, and $\left|c_m\right|^2=(2s+1)^{-1}$.
So the singlet state $|{\psi}^-_s\rangle_{AB}$ of two spin-$s$
particles $A$ and $B$ is the eigenstate corresponding to the
eigenvalue $0$ of the total spin observable of these two spin systems,   namely the state
\begin{equation}
\label{singlet} |{\psi}^-_s\rangle_{AB} = \frac{1}{\sqrt{2s + 1}} \sum_{m =
-s}^{s} (- 1)^{s - m} |m\rangle_A \otimes |- m\rangle_B,
\end{equation}
where $|- s\rangle$, $|- s + 1\rangle$, $\ldots$, $|s - 1\rangle$,
$|s\rangle$ are eigenstates of the spin observable of each of the
individual spin-$s$ system. Thus $|{\psi}^-_s\rangle_{AB}$ is a
maximally entangled state of the bipartite system $A + B$, described
by the Hilbert space ${\mathbb{C}}^{2s + 1} \otimes {\mathbb{C}}^{2s
+ 1}$.
The quantum correlations
$\langle{\psi}^-_s|\hat{a}.{\vec{S}} \otimes \hat{b}.{\vec
{S}}|{\psi}^-_s\rangle$ ( which we will denote here as $\left\langle
\alpha \beta\right\rangle$, where $\alpha$ runs through all the
eigenvalues of $\hat{a}.{\vec{ S}}$ and $\beta$ runs through all the
eigenvalues of $\hat{b}.{\vec{ S}}$) is given by
\begin{equation}
\label{correlation} \langle{\psi}^-_s|\hat{a}.{\vec S} \otimes
\hat{b}.{\vec S}|{\psi}^-_s\rangle = \left\langle \alpha
\beta\right\rangle=-\frac{1}{3} s(s+1)\hat{a}.\hat{b} \;,
\end{equation}
where $\hat{a}$ and $\hat{b}$ are the unit vectors specifying the
directions along which the spin components are measured by Alice and
Bob respectively \cite{peres93}. Note that, by virtue of being a
singlet state, $\left\langle \alpha \right\rangle = 0 =
\left\langle\beta\right\rangle$ irrespective of directions $\hat{a}
$ and $\hat{b}$. 
\section{Classical Simulation of Two Spin-1/2 Singlet State}
As the working principles of our protocols are of similar in nature
with those of Toner and Bacon \cite{toner03}, before describing our
protocols, we would like to briefly describe the protocol of Toner
and Bacon to simulate the measurement correlations on
$|{\psi}^-_{1/2}\rangle$. In this scenario, Alice and Bob's job is
to simulate the quantum correlation
$\langle{\psi}^-_{1/2}|\hat{a}.{\frac{1}{2}}{\mathbf{\sigma}}
\otimes \hat{b}.{\frac{1}{2}}{\mathbf{\sigma}}|{\psi}^-_{1/2}\rangle
= - \frac{1}{4}\hat{a}.\hat{b}$, together with the conditions that
$\langle \alpha \rangle = 0 = \langle \beta \rangle$. To start with,
Alice and Bob share two independent random variables $\hat{\lambda}$
and $\hat{\mu}$, each of which has uniform distribution on the
surface of the Bloch sphere $S_2$ in $\mathbb{R}^3$. Given the
measurement direction $\hat{a}$, Alice calculates $-
\frac{1}{2}\rm {sgn}(\hat{a}.\hat{\lambda})$, which she takes as her measurement
output $\alpha$. Note that $\rm {sgn}(x) = 1$ for all $x \ge 0$ and
$\rm {sgn}(x) = - 1$ for all $x < 0$. As $\hat{\lambda}$ is uniformly
distributed on $S_2$, for each given $\hat{a}$, $-
\frac{1}{2}\rm {sgn}(\hat{a}.\hat{\lambda})$ will take its values $\frac{1}{2}$ and $- \frac{1}{2}$ with
equal probabilities, i.e., ${\rm Prob} (\alpha = 1/2) =~ {\rm Prob}
(\alpha = - 1/2) = 1/2$ ( and hence, $\langle \alpha \rangle = 0$).
Alice then sends the one bit information $c \equiv
\rm {sgn}(\hat{a}.\hat{\lambda}) \rm {sgn}(\hat{a}.\hat{\mu})$ to Bob. Note that, 
instead of sending $\rm {sgn}(\hat{a}.\hat{\lambda})$, by sending $c$,
Alice does not allow Bob to extract any information about her output
$\alpha$. This is so because ${\rm Prob} (\alpha = 1/2 | c = 1) =~
{\rm Prob} (\alpha = - 1/2 | c = 1)$ and ${\rm Prob} (\alpha = 1/2 | c =
- 1) =~ {\rm Prob} (\alpha = - 1/2 | c = - 1)$. After receiving $c$,
and using his measurement direction $\hat{b}$, Bob now calculates
his output $\beta \equiv \frac{1}{2}\rm {sgn}[\hat{b}.(\hat{\lambda} + c
\hat{\mu})]$. Now
\begin{equation}
\label{betavalue}
\langle \beta \rangle = \frac{1}{2(4\pi)^2} \int_{\hat{\lambda} \in
S_2}~ \int_{\hat{\mu} \in S_2}~ \rm {sgn}[\hat{b}.(\hat{\lambda} +
\rm {sgn}(\hat{a}.\hat{\lambda})\rm {sgn}(\hat{a}.\hat{\mu}) \hat{\mu})]
d\hat{\lambda} d\hat{\mu}\;.
\end{equation}
Given any $\hat{\mu} \in S_2$, for each
choice of $\hat{\lambda} \in S_2$, the two values of the integrand
corresponding to $\hat{\lambda}$ and $- \hat{\lambda}$ are negative
of each other. As the distribution of $\hat{\lambda}$ on $S_2$ is
taken to be uniform, the above-mentioned observation
immediately shows that $\langle \beta \rangle = 0$. As $\beta \in
\{1/2, - 1/2\}$, therefore ${\rm Prob} (\beta = 1/2) =~ {\rm Prob} (\beta
= - 1/2) = 1/2$. In order to compute $\langle \alpha \beta \rangle$,
one should observe that Bob's output can also be written as $\beta =
\frac{1}{2}\sum_{d = \pm 1} [(1 + cd)/2] \rm {sgn}[\hat{b}.(\hat{\lambda} + d \;
\hat{\mu})]$. The following two among the four integrals ( which
appears in $\langle \alpha \beta \rangle$)
$$- \frac{1}{8(4\pi)^2} \int_{\hat{\lambda} \in S_2}~
\int_{\hat{\mu} \in S_2}~ \rm {sgn}(\hat{a}.\hat{\lambda})
\rm {sgn}[\hat{b}.(\hat{\lambda} \pm \hat{\mu})] d\hat{\lambda}
d\hat{\mu}$$ cancel each other by incorporating the inversion
$\hat{\mu} \rightarrow - \hat{\mu}$. And the rest two integrals
$$\pm \frac{1}{8(4\pi)^2}  \int_{\hat{\lambda} \in S_2}~
\int_{\hat{\mu} \in S_2}~ \rm {sgn}(\hat{a}.\hat{\lambda})
\rm {sgn}[\hat{b}.(\hat{\lambda} \pm \hat{\mu})] d\hat{\lambda}
d\hat{\mu}$$ are same and they are equal to the integral
$$\frac{1}{8(4\pi)^2}  \int_{\hat{\lambda} \in S_2}~
\int_{\hat{\mu} \in S_2}~ \rm {sgn}(\hat{a}.\hat{\lambda})
\rm {sgn}[\hat{b}.(\hat{\mu} - \hat{\lambda})] d\hat{\lambda}
d\hat{\mu}\;.$$Hence we have,
\begin{eqnarray}
\label{alphabetaaverage}
\langle \alpha \beta\rangle &\equiv& \langle{\psi}^-_{1/2}|\hat{a}.{\mathbf{\sigma}}
\otimes \hat{b}.{\mathbf{\sigma}}|{\psi}^-_{1/2}\rangle 
 \nonumber\\&=& \frac{-1}{4(4\pi)^2}\int d\lambda \rm {sgn}(\hat{a}.\hat{\lambda})\int d\mu \rm {sgn}\left[\hat{b}.(\hat\lambda+c\hat\mu)\right]\nonumber\\&=&\frac{-1}{4(4\pi)^2}\int d\lambda \rm {sgn}(\hat{a}.\hat{\lambda})\frac{1}{2}\int d\mu \{\rm {sgn}\left[\hat{b}.(\hat{\mu} +\hat{\lambda})\right]-\rm {sgn}\left[\hat{b}.(\hat{\mu} - \hat{\lambda})\right]\}\nonumber\\&+& \frac{-1}{4(4\pi)^2}\int d\mu \rm {sgn}(\hat{a}.\hat{\mu})\frac{1}{2}\int d\lambda \{\rm {sgn}\left[\hat{b}.(\hat{\mu} +\hat{\lambda})\right]+\rm {sgn}\left[\hat{b}.(\hat{\mu} - \hat{\lambda})\right]\}\nonumber\\&=&\frac{-2}{4(4\pi)^2}
\int_{\hat{\lambda} \in S_2}~ \int_{\hat{\mu} \in S_2}~
\rm {sgn}(\hat{a}.\hat{\lambda}) \rm {sgn}\left[\hat{b}.(\hat{\mu} - \hat{\lambda})\right]
d\hat{\lambda} d\hat{\mu} \nonumber\\ &=& \frac{-2}{4(4\pi)} \int_{\hat{\lambda} \in
S_2}~ \rm {sgn}(\hat{a}.\hat{\lambda}) \hat{b}.\hat{\lambda} d\hat{\lambda} 
=-\frac{1}{4}\hat{a}.\hat{b}\;.
\end{eqnarray}
     
\section{Simulation of Two Spin-$s$ Singlet Correlations for All $s$ Involving Spin Measurements-I}
Classical simulation of quantum correlations accomplished for spin-$1/2$ singlet state, requiring the optimal amount, namely, \textit{one} cbit of classical communication in the worst-case scenario, using arbitrary projective measurement on each site \cite{toner03}. It is important to know how does the amount of this classical communication change with the
change in the value of the spin $s$, in order to quantify the
advantage offered by quantum communication over the classical one.
Further, this communication cost quantifies, in terms of classical
resources, the variation of the nonlocal character of quantum
correlations with spin values.
In the present section, we give a classical protocol to exactly simulate quantum correlations implied by the spin-$s$ singlet state corresponding to all integer as well as half-integer spin values $s$. The class of measurements we consider here are only those corresponding to spin observables ( i.e., measurement of observables of the form
${\hat{a}}.{\vec{\Lambda}}$ where $\hat{a}$ is any unit vector in
$\mathbb{R}^3$ and $\vec{\Lambda} = ({\Lambda}_x, {\Lambda}_y,
{\Lambda}_z)$ with each ${\Lambda}_i$ being a $(2s + 1) \times (2s +
1)$ traceless Hermitian matrix and the all three together form the
$SU(2)$ algebra). The required amount of communication is found to be $\left\lceil \log_{2}(s+1)\right\rceil$
in the worst case scenario, where $\left\lceil x\right\rceil$ is the least integer greater than or equal to $x$. In the simulation of the
measurement of the observable $\hat{a}.{\vec{S}}$ ( where $\hat{a} \in
\mathbb{R}^3$ is the supplied direction of measurement), Alice will
have to reproduce the $2s + 1$ number of outcomes $\alpha = s, s -
1, \ldots, - s + 1, - s$ with equal probability. Similarly, Bob will
have to reproduce the $2s + 1$ number of outcomes $\beta = s, s - 1,
\ldots, - s + 1, - s$ with equal probability. We will describe our
protocol for the simulation by first giving the ones for smaller
values of the spin and then by giving the protocol for general value
of the spin.
Before describing the simulation scheme, we mention here few
mathematical results which will be frequently needed during our
discussion of the simulation scheme. Consider the unit sphere in
three dimensional Euclidean space: $S_2 = \{|{\vec r}| = 1 : {\vec r}
\in \mathbb{R}^3\}$. Let ${\hat{\lambda}}_1$, ${\hat{\lambda}}_2$,
${\hat{\mu}}_1$, ${\hat{\mu}}_2$, ${\hat{\nu}}_1$, ${\hat{\nu}}_2$
be ( mutually ) independent but uniformly distributed random variables
on $S_2$. Let $\hat{a}$ and $\hat{b}$ be given any two elements from
$S_2$. Also $\hat{z}$ be the unit vector along the $z$-axis of the
rectangular Cartesian co-ordinate axes $x$, $y$ and $z$ -- the
associated reference frame. Let us define:
$$c_k = \rm {sgn} (\hat{a}.{\hat{\lambda}}_k)~ {\rm sgn}
(\hat{a}.{\hat{\mu}}_k)~~ (k = 1, 2),$$
$$f_k = {\rm sgn} \left(\hat{z}.{\hat{\nu}}_k + p_k\right)~~ (p_k \in
(0, 1)),$$ where ${\rm sgn} :\mathbb{R} \rightarrow \{+1, -1\}$ is
the function defined as ${\rm sgn} (x) = 1$ if $x \ge 0$ and ${\rm
sgn} (x) = -1$ if $x < 0$. One can show that ( see ref.
\cite{toner03} and previous section  for the derivations):
\begin{eqnarray}
\label{alphades1} 
\langle \rm {sgn}(\hat{a}.{\hat{\lambda}}_k)\rangle &=& \frac{1}{4\pi}\int_0^{2\pi}d\phi_k \int_0^\pi \sin\theta_k d\theta_k  \rm {sgn}(\hat a.\hat \lambda_k)\nonumber\\&=&\frac{1}{2}\int_0^\pi \sin\theta_k d\theta_k \rm {sgn}(\hat a.\hat \lambda_k)=0~~(k = 1, 2),
\end{eqnarray}
and hence
\begin{equation}
\label{alphadescription} {\rm Prob} \left(\rm {sgn}
(\hat{a}.{\hat{\lambda}}_k ) = {\pm}1\right) =
\frac{1}{2}~~~ (k = 1, 2).
\end{equation}
We know from previous section:
\begin{equation}
\label{betades1} \left\langle \rm {sgn}
\left[\hat{b}.\left({\hat{\lambda}}_k +
c_k{\hat{\mu}}_k\right)\right]\right \rangle = 0~~ (k =
1, 2),
\end{equation}
so
\begin{equation}
\label{betadescription}
 \rm {Prob} \left( \rm {sgn} \left[ \hat b.(\hat \lambda_k +c_k\hat\mu_k)\right]=\pm 1\right)=\frac{1}{2}.
 \end{equation}
 By using Equation \ref{alphabetaaverage}, we get:
\begin{equation}
\label{alphabeta} \left\langle {\rm sgn}
\left(\hat{a}.{\hat{\lambda}}_k\right) \times {\rm sgn}
\left[\hat{b}.\left({\hat{\lambda}}_l +
c_l{\hat{\mu}}_l\right)\right] \right\rangle =
{\delta}_{kl}\left(\hat{a}.\hat{b}\right)~~ (k, l = 1,
2).
\end{equation}
 Also we have ( taking ${\hat{\nu}}_k = ({\rm sin}
{\theta}_k~ {\rm cos} {\phi}_k,~ {\rm sin} {\theta}_k~ {\rm sin}
{\phi}_k,~ {\rm cos} {\theta}_k)$)
\begin{equation}
\label{f+} {\rm Prob} \left(f_k = +1\right) =
\frac{1}{4\pi}\int_{{\phi}_k = 0}^{{\phi}_k = 2\pi} \int_{{\theta}_k
= 0}^{{\theta}_k = {\rm cos}^{-1} (-p_k)} {\rm sin} {\theta}_k
d{\theta}_k d{\phi}_k = \frac{1 + p_k}{2},
\end{equation}
and hence
\begin{equation}
\label{f-} {\rm Prob} \left(f_k = -1\right) = \frac{1 - p_k}{2}~~
(k = 1, 2).
\end{equation}
So \begin{equation} \label{faverage} \left\langle f_k \right\rangle
= p_k~~ (k = 1, 2).
\end{equation}
Moreover, as $f_k^2$ will always have the value $+1$, therefore
$$\left\langle f_k^2 \right\rangle = 1~~ (k = 1, 2).$$
Consequently
\begin{equation}
\label{faveragenew} \left\langle \left(1 + f_k\right)^2
\right\rangle = 2\left(1 + p_k\right)~~ (k = 1, 2)
\end{equation}
and ( as ${\hat{\nu}}_1$ and ${\hat{\nu}}_2$ are independent random
variables)
\begin{equation}
\label{f2average} \left\langle \left(1 + f_1\right)^2\left(1 +
f_2\right)^2 \right\rangle = \left\langle \left(1 + f_1\right)^2
\right\rangle  \left\langle \left(1 + f_2\right)^2
\right\rangle = 4\left(1 + p_1\right)\left(1 + p_2\right).
\end{equation}
Again, as ${\hat{\lambda}}_1$, ${\hat{\lambda}}_2$, ${\hat{\mu}}_1$,
${\hat{\mu}}_2$, ${\hat{\nu}}_1$, ${\hat{\nu}}_2$ are independent
random variables, therefore
\begin{equation}
\label{genaverage} \left\langle \left(1 + f_k\right)^2 \times {\rm
sgn} \left(\hat{a}.{\hat{\lambda}}_l\right) \times {\rm sgn}
\left[\hat{b}.\left({\hat{\lambda}}_m +
c_l{\hat{\mu}}_m\right)\right] \right\rangle = 2\left(1 +
p_k\right){\delta}_{lm}\left(\hat{a}.\hat{b}\right),
\end{equation}
and
\begin{eqnarray}
\label{gen2average} \left\langle \left(1 + f_1\right)^2\left(1 +
f_2\right)^2 \times {\rm sgn} \left(\hat{a}.{\hat{\lambda}}_k\right)
\times {\rm sgn} \left[\hat{b}.\left({\hat{\lambda}}_l +
c_l{\hat{\mu}}_l\right)\right] \right\rangle &=& \nonumber \\4\left(1 +
p_1\right)\left(1 +
p_2\right){\delta}_{kl}\left(\hat{a}.\hat{b}\right).
\end{eqnarray}
\subsection{Examples}
For each value $s$ of the spin, we can always find a positive
integer $n$ such that $2^{n - 1} < s + 1 \le 2^n$. We show here
below that the above-mentioned simulation can be done with just $n$
bits of communication if $s$ is such that $2^{n - 1} < s + 1 \le
2^n$. To give a clear picture, let us first describe our protocol
for few lower values of $s$, and after that, the general protocol
will be given. To start with, Alice and Bob fix a common reference
frame ( with rectangular Cartesian co-ordinate axes $x$, $y$ and $z$)
for them.\\
\vspace{0.2cm} {\noindent {\bf Example 1:} $2^{1 - 1} < s + 1 \le
2^1$.\\ Thus the allowed values of $s$ are $1/2$ and $1$.}\\
\vspace{0.2cm} {\underline{{\bf Case (1.1)} $s = 1/2$:}\\
Alice and Bob {\it a priori} share two independent and uniformly
distributed random variables ${\hat{\lambda}}_{1/2}$,
${\hat{\mu}}_{1/2} \in S_2$. Given the measurement direction
$\hat{a} \in S_2$, Alice calculates her output as $$\alpha = -(1/2)
{\rm sgn} (\hat{a}.{\hat{\lambda}}_{1/2}) \equiv -\alpha(1/2)
(\rm say). $$She also sends the bit value $c_{1/2} = {\rm sgn}
(\hat{a}.{\hat{\lambda}}_{1/2})~ {\rm sgn} (\hat{a}.
{\hat{\mu}}_{1/2})$ to Bob by classical communication. After
receiving this bit value and using the supplied measurement
direction $\hat{b} \in S_2$, Bob now calculates his output as $$\beta
= (1/2) {\rm sgn} [\hat{b}.({\hat{\lambda}}_{1/2} +
c_{1/2}{\hat{\mu}}_{1/2})] \equiv \beta(1/2) (\rm say).$$ It is known
that ( see equations (\ref{alphades1}) - (\ref{alphabeta})) for the
two spin-$1/2$ singlet state $|{\psi}^-_{1/2}\rangle$, $\alpha,
\beta \in \{+1/2, -1/2\}$, ${\rm Prob} (\alpha = {\pm}1/2) = {\rm
Prob} (\beta = {\pm}1/2) = 1/2$.
So 
\begin{eqnarray}
\langle \alpha \rangle &=&0,\nonumber\\ \langle \beta \rangle &=& 0,\\  \langle \alpha \beta \rangle &=&-(1/3)(1/2)(1/2 + 1)\hat{a}.\hat{b} = \langle \alpha \beta \rangle_{QM}. \nonumber
\end{eqnarray}
Thus the total number of cbits required ( we denote it
by $n_c$), for simulating the measurement correlation in the worst
case scenario, is one and the total number of shared random variable
is two: ${\lambda}_{1/2}$ and ${\mu}_{1/2}$. Thus here $n_{\lambda}
\equiv$ the total number of $\hat{\lambda}$'s $= 1$ and $n_{\mu}
\equiv$ the total number of $\hat{\mu}$'s $= 1$.\\
\vspace{0.2cm} {\underline{{\bf Case (1.2)} $s = 1$:}\\
Alice and Bob {\it a priori} share three independent and uniformly
distributed random variables ${\hat{\lambda}}_{1}$,
${\hat{\mu}}_{1}$, ${\hat{\nu}}_{1} \in S_2$. Given the measurement
direction $\hat{a} \in S_2$, Alice calculates her output as $$\alpha
= -((1 + f_1)/2) {\rm sgn} (\hat{a}.{\hat{\lambda}}_{1}) \equiv
-\alpha(1) (\rm say).$$ She also sends the bit value $c_{1} = {\rm sgn}
(\hat{a}.{\hat{\lambda}}_{1})~ {\rm sgn} (\hat{a}. {\hat{\mu}}_{1})$
to Bob by classical communication. After receiving this bit value
and using the supplied measurement direction $\hat{b} \in S_2$, Bob
now calculates his output as $$\beta = ((1 + f_1)/2) {\rm sgn}
[\hat{b}.({\hat{\lambda}}_{1} + c_{1}{\hat{\mu}}_{1})] \equiv
\beta(1) (\rm say),$$ where $f_1 = {\rm sgn} (\hat{z}.{\hat{\nu}}_{1} +
1/3)$ and $c_1 = {\rm sgn} (\hat{a}.{\hat{\lambda}}_1)~ {\rm sgn}
(\hat{a}.{\hat{\mu}}_1)$. Now, by equations (\ref{f+}) -
(\ref{faverage}), we have ${\rm Prob} (f_1 = + 1) = 2/3$, ${\rm
Prob} (f_1 = - 1) = 1/3$ and $\langle f_1 \rangle = 1/3$. Thus we
see that ( using equations (\ref{alphadescription}),
(\ref{betadescription}), the probability distribution of $f_1$, and
the fact that ${\hat{\lambda}}_1$, ${\hat{\mu}}_1$, ${\hat{\nu}}_1$
are independent random variables) $\alpha, \beta \in \{+1, 0, -1\}$
and ${\rm Prob} (\alpha = j) = {\rm Prob} (\beta = k) = 1/3$ for all
$j, k \in \{+1, 0, -1\}$. Also we have ( using equation
(\ref{genaverage}))
\begin{equation}
 \langle \alpha \beta \rangle = -(1/3) \times 1
\times (1 + 1)\hat{a}.\hat{b} = \langle \alpha \beta \rangle_{QM}.
\end{equation}
Thus here $n_c = 1$, $n_{\lambda} = 1$, $n_{\mu} = 1$, $n_{\nu}
\equiv$ the total number of $\hat{\nu}$'s $= 1$.\\

\vspace{0.4cm} {\noindent {\bf Example 2:} $2^{2 - 1} < s + 1 \le
2^2$.\\ Here the allowed values of $s$ are $3/2$, $2$, $5/2$, and
$3$.}\\
\vspace{0.2cm} {\underline{{\bf Case (2.1)} $s = 3/2$:}\\
Alice and Bob {\it a priori} share four independent and uniformly
distributed random variables ${\hat{\lambda}}_{1/2}$,
${\hat{\lambda}}_{3/2}$, ${\hat{\mu}}_{1/2}$, ${\hat{\mu}}_{3/2} \in
S_2$. Given the measurement direction $\hat{a} \in S_2$, Alice
calculates her output as $$\alpha = - [{\rm sgn}
(\hat{a}.{\hat{\lambda}}_{3/2}) + \alpha(1/2)] \equiv -\alpha(3/2)
(\rm say),$$ where $\alpha(1/2)$ involves ${\hat{\lambda}}_{1/2}$ and is
described in case (1.1) above. She also sends the two bit values $c_{k} =
{\rm sgn} (\hat{a}.{\hat{\lambda}}_{k})~ {\rm sgn} (\hat{a}.
{\hat{\mu}}_{k})$ (for $k = 1/2, 3/2$) to Bob by classical
communication. After receiving these two bit values and using the
supplied measurement direction $\hat{b} \in S_2$, Bob now calculates
his output as $$\beta = {\rm sgn} [\hat{b}.({\hat{\lambda}}_{3/2} +
c_{3/2}{\hat{\mu}}_{3/2})] + \beta(1/2) \equiv \beta(3/2) (\rm say),$$
where $\beta(1/2)$ involves ${\hat{\lambda}}_{1/2}$,
${\hat{\mu}}_{1/2}$ and is described in case (1.1) above. Using equations
(\ref{alphadescription}) and (\ref{betadescription}), and using the
fact that ${\hat{\lambda}}_{1/2}$, ${\hat{\lambda}}_{3/2}$,
${\hat{\mu}}_{1/2}$, ${\hat{\mu}}_{3/2}$ are independent and
uniformly distributed random variables on $S_2$, we have ${\rm Prob}
(\alpha = j) = {\rm Prob} (\beta = k) = 1/4$ for all $j, k \in
\{+3/2, +1/2, -1/2, -3/2\}$. Also, by using equation
(\ref{alphabeta}), we have
\begin{equation}
 \langle \alpha \beta \rangle = -(1/3)
\times (3/2) \times (3/2 + 1)\hat{a}.\hat{b} = \langle \alpha \beta
\rangle_{QM}.
\end{equation}
Thus here $n_c = 2$, $n_{\lambda} = 2$, $n_{\mu} = 2$
and $n_{\nu} = 0$.\\
\vspace{0.2cm} {\underline{{\bf Case (2.2)} $s = 2$:}\\
Alice and Bob {\it a priori} share five independent and uniformly
distributed random variables ${\hat{\lambda}}_{1/2}$,
${\hat{\lambda}}_{2}$, ${\hat{\mu}}_{1/2}$, ${\hat{\mu}}_{2}$,
${\hat{\nu}}_{2} \in S_2$. Given the measurement direction $\hat{a}
\in S_2$, Alice calculates her output as $$\alpha = - ((1 +
f_2)/2)[(3/2){\rm sgn} (\hat{a}.{\hat{\lambda}}_{2})
 + \alpha(1/2)]
\equiv -\alpha(2) (\rm say),$$ where $\alpha(1/2)$ involves
${\hat{\lambda}}_{1/2}$ and is described in case (1.1) above. She also
sends the two bit values $c_{k} = {\rm sgn}
(\hat{a}.{\hat{\lambda}}_{k})~ {\rm sgn} (\hat{a}. {\hat{\mu}}_{k})$
(for $k = 1/2, 2$) to Bob by classical communication. After
receiving these two bit values and using the supplied measurement
direction $\hat{b} \in S_2$, Bob now calculates his output as $$\beta
= ((1 + f_2)/2)[(3/2){\rm sgn} [\hat{b}.({\hat{\lambda}}_{2} +
c_{2}{\hat{\mu}}_{2})] + \beta(1/2)] \equiv \beta(2) (\rm say),$$
where $\beta(1/2)$ involves ${\hat{\lambda}}_{1/2}$,
${\hat{\mu}}_{1/2}$ and is described in case (1.1) above. Here $f_2 =
{\rm sgn} (\hat{z}.{\hat{\nu}}_{2} + 3/5)$. By using equations
(\ref{f+}) - (\ref{faverage}), we see that ${\rm Prob} (f_2 = +1) =
4/5$, ${\rm Prob} (f_2 = -1) = 1/5$ and $\langle f_2 \rangle = 3/5$.
Using these facts and the fact that ${\hat{\lambda}}_{1/2}$,
${\hat{\lambda}}_{2}$, ${\hat{\mu}}_{1/2}$, ${\hat{\mu}}_{2}$,
${\hat{\nu}}_{2}$ are independent and uniformly distributed random
variables on $S_2$, we have ${\rm Prob} (\alpha = j) = {\rm Prob}
(\beta = k) = 1/5$ for all $j, k \in \{+2, +1, 0, -1, -2\}$. Also,
by using equation (\ref{genaverage})
\begin{equation}
 \langle \alpha \beta \rangle =
-(1/3) \times 2 \times (2 + 1)\hat{a}.\hat{b} = \langle \alpha \beta
\rangle_{QM}. 
\end{equation}
Thus here $n_c = 2$, $n_{\lambda} = 2$, $n_{\mu} = 2$
and $n_{\nu} = 1$.\\

\vspace{0.2cm} {\underline{{\bf Case (2.3)} $s = 5/2$:}\\
Alice and Bob {\it a priori} share five independent and uniformly
distributed random variables ${\hat{\lambda}}_{1}$,
${\hat{\lambda}}_{5/2}$, ${\hat{\mu}}_{1}$, ${\hat{\mu}}_{5/2}$,
${\hat{\nu}}_{1} \in S_2$. Given the measurement direction $\hat{a}
\in S_2$, Alice calculates her output as $$\alpha = - [(3/2){\rm sgn}
(\hat{a}.{\hat{\lambda}}_{5/2}) + \alpha(1)] \equiv -\alpha(5/2)
(\rm say),$$ where $\alpha(1)$ involves ${\hat{\lambda}}_{1}$,
${\hat{\nu}}_1$ and is described in case (1.2) above. She also sends the
two bit values $c_{k} = {\rm sgn} (\hat{a}.{\hat{\lambda}}_{k})~
{\rm sgn} (\hat{a}. {\hat{\mu}}_{k})$ (for $k = 1, 5/2$) to Bob by
classical communication. After receiving these two bit values and
using the supplied measurement direction $\hat{b} \in S_2$, Bob now
calculates his output as $$\beta = (3/2){\rm sgn}
[\hat{b}.({\hat{\lambda}}_{5/2} + c_{5/2}{\hat{\mu}}_{5/2})] +
\beta(1) \equiv \beta(5/2) (\rm say), $$ where $\beta(1)$ involves
${\hat{\lambda}}_{1}$, ${\hat{\mu}}_{1}$, ${\hat{\nu}}_1$ and is
described in case (1.2) above. Using the fact that ${\hat{\lambda}}_{1}$,
${\hat{\lambda}}_{5/2}$, ${\hat{\mu}}_{1}$, ${\hat{\mu}}_{5/2}$,
${\hat{\nu}}_{1}$ are independent and uniformly distributed random
variables on $S_2$, equations (\ref{alphadescription}) and
(\ref{betadescription}), and the discussions in (1.2) above, we have
${\rm Prob} (\alpha = j) = {\rm Prob} (\beta = k) = 1/6$ for all $j,
k \in \{+5/2, +3/2, +1/2, -1/2, -3/2, -5/2\}$. Also, by using
equation (\ref{genaverage})
\begin{equation} 
\langle \alpha \beta \rangle = -(1/3)
\times (5/2) \times (5/2 + 1)\hat{a}.\hat{b} = \langle \alpha \beta
\rangle_{QM}.
\end{equation}
Thus here $n_c = 2$, $n_{\lambda} = 2$, $n_{\mu} = 2$
and $n_{\nu} = 1$.\\

\vspace{0.2cm} {\underline{{\bf Case (2.4)} $s = 3$:}\\
Alice and Bob {\it a priori} share six independent and uniformly
distributed random variables ${\hat{\lambda}}_{1}$,
${\hat{\lambda}}_{3}$, ${\hat{\mu}}_{1}$, ${\hat{\mu}}_{3}$,
${\hat{\nu}}_{1}$, ${\hat{\nu}}_3 \in S_2$. Given the measurement
direction $\hat{a} \in S_2$, Alice calculates her output as $$\alpha
= - ((1 + f_3)/2)[2{\rm sgn} (\hat{a}.{\hat{\lambda}}_{3}) +
\alpha(1)] \equiv -\alpha(3) (\rm say),$$ where $\alpha(1)$ involves
${\hat{\lambda}}_{1}$, ${\hat{\nu}}_1$ and is described in case (1.2)
above. Here $f_3 = {\rm sgn} (\hat{z}.{\hat{\nu}}_3 + 5/7)$. She
also sends the two bit values $c_{k} = {\rm sgn}
(\hat{a}.{\hat{\lambda}}_{k})~ {\rm sgn} (\hat{a}. {\hat{\mu}}_{k})$
(for $k = 1, 3$) to Bob by classical communication. After receiving
these two bit values and using the supplied measurement direction
$\hat{b} \in S_2$, Bob now calculates his output as $$\beta = ((1 +
f_3)/2)[2{\rm sgn} [\hat{b}.({\hat{\lambda}}_{3} +
c_{3}{\hat{\mu}}_{3})] + \beta(1)] \equiv \beta(3) (\rm say),$$ where
$\beta(1)$ involves ${\hat{\lambda}}_{1}$, ${\hat{\mu}}_{1}$,
${\hat{\nu}}_1$ and is described in case (1.2) above. Using the fact that
${\hat{\lambda}}_{1}$, ${\hat{\lambda}}_{3}$, ${\hat{\mu}}_{1}$,
${\hat{\mu}}_{3}$, ${\hat{\nu}}_{1}$, ${\hat{\nu}}_3$ are
independent and uniformly distributed random variables on $S_2$,
equations (\ref{alphadescription}) and (\ref{betadescription}), and
the discussions in (1.2) above, we have ${\rm Prob} (\alpha = j) =
{\rm Prob} (\beta = k) = 1/7$ for all $j, k \in \{+3, +2, +1, 0, -1,
-2, -3\}$. Also, by using equations (\ref{genaverage}) and
(\ref{gen2average}), we have
\begin{equation}
\langle \alpha \beta \rangle = -(1/3)
\times 3 \times (3 + 1)\hat{a}.\hat{b} = \langle \alpha \beta
\rangle_{QM}. 
\end{equation}
Thus here $n_c = 2$, $n_{\lambda} = 2$, $n_{\mu} = 2$
and $n_{\nu} = 2$.
\subsection{General Simulation Scheme}
Let us now describe the protocol for general $s$. One can always
find out uniquely a positive integer $n$ such that $2^{n - 1} < s +
1 \le 2^n$. Equivalently, given the dimension $d = 2s + 1$ of the
Hilbert space, one can always find out a unique positive integer $n$
such that $2^{n} - 1 < d \le 2^{n + 1} - 1$. Let $d = a_02^n +
a_12^{n - 1} + \ldots + a_n2^0 \equiv \underline{a_0a_1 \ldots a_n}$
be the binary representation of $d$ (where $a_0$, $a_1$, $\ldots$,
$a_n \in \{0, 1\}$). So we must have $a_0 \ne 0$. Before describing
the general simulation scheme, using the help of the above-mentioned
examples, let us describe below the scheme pictorially (see Figure
1) in terms of binary representation of the dimension of the
individual spin system. The simulation scheme, we have described in
ref. \cite{ali07} for the simulation of the measurement correlation
in two spin-$s$ singlet state, where $2s + 1 = 2^n$, corresponds to
the upper most chain
$$2^1 = \underline{10} \rightarrow 2^2 = \underline{100} \rightarrow
\ldots \rightarrow 2^{n - 1} =
\underline{1000 \ldots 00} \rightarrow 2^n = \underline{1000 \ldots
000}$$ in Figure 1. In other words, when $2s + 1 = 2^n$, given the
measurement directions $\hat{a}$, Alice will calculate her output
$-\alpha\left(\frac{2^n - 1}{2}\right) \equiv
-\alpha\left(\frac{\underline{1000 \ldots 000} - 1}{2}\right)$ as:
\begin{eqnarray*}
&-\alpha\left(\frac{\underline{1000 \ldots 000} - 1}{2}\right) =&\\
&-\left[\left(\frac{\frac{\underline{1000 \ldots 000} - 1}{2} +
\frac{1}{2}}{2}\right) {\rm sgn}
\left(\hat{a}.{\hat{\lambda}}_{\frac{\underline{1000 \ldots 000} -
1}{2}}\right) +  \alpha\left(\frac{\underline{1000 \ldots 00} -
1}{2}\right)\right]&\\
%\end{eqnarray*}
&= -[\left(\frac{\frac{\underline{1000 \ldots 000} - 1}{2} +
\frac{1}{2}}{2}\right) {\rm sgn}
\left(\hat{a}.{\hat{\lambda}}_{\frac{\underline{1000 \ldots 000} -
1}{2}}\right) + \left(\frac{\frac{\underline{1000 \ldots 00} - 1}{2}
+ \frac{1}{2}}{2}\right) {\rm sgn}
\left(\hat{a}.{\hat{\lambda}}_{\frac{\underline{1000 \ldots 00} -
1}{2}}\right) +&\nonumber\\
&\alpha\left(\frac{\underline{1000 \ldots 0}
- 1}{2}\right)]&
\end{eqnarray*}
$$\ldots$$
$$\ldots$$
\begin{eqnarray*}
&= -[\left(\frac{\frac{\underline{1000 \ldots 000} - 1}{2} +
\frac{1}{2}}{2}\right) {\rm sgn}
\left(\hat{a}.{\hat{\lambda}}_{\frac{\underline{1000 \ldots 000} -
1}{2}}\right) + \left(\frac{\frac{\underline{1000 \ldots 00} - 1}{2}
+ \frac{1}{2}}{2}\right) {\rm sgn}
\left(\hat{a}.{\hat{\lambda}}_{\frac{\underline{1000 \ldots 00} -
1}{2}}\right) +& \\
&\ldots + \left(\frac{\frac{\underline{10} -
1}{2} + \frac{1}{2}}{2}\right) {\rm sgn}
\left(\hat{a}.{\hat{\lambda}}_{\frac{\underline{10} -
1}{2}}\right)]&\\
&= -\frac{1}{2}\sum_{k = 1}^{n} 2^{n - k} {\rm sgn} \left(\hat{a}.{\hat{\eta}}_k\right),&
\end{eqnarray*}
where ${\hat{\eta}}_k = {\hat{\lambda}}_{\frac{2^k - 1}{2}}$.
Similarly for Bob. We have generalized below this scheme to
arbitrary value of $s$ (see equations (\ref{halfintegeralpha}) -
(\ref{integerbeta})).
\begin{figure}
\begin{center}
\includegraphics[width=14cm,height=18cm]{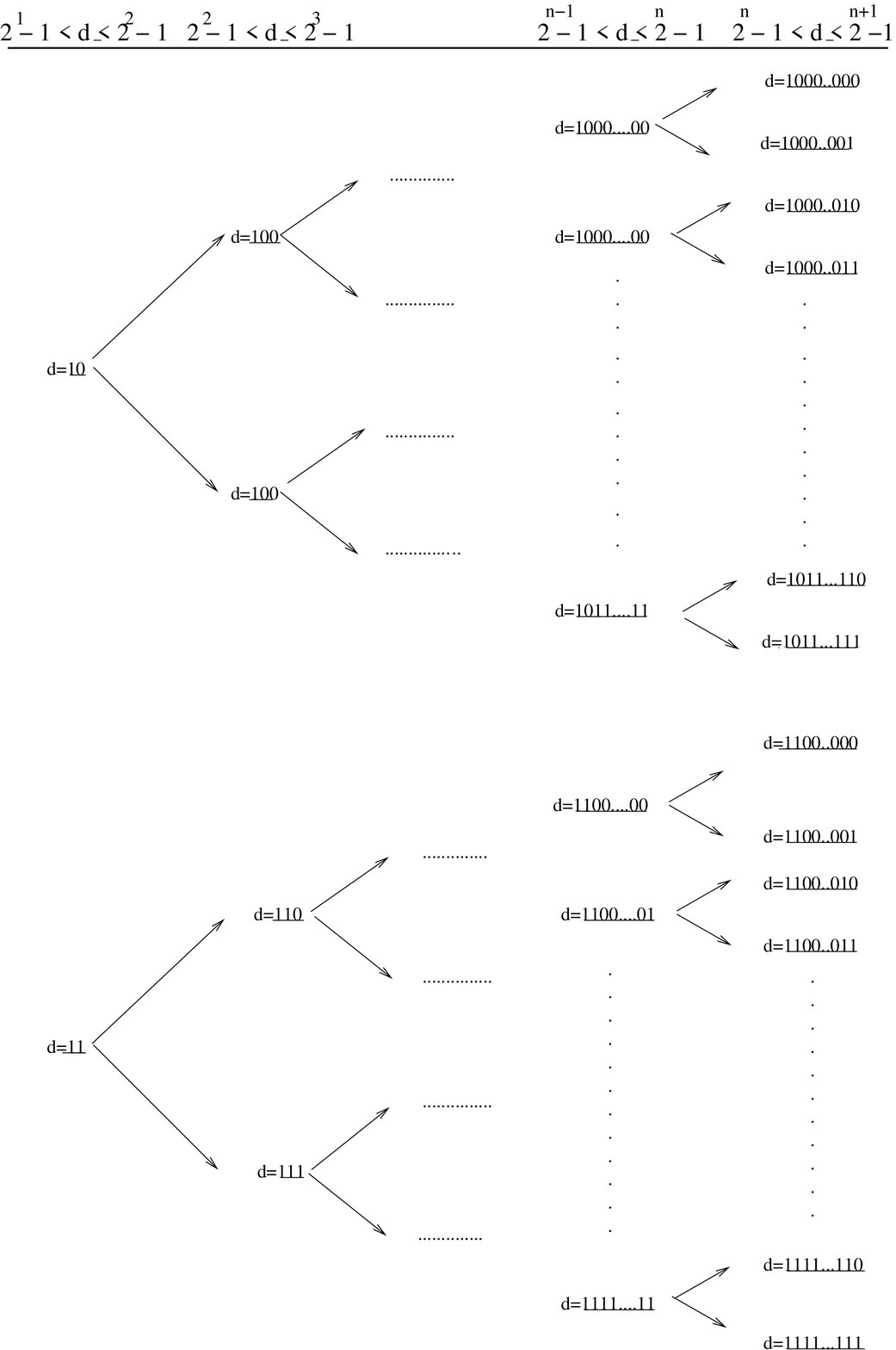}

\vspace{0.4cm}
\underline{Figure 1 :} \small{ The paths (mentioned by
concatenated arrows from left to right) of simulation for each
integer and half-integer spins $s$ such that $2^{n - 1} < s + 1 \le
2^n$}
\end{center}
\end{figure}
\vspace{0.8cm}
\newpage

To describe the general simulation, we consider the following two
cases:

\vspace{0.2cm} {\noindent \underline{\bf $s$ is a half-integer
spin:}}}
Over and above the $n - 1$ number of ${\hat{\lambda}}$'s, $n - 1$
number of ${\hat{\mu}}$'s and $(a_1 + a_2 + \ldots + a_{n - 1})$
number of ${\hat{\nu}}$'s appeared in the expression for
$\alpha\left(\frac{\underline{a_0a_1 \ldots a_{n - 1}} -
1}{2}\right)$ and $\beta\left(\frac{\underline{a_0a_1 \ldots a_{n -
1}} - 1}{2}\right)$, Alice and Bob share the random variables
${\hat{\lambda}}_{\frac{\underline{a_0a_1 \ldots a_n} - 1}{2}}$ and
${\hat{\mu}}_{\frac{\underline{a_0a_1 \ldots a_n} - 1}{2}}$, where,
it has been assumed that all these $2n + (a_1 + a_2 + \ldots a_{n -
1})$ number of random variables are independent and uniformly
distributed on $S_2$. Let us denote the set of all these $n$
${\hat{\lambda}}$'s by $S_{\lambda}$, the set of all these $n$
${\hat{\mu}}$'s by $S_{\mu}$, and the set of all these $(a_1 + a_2 +
\ldots + a_{n - 1})$ ${\hat{\nu}}$'s by $S_{\nu}$. Given the
measurement direction $\hat{a} \in S_2$, Alice calculates her output
as
 $$\alpha =
-\left[\left(\frac{\frac{\underline{a_0a_1 \ldots a_n} - 1}{2} +
\frac{1}{2}}{2}\right) {\rm sgn}
\left(\hat{a}.{\hat{\lambda}}_{\frac{\underline{a_0a_1 \ldots a_n} -
1}{2}}\right) + \alpha\left(\frac{\underline{a_0a_1 \ldots a_{n -
1}} - 1}{2}\right)\right]$$
\begin{equation}
\label{halfintegeralpha}
\equiv -\alpha\left(\frac{\underline{a_0a_1
\ldots a_n} - 1}{2}\right),
\end{equation} and she sends the $n$ cbits
\begin{equation}
\label{cbitshalf} c_{k} = {\rm sgn}
(\hat{a}.{\hat{\lambda}}_k)~ {\rm sgn} (\hat{a}.{\hat{\mu}}_k),
\end{equation}
to Bob where $k = \frac{\underline{a_0a_1 \ldots a_n} - 1}{2},
\frac{\underline{a_0a_1 \ldots a_{n - 1}} - 1}{2}, \ldots,
\frac{\underline{a_0a_1} - 1}{2}$. After receiving these $n$ cbits
and using his measurement direction $\hat{b} \in S_2$, Bob
calculates his output as
$$\beta =
\left[\left(\frac{\frac{\underline{a_0a_1 \ldots a_n} - 1}{2} +
\frac{1}{2}}{2}\right) {\rm sgn}
\left[\hat{b}.\left({\hat{\lambda}}_{\frac{\underline{a_0a_1 \ldots
a_n} - 1}{2}} + c_{\frac{\underline{a_0a_1 \ldots a_n} -
1}{2}}{\hat{\mu}}_{\frac{\underline{a_0a_1 \ldots a_n} -
1}{2}}\right)\right]\right.$$
\begin{equation}
\label{halfintegerbeta} \left.+ \beta\left(\frac{\underline{a_0a_1
\ldots a_{n - 1}} - 1}{2}\right)\right] \equiv
\beta\left(\frac{\underline{a_0a_1 \ldots a_n} - 1}{2}\right).
\end{equation}
Let $L = a_1 + a_2 + \ldots a_n$ and let $i_1$, $i_2$, $\ldots$,
$i_L$ be all those elements from $\{1, 2, \ldots, n\}$ such that
$i_1 < i_2 < \ldots < i_L$ and $a_{i_1} = a_{i_2} = \ldots = a_{i_L}
= 1$. It is then easy to see that
$$S_{\lambda} = \left\{{\hat{\lambda}}_{\frac{\underline{a_0a_1} -
1}{2}}, {\hat{\lambda}}_{\frac{\underline{a_0a_1a_2} - 1}{2}},
\ldots, {\hat{\lambda}}_{\frac{\underline{a_0a_1 \ldots a_n} -
1}{2}}\right\},$$
$$S_{\mu} = \left\{{\hat{\mu}}_{\frac{\underline{a_0a_1} -
1}{2}}, {\hat{\mu}}_{\frac{\underline{a_0a_1a_2} - 1}{2}}, \ldots,
{\hat{\mu}}_{\frac{\underline{a_0a_1 \ldots a_n} - 1}{2}}\right\},$$
$$S_{\nu} = \left\{{\hat{\nu}}_{\frac{\underline{a_0a_{i_1}} -
1}{2}}, {\hat{\nu}}_{\frac{\underline{a_0a_{i_1}a_{i_2}} - 1}{2}},
\ldots, {\hat{\nu}}_{\frac{\underline{a_0a_{i_1} \ldots a_{i_L}} -
1}{2}}\right\}.$$

\vspace{0.2cm} {\noindent \underline{\bf $s$ is an integer spin:}}}

Over and above the $n - 1$ number of ${\hat{\lambda}}$'s, $n - 1$
number of ${\hat{\mu}}$'s and $(a_1 + a_2 + \ldots + a_{n - 1})$
number of ${\hat{\nu}}$'s appeared in the expression for
$\alpha\left(\frac{\underline{a_0a_1 \ldots a_{n - 1}} -
1}{2}\right)$ and $\beta\left(\frac{\underline{a_0a_1 \ldots a_{n -
1}} - 1}{2}\right)$, Alice and Bob share the random variables
${\hat{\lambda}}_{\frac{\underline{a_0a_1 \ldots a_n} - 1}{2}}$ and
${\hat{\mu}}_{\frac{\underline{a_0a_1 \ldots a_n} - 1}{2}}$, where,
it has been assumed that all these $2n + (a_1 + a_2 + \ldots a_{n -
1})$ number of random variables are independent and uniformly
distributed on $S_2$. Given the measurement direction $\hat{a} \in
S_2$, Alice calculates her output as
\begin{eqnarray}
&\alpha = -\left(\frac{1 +
f_{\frac{\underline{a_0a_1 \ldots a_n} -
1}{2}}}{2}\right)\times& \nonumber\\&
\left[\left(\frac{\frac{\underline{a_0a_1 \ldots
a_n} - 1}{2} + 1}{2}\right) {\rm sgn}
\left(\hat{a}.{\hat{\lambda}}_{\frac{\underline{a_0a_1 \ldots a_n} -
1}{2}}\right)+ \alpha\left(\frac{\underline{a_0a_1 \ldots a_{n -
1}} - 1}{2}\right)\right]&\nonumber
\end{eqnarray}
\begin{equation}
\label{integeralpha}
\equiv -\alpha\left(\frac{\underline{a_0a_1
\ldots a_n} - 1}{2}\right),
\end{equation} and she sends the $n$ cbits
\begin{equation}
\label{cbitsinteger} c_{k} = {\rm sgn}
(\hat{a}.{\hat{\lambda}}_k)~ {\rm sgn} (\hat{a}.{\hat{\mu}}_k),
\end{equation}
to Bob where $k = \frac{\underline{a_0a_1 \ldots a_n} - 1}{2},
\frac{\underline{a_0a_1 \ldots a_{n - 1}} - 1}{2}, \ldots,
\frac{\underline{a_0a_1} - 1}{2}$. After receiving these $n$ cbits
and using his measurement direction $\hat{b} \in S_2$, Bob
calculates his output as
\begin{eqnarray}
&\beta =
\left(\frac{1 + f_{\frac{\underline{a_0a_1 \ldots a_n} -
1}{2}}}{2}\right)\times& \nonumber \\ &\left[\left(\frac{\frac{\underline{a_0a_1 \ldots
a_n} - 1}{2} + 1}{2}\right) {\rm sgn}
\left[\hat{b}.\left({\hat{\lambda}}_{\frac{\underline{a_0a_1 \ldots
a_n} - 1}{2}} + c_{\frac{\underline{a_0a_1 \ldots a_n} -
1}{2}}{\hat{\mu}}_{\frac{\underline{a_0a_1 \ldots a_n} -
1}{2}}\right)\right]\right.&\nonumber
\end{eqnarray}
\begin{equation}
\label{integerbeta} \left.+ \beta\left(\frac{\underline{a_0a_1
\ldots a_{n - 1}} - 1}{2}\right)\right]
 \equiv \beta\left(\frac{\underline{a_0a_1 \ldots
a_n} - 1}{2}\right).
\end{equation}
Here \begin{equation} \label{fvalue} f_{\frac{\underline{a_0a_1
\ldots a_n} - 1}{2}} = {\rm sgn}
\left(\hat{z}.{\hat{\nu}}_{\frac{\underline{a_0a_1 \ldots a_n} -
1}{2}} + \frac{\underline{a_0a_1 \ldots a_n}-2}{\underline{a_0a_1
\ldots a_n}}\right).
\end{equation}
Let $L = a_1 + a_2 + \ldots
a_n$ and let $i_1$, $i_2$, $\ldots$, $i_L$ be elements from $\{1, 2,
\ldots, n\}$ such that $i_1 < i_2 < \ldots < i_L$ and $a_{i_1} =
a_{i_2} = \ldots = a_{i_L} = 1$. It is then easy to see that
$$S_{\lambda} = \left\{{\hat{\lambda}}_{\frac{\underline{a_0a_1} -
1}{2}}, {\hat{\lambda}}_{\frac{\underline{a_0a_1a_2} - 1}{2}},
\ldots, {\hat{\lambda}}_{\frac{\underline{a_0a_1 \ldots a_n} -
1}{2}}\right\},$$
$$S_{\mu} = \left\{{\hat{\mu}}_{\frac{\underline{a_0a_1} -
1}{2}}, {\hat{\mu}}_{\frac{\underline{a_0a_1a_2} - 1}{2}}, \ldots,
{\hat{\mu}}_{\frac{\underline{a_0a_1 \ldots a_n} - 1}{2}}\right\},$$
$$S_{\nu} = \left\{{\hat{\nu}}_{\frac{\underline{a_0a_{i_1}} -
1}{2}}, {\hat{\nu}}_{\frac{\underline{a_0a_{i_1}a_{i_2}} - 1}{2}},
\ldots, {\hat{\nu}}_{\frac{\underline{a_0a_{i_1} \ldots a_{i_L}} -
1}{2}}\right\}.$$

\vspace{0.2cm} The way we have defined
$\alpha\left(\frac{\underline{a_0a_1 \ldots a_n} - 1}{2}\right)$ as
well as $\beta\left(\frac{\underline{a_0a_1 \ldots a_n} -
1}{2}\right)$ (see examples (1.1) - (2.4) as well as equations
(\ref{halfintegeralpha}), (\ref{halfintegerbeta}),
(\ref{integeralpha}) and (\ref{integerbeta})), one can show
recursively that
$${\rm Prob} \left(\alpha\left(\frac{\underline{a_0a_1 \ldots a_n} -
1}{2}\right) = j\right) = {\rm Prob}
\left(\beta\left(\frac{\underline{a_0a_1 \ldots a_n} - 1}{2}\right)
= k\right)$$  $$= \frac{1}{\underline{a_0a_1 \ldots a_n}}$$ for $j, k \in
\{(\underline{a_0a_1 \ldots a_n} - 1)/2, (\underline{a_0a_1 \ldots
a_n} - 3)/2, \ldots, -(\underline{a_0a_1 \ldots a_n} - 3)/2,\\
-(\underline{a_0a_1 \ldots a_n} - 1)/2\}$ and also
$$\langle \alpha \beta \rangle = \left\langle -\alpha\left(\frac{\underline{a_0a_1 \ldots a_n} -
1}{2}\right) \times \beta\left(\frac{\underline{a_0a_1 \ldots a_n} -
1}{2}\right) \right\rangle$$
$$= -\frac{1}{3} \times
\frac{\underline{a_0a_1 \ldots a_n} - 1}{2} \times
\left(\frac{\underline{a_0a_1 \ldots a_n} - 1}{2} +
1\right)\left(\hat{a}.\hat{b}\right) = \langle \alpha \beta
\rangle_{QM}$$

Thus we see that for any given value of the spin $s$ (integer or
half-integer) for which $2^n - 1 < d = 2s + 1 \le 2^{n + 1} - 1$
(hence $d$ has the binary representation $d = \underline{a_0a_1
\ldots a_n}$ where $a_0$, $a_1$, $\ldots$, $a_n \in \{0, 1\}$ and
$a_0 \ne 0$), Alice and Bob can simulate, in the worst case
scenario,  the measurement correlation in the two spin-$s$ singlet
state $|{\psi}^-_s\rangle$ for performing measurement of arbitrary
spin observables by using only $n = \lceil{\rm log}_2 (s + 1)\rceil$
bits of communication if they {\it a priori} share $2n + (a_1 + a_2
+ \ldots a_n)$ number of independent and uniformly distributed
random variables on $S_2$.

For any maximally entangled state $|{\psi}_{max}\rangle$ of two
spin-$s$ systems, we know that there exists a $(2s + 1) \times (2s +
1)$ unitary matrix $U$ such that $|{\psi}_{max}\rangle = (U \times
I)|{\psi}^-_s\rangle$. Our protocol works equally well for those two
spin-$s$ maximally entangled state $|{\psi}_{max}\rangle$ for each
of which the above-mentioned unitary matrix $U$ induces a rotation
in $\mathbb{R}^3$, as in those cases, both Alice and Bob can
perform the protocol for the spin-£s£ singlet state
$|{\psi}^-_s\rangle$ for the rotated input vectors $\hat{a}$ and
$\hat{b}$ and, hence, they will achieve their goal.

\section{Simulation of Two Spin-$s$ Singlet Correlations for All $s$ Involving Spin Measurement-II}
In this section, we give a classical
protocol to simulate the measurement correlation in a singlet state
of two spin-$s$ systems, considering only measurement of spin
observables where $2s + 1 = P^{n}$, $P$ and $n$ being any positive
integers. Thus, this protocol also classically simulates all spin $s$ singlet correlations, restricted to spin measurements. One can always find out a unique positive integer $m$ such that $2^m -1<P\leq 2^{m+1}-1$. Let $P=a_0 2^m +a_1 2^{m-1}+\ldots+a_m 2^0\equiv\underline{a_0a_1 \ldots a_m}$
be the binary representation of $P$ (where $a_0$, $a_1$, $\ldots$,
$a_m \in \{0, 1\}$). So we must have $a_0 \ne 0$.\\
As $s$ can only take integral or half-integral values,
the allowed values of $s$ form the infinite sequence
$s_{n = 1} = (P-1)/2$, $s_{n = 2} = (P^{2}-1)/2$, $s_{n = 3} = (P^{3}-1)/2$, etc.\\ For $s_n ={P^{n-1} - 1}/{2}$, our protocol requires 
\begin{equation}
nm=\left[log_2P\right]^{-1}\left[log_2(2s+1)\right](\left\lceil log_2(P+1)\right\rceil-1)
\end{equation}
cbits of
communication and 
\begin{equation}
n(2m+\sum_{i=1}^{m} a_i)=\left[log_2P\right]^{-1}\left[log_2(2s+1)\right]\left[2(\left\lceil log_2(P+1)\right\rceil-1)+\sum_{i=1}^{m}a_i\right]
\end{equation}
 number of independent and uniformly distributed shared random variables.\\ 
 To give a clear picture, let us first describe protocol for $2s+1=2^n$ and $2s+1=3^n$, and after that, the general protocol will be given.
 
 \subsection{Case ~ $ 2s+1=2^n$}
In the simulation of the
measurement of the observable $\hat{a}.{\bf S}$ (where $\hat{a} \in
\mathbb{R}^3$ is the supplied direction of measurement), Alice will
have to reproduce the $2^n$ number of outcomes $\alpha = 2^{n - 1} -
1/2, 2^{n - 1} - 3/2, \ldots, - 2^{n - 1} + 1/2$ with equal
probability. If we consider the series $-\frac{1}{2} \sum_{k =
1}^{n} g(k) 2^{n - k}$, where, for each $k$, $g(k)$ can be either
$1$ or $- 1$, it turns out that the series can only take the
above-mentioned $2^n$ different values of $\alpha$. The
probability distribution of these different values of the series
will depend on that of the $n$-tuple $\{g(1), g(2), \ldots, g(n)\}$.
In order to make this probability distribution an uniform one ( which
is essential here for the simulation purpose), we choose here $g(k)=\rm {sgn}(\hat{a}.\hat{\lambda}_k)$ for each $k$, where $\hat{a}$
is the measurement direction for Alice while all share random are independent and uniformly distributed random variables on $S_2$. We have seen in the above-mentioned Toner
and Bacon protocol that if Alice and Bob share the two independent
and uniformly distributed random variables ${\hat{\lambda}}_k
\in S_2$ and ${\hat{\mu}}_k \in S_2$, then the random variable
$r_k \equiv \rm {sgn}[\hat{b}.({\hat{\lambda}}_k +
\rm {sgn}(\hat{a}.{\hat{\lambda}}_k) \rm {sgn}(\hat{a}.{\hat{\mu}}_k)
{\hat{\mu}}_k)]$ is uniformly distributed over $\{1, - 1\}$.
Hence, as above, the quantity $\frac{1}{2} \sum_{k = 1}^{n} 2^{n -
k} \rm {sgn}[\hat{b}.({\hat{\lambda}}_k + r_k {\hat{\mu}}_k)]$
will have $2^n$ different values $\beta = 2^{n - 1} - 1/2, 2^{n - 1}
- 3/2, \ldots, - 2^{n - 1} + 1/2$ all with equal probabilities. But
the interesting point to note is that in the calculation of the
average (over the independent but uniformly distributed random
variables ${\hat{\lambda}}_1$, ${\hat{\lambda}}_2$,
$\ldots$, ${\hat{\lambda}}_n$, ${\hat{\mu}}_1$,
${\hat{\mu}}_2$, $\ldots$, ${\hat{\mu}}_n$) of the product
$\alpha \beta$, there will be no contribution from cross terms like
$\rm {sgn}(\hat{a}.{\hat{\lambda}}_k)
\rm {sgn}[\hat{b}.(\hat{{\lambda}_l} + r_l \hat{{\mu}_l})]$ if $k
\ne l$.
The protocol proceeds as follows:\\ Alice outputs
\begin{equation}
\alpha=-\frac{1}{2}\sum^{n}_{k=1}2^{n-k}\rm {sgn}(\hat{a}.{\hat{\lambda}}_k).
\end{equation}
Alice sends $n$ cbits $c_{1}, c_{2}, \dots, c_{n}$ to Bob where
$c_{k} =
\rm {sgn}(\hat{a}.{\hat{\lambda}}_k)\rm {sgn}(\hat{a}.{\hat{\mu}}_k)$
for $k = 1, 2, \ldots, n$, where ${\hat{\lambda}}_1$,
${\hat{\lambda}}_2$, $\ldots$, ${\hat{\lambda}}_n$,
${\hat{\mu}}_1$, ${\hat{\mu}}_2$, $\ldots$,
${\hat{\mu}}_n$ are $2n$ independent shared random variables
between Alice and Bob, each being uniformly distributed on $S_2$. Thus we see that, in terms of shared randomness, $\lambda = ({\hat{\lambda}}_1, {\hat{\lambda}}_2, \ldots, {\hat{\lambda}}_n, {\hat{\mu}}_1, {\hat{\mu}}_2, \ldots, {\hat{\mu}}_n)$ is the shared random variable between Alice and Bob. After receiving these $n$ cbits from Alice, Bob outputs 
\begin{equation}
\beta = \frac{1}{2}\sum^{n}_{k = 1} 2^{n -
k}\rm {sgn}[\hat{b}.({\hat{\lambda}}_k + c_{k} {\hat{\mu}}_k)].
\end{equation}
It follows immediately from the discussion in the last paragraph 
that
\begin{eqnarray}
\langle \alpha \beta \rangle =& \frac{-1}{4}\frac{1}{(4\pi)^{2n}}\sum^{n}_{k =1}2^{2n - 2k}
\int d{\hat{\lambda}}_1 \ldots d{\hat{\lambda}}_{k -
1}d{\hat{\lambda}}_{k + 1} \ldots
d{\hat{\lambda}}_n  d{\hat{\mu}}_1 \ldots d{\hat{\mu}}_{k -
1}\nonumber\\ & d{\hat{\mu}}_{k + 1}  \ldots  d{\hat{\mu}}_n 
%\begin{equation}
\label{finalcor1} \int_{({\hat{\lambda}}_k, {\hat{\mu}}_k)
\in S_2 \times S_2} \rm {sgn}(\hat{a}.{\hat{\lambda}}_k)
\rm {sgn}[\hat{b}.({\hat{\lambda}}_{k} + d_{k}{\hat{\mu}}_{k})]
d{\hat{\lambda}}_k d{\hat{\mu}}_k.\nonumber\\
\end{eqnarray}
It follows from the discussion in section 4 regarding Toner and
Bacon's work that $$\langle \alpha \beta \rangle = -
\frac{1}{4}\sum^{n}_{k = 1} 2^{2n - 2k} \hat{a}.\hat{b}.$$  Summing
the geometric series $(\sum^{n}_{k = 1} 2^{2n - 2k}=\frac{2^{2n}-1}{3})$and using $(2s + 1) = 2^{n}$ we finally get
\begin{equation}
\label{finalform} \langle \alpha \beta \rangle = - \frac{1}{3} s (s
+ 1) \hat{a}.\hat{b}\;.
\end{equation}
This protocol exactly simulates quantum mechanical probability
distribution for particular types of projective measurements, namely
the spin measurement, on the spin $s$ singlet state with $2s + 1 =
2^{n}$ for positive integer $n$. The above protocol applies to
infinite, although sparse, subset of the set of all spins ( i.e., all
integral and half integral values ). The most important finding is
that the amount of communication goes as $log_{2}(2s + 1)$ or as
$log_{2} s$ for $s \gg 1$. Our protocol works equally for any two
spin-$s$ maximally entangled state as that can be locally unitarily
connected to the singlet state.
Our result provides the amount of classical communication in the
worst case scenario if we consider only measurement of spin
observables on both sides of a two spin-$s$ singlet state with the
restriction that the dimension $2s + 1$ of each subsystem must be a
positive integral power of $2$, and just $n = log_2 (2s + 1)$ bits of
communication from Alice to Bob is sufficient.
\subsection{Case~ $ 2s+1=3^{n}$} 
In  earlier example, it was shown that only ${\rm log}_2 (2s + 1)$ bits of communication is needed, in the worst case scenario, to simulate the measurement correlation of two spin-$s$ singlet state for performing only measurement of spin observables on each site, where $s$ is a
half-integer spin satisfying $2s + 1 = 2^n$. Thus these spin values
do not include integer spins. In the present example we give a
classical protocol to simulate the measurement correlation in a
singlet state of two spin-$s$ systems, considering only ( as above)
measurement of spin observables  where $2s + 1 = 3^{n}$, $n$ being
any positive integer. Note that the allowed values of $s$ form the
infinite sequence of integer spins: $s_{n = 1} = 1$, $s_{n = 2} =
4$, $s_{n = 3} = 13$, etc. For $s_n = 3^n/2 - 1/2$, our protocol
requires $n$ cbits of communication and $3n$ number of independent
and uniformly distributed shared random variables.
In the simulation of the measurement of the observable $\hat{a}.{\bf S}$ ( where $\hat{a} \in
\mathbb{R}^3$ is the supplied direction of measurement), Alice will
have to reproduce the $3^n$ number of outcomes $\alpha = 3^n/2 -
1/2, 3^n/2 - 3/2, \ldots, - 3^n/2 + 3/2, - 3^n/2 + 1/2$ with equal
probability. If we consider the series $- \sum_{k = 1}^{n} 3^{n - k}
g_k$, where, for each $k$, $g_k$ can be either $1$, $0$ or $- 1$, it
turns out that the series can only take the above-mentioned $3^n$
different values of $\alpha$. More specifically, if we vary $(g_1,
g_2, \ldots, g_n)$ over all the members of the $3^n$ - element set
$\{1, 0, -1\}^n$, the series $- \sum_{k = 1}^{n} 3^{n - k} g_k$ will
have the corresponding $3^n$ different values $3^n/2 - 1/2, 3^n/2 -
3/2, \ldots, - 3^n/2 + 3/2, - 3^n/2 + 1/2$. So we have to choose
$g_k$'s in such a way that all the elements $(g_1, g_2, \ldots,
g_n)$ are equally probable. In this direction, we choose Alice's
output as
\begin{equation}
\label{alphachoice3} \alpha = - \sum_{k = 1}^{n} 3^{n - k}
[\frac{1}{2}(1+f_{k})\rm {sgn}(\hat{a}.\hat{\lambda_{k}})],
\end{equation}
where
\begin{eqnarray}
	f_{k}=\rm {sgn}(\hat{z}.\hat{\nu_{k}}+\frac{1}{3}).
\end{eqnarray}
 ${\hat{\lambda}}_1$, ${\hat{\lambda}}_2$, $\ldots$,
${\hat{\lambda}}_n$ ${\hat{\nu}}_{1}$,  $\ldots$,
${\hat{\nu}}_{n}$, are independent and uniformly distributed shared
(between Alice and Bob) random variables on the unit sphere $S_2$ in
$\mathbb{R}^3$. 
So, by this prescription, Alice's output $\alpha$ can only take the
$3^n$ different values $3^n/2 - 1/2$, $3^n/2 - 3/2$, $\ldots$,
$-3^n/2 + 3/2$, $-3^n/2 + 1/2$ with equal probabilities. This is so
because, for given any $\hat{a} \in S_2$, the $n$-tuple
$\frac{1}{2}(1+f_{1})\rm {sgn}(\hat{a}.\hat{\lambda_{1}}),
\frac{1}{2}(1+f_{2})\rm {sgn}(\hat{a}.\hat{\lambda_{2}})), \ldots,
\frac{1}{2}(1+f_{n})\rm {sgn}(\hat{a}.\hat{\lambda_{n}})$ will be uniformly
distributed on $\{1, 0, -1\}^n$ due to the fact that
${\hat{\lambda}}_1$, ${\hat{\lambda}}_2$, $\ldots$,
${\hat{\lambda}}_n$ and ${\hat{\nu}}_{1}$, $\ldots$,
${\hat{\nu}}_{n}$, are independent and uniformly distributed random
variables on $S_2$. The average value of $f_{k}$ is
\begin{eqnarray}
	\left\langle f_{k} \right\rangle&=&\frac{1}{(4\pi)^{3}}\int d\nu_{k}  \rm {sgn}(\hat{z}.\hat{\nu_{k}}+\frac{1}{3})\nonumber\\
	&=&\frac{1}{3}.
\end{eqnarray}
So
\begin{eqnarray}
	\left\langle \alpha\right\rangle=-\sum_{k=1}^{n}3^{n-k}\left\langle f_{k}\right\rangle \left\langle \rm {sgn}(\hat{a}.\hat{\lambda_{k}})\right\rangle=-\sum_{k=1}^{n}3^{n-k}\frac{2}{3}\left\langle \rm {sgn}(\hat{a}.\hat{\lambda_{k}})\right\rangle=0,
\end{eqnarray}
because$$\left\langle \rm {sgn}(\hat{a}.\hat{\lambda_{k}})\right\rangle=0.$$
To start with, Alice and Bob share the $3n$ number of independent
but uniformly distributed (on $S_2$) random variables
${\hat{\lambda}}_1$, ${\hat{\lambda}}_2$, $\ldots$,
${\hat{\lambda}}_n$, ${\hat{\mu}}_1$, ${\hat{\mu}}_2$, $\ldots$,
${\hat{\mu}}_{n}$ and ${\hat{\nu}}_{1}$, ${\hat{\nu}}_{2}$, $\ldots$,
${\hat{\nu}}_{n}$. Given $\hat{a}$ from $S_2$, Alice produces her
output $\alpha$, as described in equation (\ref{alphachoice3}). Alice
then sends the $n$ cbits $c_{1}, c_{2}, \dots, c_{n}$ to Bob where
$c_{k} =
\rm {sgn}(\hat{a}.{\hat{\lambda}}_k)\rm {sgn}(\hat{a}.{\hat{\mu}}_{k})$ and
for $k= 1, 2, \ldots, n$. After receiving these $n$ cbits from Alice, and
his measurement direction $\hat{b} \in S_2$, Bob outputs
\begin{equation}
\beta=\sum_{k=1}^{n}3^{n-k}\{\frac{1}{2}(1+f_{k})\rm {sgn}[\hat{b}.(\hat{\lambda_{k}}+c_{k}\hat{\mu_{k}})]\}.
\end{equation}
Note that each of $c_1$, $c_2$, $\ldots$, $c_{n}$, described above,
can only have two values $1$ or $-1$ and these values occur with
equal probabilities. So Bob will get no extra information about
Alice's output $\alpha$ after having the $n$ values $c_1$, $c_2$,
$\ldots$, $c_{n}$.
 As ${\hat{\lambda}}_1$, ${\hat{\lambda}}_2$,
$\ldots$, ${\hat{\lambda}}_n$, ${\hat{\mu}}_1$, ${\hat{\mu}}_2$,
$\ldots$, ${\hat{\mu}}_{n}$, ${\hat{\nu}}_{1}$, ${\hat{\nu}}_{2}$,$\ldots$,
${\hat{\nu}}_{n}$,  are independent and uniformly
distributed on $S_2$, therefore the unit vector
$$\frac{{\hat{\lambda}}_k +
\rm {sgn}(\hat{a}.{\hat{\lambda}}_k)\rm {sgn}(\hat{a}.{\hat{\mu}}_{k}){\hat{\mu}}_{k}}{|{\hat{\lambda}}_k +
\rm {sgn}(\hat{a}.{\hat{\lambda}}_k)\rm {sgn}(\hat{a}.{\hat{\mu}}_{k}){\hat{\mu}}_{k}|}$$
is uniformly distributed on $S_2$. Therefore, for given any $\hat{b}
\in S_2$, the quantity $\hat{b}.({\hat{\lambda}}_k +
c_{k}{\hat{\mu}}_{k})$
% =\hat{b}.(\rm {sgn}(\hat{a}.{\hat{\lambda}}_k)\rm {sgn}(\hat{a}.{\hat{\mu}}_{2k -
%1}){\hat{\mu}}_{2k - 1} +
%\rm {sgn}(\hat{a}.{\hat{\lambda}}_k)\rm {sgn}(\hat{a}.{\hat{\mu}}_{2k}){\hat{\mu}}_{2k})$
is uniformly distributed on $S_2$. Hence, we see that Bob's output
$\beta$ can only take the values $3^n/2 - 1/2$, $3^n/2 - 3/2$,
$\ldots$, $-3^n/2 + 3/2$, $3^n/2 + 1/2$ with equal probabilities. So
$\langle \beta \rangle = 0$.
The correlation $\langle \alpha \beta \rangle$ is given by
\begin{eqnarray}
\left\langle \alpha \beta \right\rangle&=-E\{\sum_{k=1}^{n}3^{n-k}[\frac{1}{2}(1+f_{k})\rm {sgn}(\hat{a}.\hat{\lambda_{k}})]\nonumber\\                  &\sum_{l=1}^{n}3^{n-l}[\frac{1}{2}(1+f_{l})\rm {sgn}(\hat{b}.(\hat{\lambda_{l}}+c_{l}\hat{\mu_{l}}))]\},
\end{eqnarray}
where $E\left\{x\right\}$ denote average value of $x$
\begin{eqnarray}
E\left\{x\right\} &=&
-\frac{1}{(4\pi)^{3n}}\int_{{\hat{\lambda}}_1} \in S_2}\int_{{\hat{\mu}}_1 \in S_2}\int_{{\hat{\nu}}_{1}  \ldots  \\ &&
\int_{{\hat{\lambda}}_n \in S_2}\int_{{\hat{\mu}}_{n}\in S_2} \int_{{\hat{\nu}}_{n}\in S_2} 
 d{{\hat{\lambda}}_1}  d{{\hat{\mu}}_1} d{{\hat{\nu}}_{1}}\ldots  d{{\hat{\lambda}}_n}  d{{\hat{\mu}}_n} d{{\hat{\nu}}_{n}}  x \nonumber
\end{eqnarray}
We know from Toner and Bacon \cite{toner03}
\begin{eqnarray}
	\left\langle \rm {sgn}(\hat{a}.\hat{\lambda_{k}})\rm {sgn}[\hat{b}.(\hat{\lambda_{l}}+c_{l}\hat{\mu_{l}})]\right\rangle=\hat{a}.\hat{b}\delta_{kl}
\end{eqnarray}
and, it is easy show that
\begin{eqnarray}
	\{\frac{1}{2}(1+f_{k})\}^{2}=\frac{1}{2}(1+f_{k}).
\end{eqnarray}
So, 
%by using Eqs. (10) and (11)
\begin{eqnarray}
	\left\langle \alpha \beta\right\rangle &=&-\sum_{k=1}^{n}3^{n-k}\frac{1}{2}(1+\left\langle g_{k}\right\rangle)\left\langle \rm {sgn}(\hat{a}.\hat{\lambda_{k}})\rm {sgn}[\hat{b}.(\hat{\lambda_{k}}+c_{k}\hat{\mu_{k}})]\right\rangle\\
	&=&-\frac{2}{3}\hat{a}.\hat{b}\sum_{k=1}^{n}3^{n-k}.\nonumber
\end{eqnarray}
Thus we see that by  summing the geometric series and using $2s+1=3^{n}$ we get
\begin{equation}
\label{begincor4} \langle \alpha\beta \rangle = -\frac{1}{3}s(s +
1)\hat{a}.\hat{b}.
\end{equation}
Thus we see that the above-mentioned protocol exactly simulates
quantum mechanical probability distribution for particular types of
projective measurements, namely the spin measurement, on the spin
$s$ singlet state with $2s + 1 = 3^{n}$ for positive integer $n$.
The above protocol applies to infinite, although sparse, subset of
the set of all spins (i.e., all integral and half integral values).
The most important finding is that the amount of communication goes
as $(2/{\rm log}_2 3){\rm log}_{2}(2s + 1)$.
Our result provides the amount of classical communication in the
worst case scenario if we consider only measurement of spin
observables on both sides of a two spin-$s$ singlet state with the
restriction that the dimension $2s + 1$ of each subsystem must be a
positive integral power of $3$, and just $n = (1/{\rm log}_2 3){\rm
log}_2 (2s + 1)$ bits of communication from Alice to Bob is
sufficient.

\subsection{General Simulation Scheme $(2s+1=P^n)$}
In the simulation of the measurement of the observable $\hat{a}.{\bf S}$ (where $\hat{a} \in
\mathbb{R}^3$ is the supplied direction of measurement), Alice will
have to reproduce the $P^n$ number of outcomes $\alpha = P^n/2 -
1/2, P^n/2 - 3/2, \ldots, - P^n/2 + 3/2, - P^n/2 + 1/2$ with equal
probability. If we consider the series $- \sum_{k = 1}^{n} P^{n - k}
g_k$, where, $g_1=P/2-1/2$, $g_2=P/2 - 3/2$, $\ldots$,$g_{n-1}=- P/2 + 3/2$, $g_n= - P/2 + 1/2$, it turns out that the series can only take the above-mentioned $P^n$ different values of $\alpha$. More specifically, if we vary $(g_1, g_2, \ldots, g_n)$ over all the members of the $P^n$ - element set
$P/2 -1/2, P/2 - 3/2, \ldots, - P/2 + 3/2, - P/2 + 1/2$, the series $- \sum_{k = 1}^{n} P^{n - k} g_k$ will
have the corresponding $P^n$ different values $P^n/2 - 1/2, P^n/2 -
3/2, \ldots, - P^n/2 + 3/2, - P^n/2 + 1/2$. So we have to choose
$g_k$'s in such a way that all the elements $(g_1, g_2, \ldots,
g_n)$ are equally probable. In this direction, we choose Alice's
output as
\begin{equation}
\label{alphachoice} \alpha = - \sum_{k = 1}^{n} P^{n - k}\alpha_k(\frac{P-1}{2}),
\end{equation}
where $\alpha_k(\frac{P-1}{2})$ is obtained from equations (\ref{halfintegeralpha}), (\ref{integeralpha}) and all  share random in $\alpha_k(\frac{P-1}{2}) $ is independent of all share random in $\alpha_l(\frac{P-1}{2}) $ if $k\neq l$.
We see
\begin{equation}
\langle \alpha_k(\frac{P-1}{2}) \rangle =0\Rightarrow  \langle \alpha  \rangle=0
\end{equation}
Alice then sends the $nm$ cbits $c_{11},c_{12},\ldots,c_{1m},\ldots,c_{k1},c_{k2},\ldots,c_{km},c_{n1},c_{n2},\\ \ldots,c_{nm}$ to Bob where $c_{kj}=\rm {sgn}(\hat{a}.\hat{\lambda}_{kj})\rm {sgn}(\hat{a}.\hat{\mu}_{kj})$ and for $k=1,2,\ldots,n$ and $j=1,2,\ldots,m$. After receiving these $nm$ cbits from Alice, and his measurement direction $\hat{b}\in S_{2}$, Bob outputs
\begin{equation}
\label{betachoice} \beta =  \sum_{k = 1}^{n} P^{n - k}\beta_k(\frac{P-1}{2}).
\end{equation}
Note that each of $c_{kj}$($k=1,2,\ldots,n$ and $j=1,2,\ldots,m$), described above, can only have two values $1$ or $-1$ and these values occur with equal probabilities. So Bob will get no extra information about Alice's output $\alpha$ after having the $nm$ values all cbits.
We can see 
\begin{equation}
\langle \beta_k(\frac{P-1}{2}) \rangle =0\Rightarrow  \langle \beta  \rangle=0.
\end{equation}
The correlation $\langle \alpha \beta \rangle $ is given by
\begin{eqnarray}
\langle \alpha \beta \rangle &=&-\langle \sum_{k=1}^{n}P^{n-k}\alpha_k(\frac{P-1}{2})\sum_{l=1}^{n}P^{n-l}\beta_l(\frac{P-1}{2})\rangle \nonumber\\ &=&-\sum_{k,l=1}^{n}P^{2n-k-l}\langle\alpha_k(\frac{P-1}{2})\beta_l(\frac{P-1}{2})\rangle \nonumber \\ &=&-\sum_{k,l=1}^{n}P^{2n-k-l}\times (\frac{1}{3})(\frac{P-1}{2})(\frac{P-1}{2}+1)\delta_{kl}\hat{a}.\hat{b}\nonumber\\ &=&-\frac{1}{3}\frac{P^2-1}{4}\hat{a}.\hat{b}\sum_{k=1}^{n}P^{2(n-k)}= -\frac{1}{3}\frac{P^2-1}{4}\hat{a}.\hat{b}\frac{P^{2n}-1}{P^2-1}\nonumber \\ &=& -\frac{1}{3}s(s+1)\hat{a}.\hat{b}
\end{eqnarray}
Thus we see that the above-mentioned protocol exactly simulates
quantum mechanical probability distribution for particular types of
projective measurements, namely the spin measurement, on the spin
$s$ singlet state with $2s + 1 = P^{n}$ for positive integer $n$.
The above protocol applies to  all spins values. 
The most important finding is that the amount of communication goes
as $$nm=\left[log_2P\right]^{-1}\left[log_2(2s+1)\right]\left[\left\lceil log_2(P+1)\right\rceil-1\right].$$
Note that, to get the number of cbits to be communicated for the case $n=1$ that is $2s+1=P$, we use the protocol given in the previous section (2.5) as these two protocols coincide for $n=1$.\\
For any maximally entangled state $|{\psi}_{max}\rangle$ of two
spin-$s$ systems, we know that there exists a $(2s + 1) \times (2s +
1)$ unitary matrix $U$ such that $|{\psi}_{max}\rangle = (U \times
I)|{\psi}^-_s\rangle$. Our protocol works equally well for those two
spin-$s$ maximally entangled state $|{\psi}_{max}\rangle$ for each
of which the above-mentioned unitary matrix $U$ induces a rotation
in ${I\!\!\!\!R}^3$, as in those cases, both Alice and Bob can
perform the protocol for the spin-£s£ singlet state
$|{\psi}^-_s\rangle$ for the rotated input vectors $\hat{a}$ and
$\hat{b}$ and, hence, they will achieve their goal.
Our result provides the amount of classical communication in the
worst case scenario if we consider only measurement of spin
observables on both sides of a two spin-$s$ singlet state where the dimension $2s + 1$ of each subsystem must be a
positive integral power of $P$, and just $nm=\left[log_2P\right]^{-1}\left[log_2(2s+1)\right]\left[\left\lceil log_2(P+1)\right\rceil-1\right]$ bits of communication from Alice to Bob is sufficient. We are not sure whether our protocol is optimal. We see that for high spin the number of cbits close to $log_2(2s+1)$ like the previous protocol. But sometimes in the lower spin this model requires less resource . For example the previous model predicts for simulation of singlet state in case $s=4$ \textit{three} cbits is enough but this protocol is doing that by \textsl{two} cbits.\\
It should be noted that if we consider most general projective
measurements on both the sides of a maximally entangled state of two
qudits, with $d = 2^n$, it is known that (see \cite{brassard99})
Alice would require at least of the order of $2^n$ bits of
communication to be sent to Bob, in the worst case scenario when $n$
is large enough. But for general $d$, $log_2 d$ can be shown to be a
lower bound on the average amount of classical communication that
one would require to simulate the maximally entangled correlation of
two qudits considering most general type of projective measurements
 \cite{barrett06}. It is also known that $log_2 d$ bits of classical
communication on average is sufficient to simulate the measurement
correlation of a maximally entangled state of two qudits, when both
Alice and Bob consider only measurement of traceless binary
observables \cite{degorre07}. It thus seems that even if simulation
of maximally entangled correlation in the most general case of
projective measurement is a hard problem, and one would require to
send classical communication at least of the order of the dimension
( for large dimensional case), there is still some room to search for
efficient simulation protocols in lower dimensions.

\section{Classical Simulation of Non maximally Entangled State}
In this section, we present a protocol, based on \cite{toner03}, to simulate pure two qubits non maximally entangled state in the special case by using two cbits.\\
Toner and Bacon \cite{toner03} by using the classical teleportation protocol, obtained a protocol to simulate joint projective measurements on partially entangled states of two qubits, which uses \textit{two} bits of communication: Alice first simulate her measurement and determines the post measurement state of Bob's qubit; Alice and Bob then execute the classical teleportation protocol. However, whether two cbits communication is optimal is still an open problem. N. Brunner, N. Gisin and V. Scarani \cite{brunnergisin} showed there exist non maximally entangled two qubit states which cannot be simulated using one cbit of communication. They demonstrated that, in order to simulate the correlations of a pure non-maximally entangled state, required amount of communication can be strictly larger than that needed to simulate the maximally entangled state. They proved that a single nonlocal bit of communication or a single use of the non local machine (NLM) do not provide enough non locality to simulate non-maximally entangled states of two qubits, while, these resources are enough to simulate the singlet. This result is striking, but is not something totally unexpected, a few previous hints being present in the literature: Eberhard proved that non-maximally entangled states require lower detection efficiencies than maximally entangled ones, in order to close the detection loophole \cite{eberhard93}; Bell inequalities have been found whose largest violation is given by a non-maximally entangled state \cite{acin02}; and it is also known that some mixed entangled states admit a local variable model, even for the most general measurements \cite{barrett02}.\\Two spatially separate parties, Alice and Bob, each have a spin-$\frac{1}{2}$ particle, or qubit. The general non maximally entangled state is
\begin{equation}
|\psi (\gamma)\rangle =\cos(\gamma)|01\rangle +\sin(\gamma)|10\rangle
\end{equation}
with $\cos\gamma > \sin\gamma$ $(0<\gamma<\frac{\pi}{4})$. Up to local operations, this is the most general pure state of two qubits.
The spin states $|0\rangle$, $|1\rangle$ are defined with respect to a local set of coordinate axes: $|0\rangle$ ($|1\rangle$) corresponds to spin along the local $+\hat{z}$ ($-\hat{z}$) direction. Alice and Bob each measure their own particle's spin along a direction parametrized by a three-dimensional unit vector: Alice measures along $\hat{a}$, Bob along $\hat{b}$. Alice and Bob obtain results, $\alpha \in\left\{+1,-1\right\}$ and $\beta \in \left\{+1,-1\right\}$, with probability:
\begin{eqnarray}
p(\alpha=\pm1)=\frac{1}{2}\left[1\pm \cos (2\gamma) a_{z}\right],\\
p(\beta=\pm1)=\frac{1}{2}\left[1\mp \cos (2\gamma) b_{z}\right].
\end{eqnarray}
The quantum expectation values of Alice's output and Bob's output will be:
\begin{eqnarray}
\left\langle \alpha\right\rangle=&\cos(2\gamma)a_{z}\nonumber\\
\left\langle \beta\right\rangle=&-\cos(2\gamma)b_{z},
\end{eqnarray}
and their joint probabilities are correlated such that
\begin{eqnarray}
\left\langle \alpha\beta\right\rangle=-a_{z}b_{z}+\sin(2\gamma)(a_{x}b_{x}+a_{y}b_{y}).
\end{eqnarray}
In the case where the particles are moving along $\pm\hat{z}$ direction, while Bob's input ($\hat{b}$) is in the X-Y plane, these becomes
\begin{eqnarray}
\label{nonmaximum}
\left\langle \alpha\right\rangle&=&\cos(2\gamma)a_{z}\nonumber\\
\left\langle \beta\right\rangle&=&0\nonumber\\
\left\langle\alpha\beta\right\rangle&=&\sin(2\gamma)(a_{x}b_{x}+a_{y}b_{y}).
\end{eqnarray}
We now describe our protocol for simulation this special case. Alice and Bob share three random variables $\hat{\lambda}_{0}$, $\hat{\lambda}_{1}$ and $\hat{\lambda}_{2}$ which are real three dimensional unit vectors. They are chosen independently and distributed uniformly over the unit sphere.\\ The protocol proceeds as follows:\\
(1) Alice calculates her outputs as:
\begin{equation}
\alpha=\rm {sgn}(\hat{a}.\hat{\lambda_{0}}+a_{z}\cos 2\gamma).
\end{equation}
Now, we show:
\begin{eqnarray}
\left\langle \alpha\right\rangle&=&\frac{1}{4\pi}\int d\lambda_{0}\rm {sgn}(\hat{a}.\hat{\lambda_{0}}+a_{z}\cos 2\gamma )\nonumber \\&=&\frac{1}{2}\int_0^\pi sin\theta_0 d\theta_0  \rm {sgn}(\cos\theta_0 +a_{z}\cos 2\gamma)\\&=&a_{z}\cos 2\gamma \nonumber
\end{eqnarray}
(2) Alice sends two bits $c_{1}$ and $c_{2}\in\left\{+1,-1\right\}$ to Bob where 
\begin{eqnarray}
c_{1}&=&\rm {sgn}(\hat{a}.\hat{\lambda}_{0}) \rm {sgn}(\hat{a}.\hat{\lambda}_{1}+l)\nonumber\\
c_{2}&=&\rm {sgn}(\hat{a}.\hat{\lambda}_{0}) \rm {sgn}(\hat{a}.\hat{\lambda}_{2}+l),
\end{eqnarray}
where, $l$ is introduced later.
Bob cannot obtain any information about Alice's output $\alpha$ from the communications. This is so because 
\begin{eqnarray}
\left\langle c_{k}\right\rangle&=&\left\langle \rm {sgn}(\hat{a}.\hat{\lambda}_{0})\right\rangle \left\langle \rm {sgn}(\hat{a}.\hat{\lambda}_{k}+l)\right\rangle=0,\\ \rm since &&\left\langle \rm {sgn}(\hat{a}.\hat{\lambda}_{0})\right\rangle=0. ~~~~   (k=1,2)\nonumber
\end{eqnarray}
(3) After receiving $c_{1}$ and $c_{2}$, Bob now calculates his output:
\begin{equation}
\beta=\rm {sgn}[\hat{b}.(c_{1}\hat{\lambda}_{1}+c_{2}\hat{\lambda}_{2})].
\end{equation}
By using discussion in previous sections, we obtain:
\begin{eqnarray}
\beta &=&\frac{c_1}{2}[\rm {sgn}(\hat{b}.(\hat{\lambda}_{1}+\hat{\lambda}_{2}) -\rm {sgn}(\hat{b}.(\hat{\lambda}_{2}-\hat{\lambda}_{1}))]\nonumber\\&+&\frac{c_2}{2}[\rm {sgn}(\hat{b}.(\hat{\lambda}_{1}+\hat{\lambda}_{2}) +\rm {sgn}(\hat{b}.(\hat{\lambda}_{2}-\hat{\lambda}_{1}))].
\end{eqnarray}
We now prove that:
\begin{eqnarray*}
&&\left\langle \beta\right\rangle = \\&&\frac{1}{2}\left\langle \rm {sgn}(\hat{a}.\hat{\lambda}_{0})\right\rangle \left\langle \rm {sgn}(\hat{a}.\hat{\lambda}_{1}+l)\left[\rm {sgn}(\hat{b}.(\hat{\lambda}_{1}+\hat{\lambda}_{2})-\rm {sgn}(\hat{b}.(\hat{\lambda}_{2}-\hat{\lambda}_{1}))\right]\right\rangle
\end{eqnarray*}
\begin{eqnarray}
&+&\frac{1}{2}\left\langle \rm {sgn}(\hat{a}.\hat{\lambda}_{0})\right\rangle \left\langle \rm {sgn}(\hat{a}.\hat{\lambda}_{2}+l)\left[\rm {sgn}(\hat{b}.(\hat{\lambda}_{1}+\hat{\lambda}_{2})-\rm {sgn}(\hat{b}.(\hat{\lambda}_{2}-\hat{\lambda}_{1}))\right]\right\rangle\nonumber\\&=&0.
\end{eqnarray}
This is true because $$\left\langle \rm {sgn}(\hat{a}.\hat{\lambda}_{0})\right\rangle=0.$$ \\The joint expectation value $\left\langle \alpha \beta\right\rangle$ can be calculated as follows:
\begin{eqnarray}
\left\langle \alpha\beta \right\rangle &=& \frac{1}{(4\pi)^{3}}\int \rm {sgn}(\hat{a}.\hat{\lambda_{0}}+a_{z}\cos 2\gamma)\rm {sgn}(\hat{a}.\hat{\lambda}_{0}) d\lambda_{0}\times \nonumber \\ &&\{\frac{1}{2} \int \rm {sgn}(\hat{a}.\hat{\lambda}_{1}+l)[\rm {sgn}(\hat{b}.(\hat{\lambda}_{1}+\hat{\lambda}_{2}) -\rm {sgn}(\hat{b}.(\hat{\lambda}_{2}-\hat{\lambda}_{1}))]d\hat{\lambda}_{1}d\hat{\lambda}_{2}\nonumber\\ &+&\frac{1}{2} \int \rm {sgn}(\hat{a}.\hat{\lambda}_{2}+l)[\rm {sgn}(\hat{b}.(\hat{\lambda}_{1}+\hat{\lambda}_{2}) +\rm {sgn}(\hat{b}.(\hat{\lambda}_{2}-\hat{\lambda}_{1}))]d\hat{\lambda}_{1}d\hat{\lambda}_{2}\}.\nonumber\\
\end{eqnarray}

First, we integrate over $\lambda_0$:
\begin{eqnarray}
\label{b}
&&\frac{1}{(4\pi)}\int \rm {sgn}(\hat{a}.\hat{\lambda_{0}}+a_{z}\cos 2\gamma)\rm {sgn}(\hat{a}.\hat{\lambda}_{0}) d\lambda_{0}\nonumber \\ &&=\frac{1}{2}\int_0^\pi sin\theta_0 d\theta_0 \rm {sgn}(cos\theta_0 +cos(2\gamma) a_z) \nonumber \\&&=1-|cos(2\gamma)a_z|.
\end{eqnarray}
Next, we know from \cite{toner03} 
\begin{eqnarray}
\label{a}
\int d\hat{\lambda}_1 \rm {sgn}(\hat{a}.\hat{\lambda}_{1}+l)\int d\hat{\lambda}_2 \rm {sgn}(\hat{b}.(\hat{\lambda}_{1}+\hat{\lambda}_{2}))&=&\nonumber\\
-\int d\hat{\lambda}_1 \rm {sgn}(\hat{a}.\hat{\lambda}_{1}+l)\int d\hat{\lambda}_2 \rm {sgn}(\hat{b}.(\hat{\lambda}_{1}-\hat{\lambda}_{2}))&=&\nonumber\\
\int d\hat{\lambda}_2 \rm {sgn}(\hat{a}.\hat{\lambda}_{2}+l)\int d\hat{\lambda}_1 \rm {sgn}(\hat{b}.(\hat{\lambda}_{1}+\hat{\lambda}_{2}))&=&\nonumber\\
\int d\hat{\lambda}_2 \rm {sgn}(\hat{a}.\hat{\lambda}_{2}+l)\int d\hat{\lambda}_1 \rm {sgn}(\hat{b}.(\hat{\lambda}_{1}-\hat{\lambda}_{2})).\nonumber\\
\end{eqnarray}
So, by using equations (\ref{b})and (\ref{a}), we have:
\begin{eqnarray}
\left\langle \alpha \beta \right\rangle&=&\frac{2}{(4\pi)^2}\left[1-|\cos(2\gamma)a_z|\right]\int d\hat{\lambda}_1 \rm {sgn}(\hat{a}.\hat{\lambda}_{1}+l) \nonumber \\ &&\int d \lambda_2 \rm {sgn}(\hat{b}.(\hat{\lambda}_{1}+\hat{\lambda}_{2}))
\end{eqnarray}
Now, we see from \cite{toner03}:
\begin{eqnarray}
\label{c}
\int d\hat{\lambda}_1 \rm {sgn}(\hat{a}.\hat{\lambda}_{1}+l)\int d \lambda_2 \rm {sgn}(\hat{b}.(\hat{\lambda}_{1}+\hat{\lambda}_{2}))=(1-l^2)(\hat{a}.\hat{b}).
\end{eqnarray}
Now, from equations (\ref{b}) and (\ref{c})
\begin{eqnarray}
\left\langle \alpha\beta \right\rangle = (1-|a_{z}cos2\gamma |)\times[1-(l^2)](\hat{a}.\hat{b})\equiv sin2\gamma(\hat{a}.\hat{b})
\end{eqnarray}
We can obtain the value $l$ from above equation for getting quantum correlation. Therefore:
\begin{eqnarray}
l=\sqrt{1-\frac{\sin2\gamma}{1-|a_z\cos2\gamma|}}~~~1\neq|a_z\cos2\gamma|~~ \mbox{because $\gamma\neq\frac{\pi}{4}$}.
\end{eqnarray}
From above equation, we can check $0\leq l\leq 1$. But, term under $\sqrt{.}$ must be always positive. Therefore;
\begin{eqnarray}
\frac{\sin2\gamma}{1-|a_z\cos2\gamma|}\leq 1\Rightarrow ~ \sin2\gamma +|a_z\cos2\gamma|\leq1.
\end{eqnarray}

This is in conformity  with quantum correlation between two parties' outputs with nonmaximall entangled state (see Eq(\ref{nonmaximum})).
\section{Summary and Conclusion}
John Bell pointed out that the correlations resulting from quantum theory cannot be reproduced by any classical local realistic theory. It follows that quantum correlations on space like separated systems cannot be reproduce classically. If, however, the systems are time like separate, then classical simulation is possible, albeit at the expense of some communication, but \textit{how much} is required? In particular, suppose a number of spatially separate parties share an entangled quantum state, and each makes a local measurement on their component. Then quantum correlations are manifest in the joint probability distribution of the parties' choice of measurement. If this probability distribution cannot be reproduced by a classical local realistic theory, then it violates some generalized Bell inequality. This means some communication between the parties is required to reproduce the probability distribution, but Bell inequality violation does nothing to \textit{quantify} how much. More generally, entanglement is a resource for performing information processing tasks, and an important goal of quantum information theory is to demarcate it from classical resources, such as shared randomness and classical communication channels. What classical resources are required to reproduce the joint probability distributions arising from local measurement on shared quantum states?\\Until now, an exact classical simulation of quantum correlations, for {\it all} possible projective measurements, is accomplished only for spin $s = 1/2$ singlet state, requiring 1 cbit of classical communication \cite{toner03}. It is important to know how does the amount of this classical communication change with the change in the value of the spin $s$, in order to quantify the advantage offered by quantum communication over the classical one. We introduced \textit{two} classical protocol to exactly simulate quantum correlations implied by a spin-$s$ singlet state for the infinite sequence of spins.\\
Our protocols provide the amount of classical communication in the
worst case scenario if we consider only measurement of spin
observables on both sides of a two spin-$s$ singlet state for all
the values of $s$ -- just $n=\lceil{\rm log}_2 (s + 1)\rceil$ ( for the first  protocol),$n = \left[log_2P\right]^{-1}\left[log_2(2s+1)\right]\left[\left\lceil log_2(P+1)\right\rceil-1\right]$ ( for the second protocol) bits
of communication from Alice to Bob is sufficient. We are unable to show whether our protocols are optimal(in the sense of using minimum number of classical communication).  We obtained that for higher spin the number of cbits close to $log_2(2s+1)$ in both protocol but sometimes in the lower spin second model ($2s+1=P^n$) is more optimal than first protocol. For example the first model predict for simulation singlet state in case $s=4$ \textit{three} cbits is enough but second model is doing that by \textsl{two} cbits.\\
Figure 2 depicts the dependence of the number of cbits required to be communicated to simulate the spin-s singlet state, on the spin value $s$, based on the general protocol involving all spins. It is interesting to see that for all $s$ values satisfying $2^n\leq s+1< 2^{n+1}$ the required communication does not vary and equals $n$ cbits. As the spin value crosses $2^{n+1}$, the required number of cbits takes a ``quantum jump" from $n$ to $n+1$. It will be interesting to find an explanation of this behavior and is one of the problems we want to tackle in future.\\
On the other hand, if we consider most general projective measurements on both the sides of a maximally entangled state of two qudits, with $d=2^{n}$, it is known that (see \cite{brassard99})
Alice would require at least of the order of $2^n$ bits of
communication to be sent to Bob, in the worst case scenario when $n$
is large enough. But for general $d$, ${\rm log}_2 d$ can be shown
to be a lower bound on the average amount of classical communication
that one would require to simulate the maximally entangled
correlation of two qudits considering most general type of
projective measurements \cite{barrett06}. It is also known that $\log_{2}(2s+1)$ bits of classical communication on average is sufficient to simulate the measurement correlation of a maximally entangled state of two qudits, when both Alice and Bob consider only measurement of traceless binary observables \cite{degorre07}. So, in the worst case
scenario, one would require at least ${\rm log}_2 d$ number of bits
of communication for simulating measurement correlation of the
two-qudit maximally entangled state, where the measurement can be
arbitrary but projection type. If one can show that ${\rm log}_2 d$
is again a lower bound for considering measurement of spin
observables only ( which we believe to be true), our simulation
scheme will turn out to be optimal.\\We also obtained a classical protocol for simulation quantum correlations between two party whom share a nonmaximum entangled state two qubits by using two cbits. Our model is working in the special case that one of Alice's input or Bob's input is on X-Y plane. This protocol is in agreement with result in      \cite{brunnergisin} that strictly more resources are needs to simulate non maximally entangled states than to simulate the singlet state.
\begin{figure}
\begin{center}
\includegraphics[width=16cm,height=12cm]{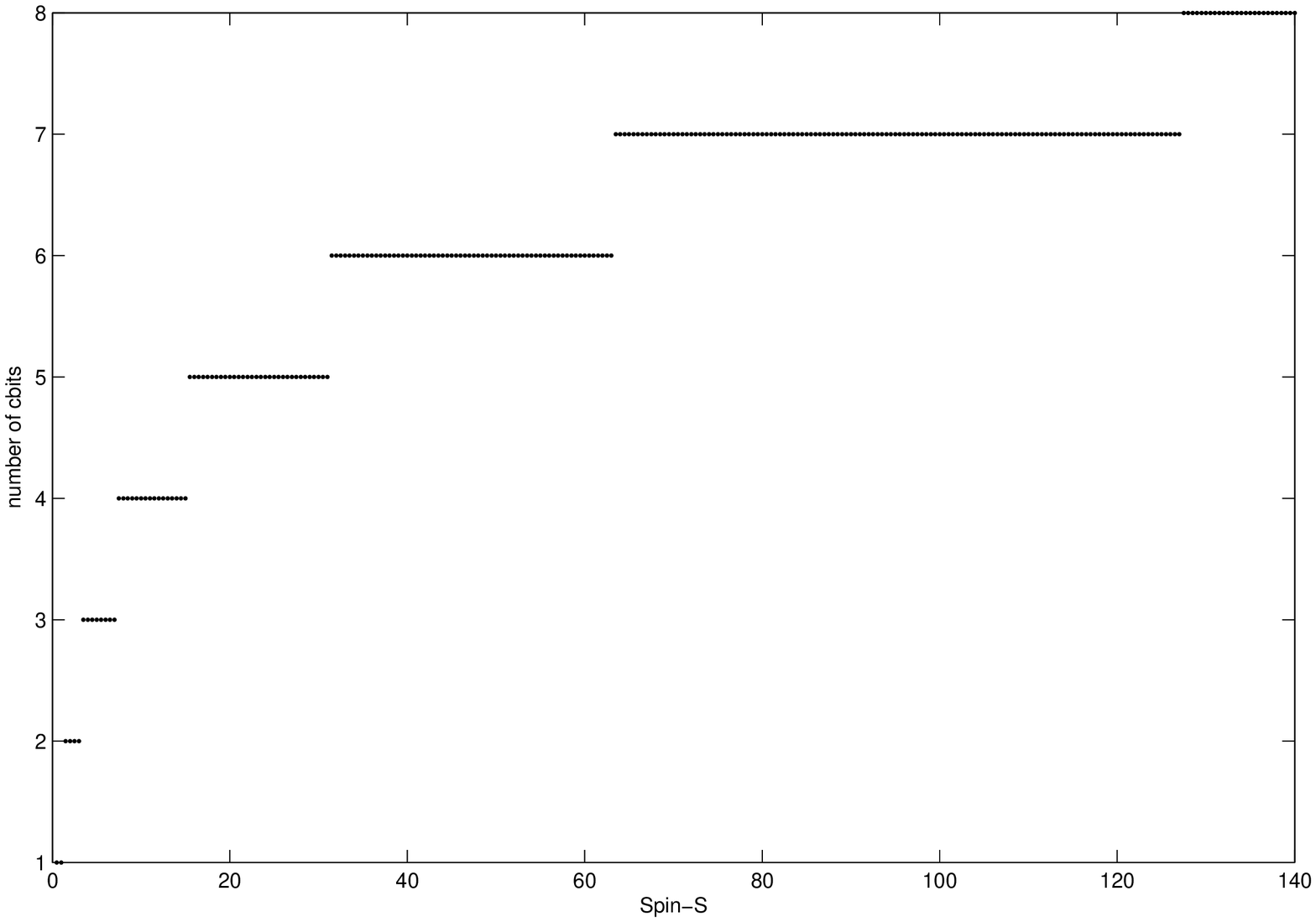}
\vspace{0.4cm}

\underline{Figure 2 :} \small{~The number of cbits required for simulation of spin-s singlet state}
\end{center}
\end{figure}
\vspace{0.8cm}
\newpage

\chapter{Quantum Correlations In Successive  Spin Measurements}}
\section{Introduction}
Quantum Mechanics (QM) is known to be non local-realistic as well as
contextual \cite{peres93}. All theories and experiments to test these aspects
of QM are mainly based on the multipartite quantum systems in
entangled states. Although this scenario is inevitable for the tests
of non locality, it is not obligatory for testing realism and
contextuality. In this chapter we propose and analyze a particular
scenario to account for the deviations of QM from `realism' ( defined
below), which involves correlations in the outputs of successive
measurements of noncommuting operators on a spin-$s$ state.

The correlations for successive measurements have been used
previously by Popescu \cite{popescu95} in the context of nonlocal quantum
correlations, in order to analyze a class of Werner states which are
entangled but do not break (bipartite) Bell-type inequality.
Although local HVT can simulate the quantum correlations between the
outputs of single ideal measurement on each part of the system, it
fails to simulate the correlations of the second measurements on
each part. Leggett and Garg have used consecutive measurements to
challenge the applicability of QM to macroscopic phenomena \cite{leggettgarg}.
While the temporal Bell inequalities, considered in refs. \cite{leggettgarg} (see also \cite{paz93}), are for histories, we deal here with Bell-type inequalities with predetermined measurement values at different times. The temporal Bell inequalities deal with measurement of the same observable at different times, whereas we deal here with different measurements at different times. Finally there is a large literature on the problem of information of a quantum state that can be obtained by measuring the same operator successively on a single
system. The research in this area is elegantly summarized in \cite{alter}.
Bell-type inequality with successive measurements was first
considered by Brukner et al. \cite{vedral}. They have derived CHSH-type
inequality \cite{chsh} for two successive measurements on an arbitrary
state of a single qubit and have shown that every such state would
violate that inequality for proper choice of the measurement
settings. They have shown that the quantum mechanical correlation
for three successive measurements, for any single qubit input state
is the product of two consecutive correlations each of which is the
correlation of two consecutive measurements -- a scenario quite
uncommon for spatial correlations. As an application of  their
approach, they have used the correlations in two successive
measurements to overcome the limitations in RAM of a computer to
calculate a Boolean function whose input bits are supplied
sequentially in time. \\ In this chapter we consider and analyze the correlations between the outputs of successive measurements for a general spin $S$ state as against the general qubit state. We show that, for $S>\frac{1}{2}$, the quantum mechanical correlation for three successive measurements is not a product of two successive correlations, that is, the correlations in two successive measurements. We show that for $S=\frac{1}{2}$, the correlation between the outputs of measurements from $n-k$ to $n$ (last $k$ out of $n$ successive measurements) $k=0,1,\ldots,n-1$, depend on the measurement prior to $n-k$, when $k$ is even, while for odd $k$, these correlations are independent of the outputs of measurements prior to $n-k$. Further, we show that all qubit states break the Bell type inequalities corresponding to $n$ successive measurements, where $n$ is any finite number. Finally, we give a protocol to classically simulate the quantum correlation of $n$ successive measurements in qubit case giving a measure of quantum correlation.\\ The chapter is organized as follows. In Section 2 we describe the basic scenario in detail. Section 3 formulates the implications of hidden variable theory (HVT) for this scenario in terms of Bell-type inequalities. Section 4 evaluates these inequalities for mixed input states of single spin-$s$ system for two and three successive measurements (considering various values of $s$). Section 5 deals with $n$ successive measurements on spin-$1/2$ system. In section 6 we give a protocol to simulate the correlations between $n$
successive measurements on a spin-$1/2$ system. Finally we conclude
with  summary and comments in Section 7. Mathematical details are
relegated to Appendices A and B.

\section{Basic Scenario}
Consider the following sequence of measurements. A quantum particle
with spin $s$, prepared in the initial state $\rho_0$, is sent
through a cascade of Stern-Gerlach (SG) measurements for the spin
components along the directions given by the unit vectors $\h a_1,
\h a_2, \h a_3, \ldots, \h a_n$ ({\it i.e.}, measurement of
observables of the form $\vec{S}.\hat{a}$, where $\vec{S} = (S_x,
S_y, S_z)$ is the vector of spin angular momentum operators $S_x$,
$S_y$, $S_z$ and $\hat{a}$ is a unit vector from $\mathbb{R}^3$).
Each measurement has $2s+1$ possible outcomes. For the $i$-th
measurement, we denote these outcomes (eigenvalues) by $\al_i \in \{
s, s-1, \ldots, -s\}$. We denote by $\langle \al_i \rangle$ the
quantum mechanical (ensemble) average $\langle \vec{S} \cdot \h a_i
\rangle,$ by $ \langle \al_i \al_j \rangle$ the average $\langle
(\vec{S} \cdot \h a_i) (\vec{S} \cdot \h a_j) \rangle$ etc.

Each of the $(2s+1)^n$ possible outcomes, which one gets after
performing $n$ consecutive measurements, corresponds to a particular
combination of the results of the measurements at previous $n - 1$
steps and the result of the measurement at the $n$-th step. The
probability of each of these $(2s + 1)^n$ outcomes is the joint
probability for such combinations. Note that even though the spin
observables $\vec{S}\cdot\hat{a}_1$, $\vec{S}\cdot\hat{a}_2$, $\ldots$,
$\vec{S}\cdot\hat{a}_n$, whose measurements are being performed at times
$t_1$, $t_2$, $\ldots$, $t_n$ respectively (with $t_1 < t_2 < \ldots
< t_n$) do not commute, above-mentioned joint probabilities for the outcomes are
well defined because each of these spin observables act on different
states \cite{fine82} \cite{anderson05} \cite{ballentine}. We emphasize that this is the joint probability for
the results of $n$ actual measurements and not a joint probability
distribution for hypothetical simultaneous values of $n$
noncommuting observables. Moreover, various sub-beams ({\it i.e.},
wave functions) emerging from every Stern-Gerlach apparatus
(corresponding to $(2s+1)$ outcomes) in every stage of measurement
are separated without any overlap or recombination between them. In
other words, the eigen wave packet ${\psi}_{s - j, t_i, \hat{a}_i}
(x)$, corresponding to the eigen value $s - j$ of the observable
$\vec{S}\cdot\hat{a}_i$, measured at time $t_i$, will not have any part
in the regions where the SG setups, for measurement of the
observables $\vec{S}\cdot\hat{a}_{i + 1, s }$, $\vec{S}\cdot\hat{a}_{i +
1, s - 1}$, $\ldots$, $\vec{S}\cdot\hat{a}_{i + 1, s - j + 1}$,
$\vec{S}\cdot\hat{a}_{i + 1, s - j - 1}$, $\ldots$, $\vec{S}\cdot\hat{a}_{i
+ 1, - s}$, are situated. We further assume that, between two
successive measurements, the spin state does not change with time
i.e. $\vec{S}$ commutes with the interaction Hamiltonian, if any.
Also, throughout the string of measurements, no component ({\it
i.e.}, sub-beam) is blocked. It is to be mentioned here that the
time of each of the measurements  are measured by a common clock.

\section{Implications of HVT}
HVT assumes that in every possible state of the system, all
observables have well defined (sharp) values \cite{redheadbook}. On the
measurement of an observable in a given state, the value possessed
by the observable in that state ( and no other value) results. To
gain compatibility with QM and the experiments, a set of `hidden'
variables is introduced which is denoted collectively by $\la$. For
given $\la$, the values of all observables are specified as the
values of appropriate real valued functions defined over the domain
$\Lambda$ of possible values of hidden variables. For the spin
observable $\vec{S} \cdot \h a$, we denote the value of $\vec{S}
\cdot \hat a$ in the QM (spin) state $| \psi \rangle$ by $\al$.
Considered as a function, $\al : \Lambda \ra I\!\!R$ , we represent
the value of $\vec{S} \cdot \h a$ when the hidden variables have the
value $\la$ by $\al(\la)$. More generally, we may require that a
value of $\la$ gives the probability density $p(\al | \la)$ over the
values of $\al$ rather than specifying the value of $\al$
(stochastic HVT). We denote the probability density function for the
hidden variables in the state $| \psi \rangle$ by $\rho_{\psi}$ ($\rho_{\psi}(\la) d\la$ measures the probability that the collective
hidden variable lies in the range $\la$ to $\la + d \la$). Then the
average value of $\vec{S} \cdot \h a$ in the state $|\psi \rangle$
is
\begin{eqnarray}
\label{(3.1)}
 \langle \al \rangle = \int_\Lambda \al(\la) \rho_\psi(\la) d\la,
\end{eqnarray}
where the integration is over $\Lambda$ defined above. In the general case (SHVT)
\benr
\label{(3.2)}
 \langle \al \rangle = \int_\Lambda \al p(\al | \la) \rho_\psi(\la) d\la. 
\eenr
We now analyze the consequences of SHVT for our scenario. In general, the outputs of $k$th and $l$th
experiments may be correlated so that,
\benr
\label{(3.3)} 
p(\al_i, \h a_k \& \al_j,
\h a_\ell)\neq p(\al_i; \h a_k)p(\al_j; \h a_l).
\eenr
However, in SHVT we suppose that these correlations have a common
cause represented by a stochastic hidden variable $\lambda$ so that
\benr
\label{(3.4)}
 p(\al_i, \h a_k \& \al_j, \h a_\ell | \la) = p(\al_i, \h a_k|\la) p(\al_j, \h a_l | \la).
\eenr
As a consequence of equation (\ref{(3.4)}), the probability $p(\al_i, \h a_k|\la)$ obtained in a measurement ($\vec{S}\cdot\hat{a}_k$ say) performed at time $t_k$ is independent of any other measurement ($\vec{S}\cdot\hat{a}_l$ say) made at some earlier or later time $t_l$. This is called locality in time \cite{leggettgarg} \cite{vedral}.\\One should note that for a two dimensional quantum mechanical system, one can always assign values ( deterministically or
probabilistically ) to the observables with the help of a HVT. Once
the measurement is done, the system will be prepared in an output
state ( namely, an eigenstate of the observable), and the earlier HVT
may or may not work to reproduce the values of the observables to be
measured on that output state ( prepared after the first
measurement). In the present paper, we have considered possibility
of existence of a HVT for every input qubit state which can
reproduce the measurement outcomes of $n$ successive measurements.\\ Equation (\ref{(3.4)}) is the crucial equation expressing the fundamental implication of SHVT to the successive measurement scenario. We now obtain the Bell type inequalities from equation (\ref{(3.4)}) which can be compared with QM. Here we assume that in HVT all probabilities corresponding to outputs of measurements account for the possible changes in the values of the observable being measured, ( due to the interaction of the measuring device and the system), occurring in the previous measurements.\\
Now $\langle \al_i  \al_j \rangle$ is the expectation value of
obtaining the outcome $\al_i$ in the measurement of the observable
$\vec{S}.\hat{a}_i$ at time $t_i$ as well as the outcome $\al_j$ in
the measurement of the observable $\vec{S}.\hat{a}_j$ at later time
$t_j$. Due to the HVT, we must have ( dropping $\h a_k, \h a_\ell)$
\benr
\label{(3.5)}
\langle \al_i  \al_j \rangle = \int \rho(\la) E(\al_i, \al_j, \la) d\la,
\eenr
where
\begin{eqnarray}
\label{(3.6)}
 E(\al_i, \al_j, \la)&=& \sum_{\al_i, \al_j} \al_i \al_j p(\al_i, \al_j | \la)= \sum_{\al_i} \al_i p(\al_i | \la) \sum_{\al_j} \al_j p(\al_j | \la)\nonumber\\
& =& E(\al_i, \la) E(\al_j,\la)
\end{eqnarray}
by equation (\ref{(3.4)}). Now let us consider the case of two successive
measurements, with options $\h a_1 ,\h a_1'$ and $\h a_2, \h a_2'$
respectively for measuring spin components. In each run of the
experiment, a random choice between $\{ \h a_1, \h a_1'\}$ and $\{
\h a_2, \h a_2'\}$ is made. Define $ \th_i$ ($ i=1,1' $) to be the
angle between $\h a_i$ and the positive $z$-axis, $ \th_{ij}$ ($
i=1,1' $ and $ j =2, 2' $) is the angle between $\h a_j$ and $\h
a_i$. Using condition (\ref{(3.6)}) and the result \cite{shimoni} \cite{jarrett}
$$ - 2s^2 \le xy + xy' + x'y - x'y' \le 2s^2,~~~~~
 x, y, x', y' \in \{-s, -s + 1, \ldots, s - 1, s\},$$
we obtain
\benr
\label{(3.6*)}
 -2s^2 \le E(\al_1, \al_2, \la) + E(\al_1, \al_2', \la) + E(\al_1', \al_2, \la) - E(\al_1', \al_2', \la) \le 2s^2.
\eenr
Multiplying by $\rho(\la) d\la$ and integrating over $\Lambda$, we
get the CHSH-type inequality \cite{chsh} (involving the hidden variable
$\la$) corresponding to performing two successive measurements of
spin-$s$ observables on a spin-$s$ initial state:
\benr
\label{(3.7)}
 |\lan BI \ran| =\frac{1}{2} |\lan \al_1 \al_2 \ran + \lan \al_1 \al_2'\ran + \lan \al_1' \al_2\ran - \lan \al_1' \al_2'\ran | \le s^2. 
\eenr
Similarly, using the algebraic fact
$$ - 2s^3 \le xyz' + xy'z + x'yz - x'y'z' \le 2s^3,$$ where
 $$x, y, z, x', y', z' \in \{-s, -s + 1, \ldots, s - 1, s\}$$
and\footnote{This is obtained by using equation (\ref{(3.4)}) and the
similar argument as has been used in deriving equation (\ref{(3.6)}).}
$$ E(\al_i, \al_j, \al_k, \la) = E(\al_i, \la) E(\al_j, \la) E(\al_k, \la),$$ we can prove Mermin-Klyshko Inequality (MKI) \cite{mermin90}, \cite{klyshko} for three successive measurements,
\benr
\label{(3.8)}
 |\lan MKI\ran| = \frac{1}{2}| \lan \al_1 \al_2 \al_3'\ran + \lan \al_1 \al_2' \al_3 \ran + \lan \al_1' \al_2 \al_3 \ran - \lan \al_1' \al_2' \al_3'\ran | \le s^3 .
\eenr 
Let $|  \lan MKI'\ran | \le s^3$, where $|\lan MKI'\ran| $ is
obtained from equation (\ref{(3.8)}) by interchanging primes with non-primes
in MKI. It is easily shown that
\benr
\label{(3.9)}
| \lan SI \ran| = | \lan MKI \ran +  \lan MKI'\ran| \le| \lan MKI \ran| + | \lan MKI'\ran|  \le 2s^3.
\eenr
This is the Svetlichny  inequality (SI) \cite{svetlinchi},\cite{seevsvet},\cite{cgprs02}.

For $n$ successive measurements on spin $s$ system, we define the MK polynomials recursively as follows:
\benr
\label{(3.10)}
M_1 = \al_1, ~M_1' = \al_1', 
\eenr
\benr
\label{(3.11)}
 M_n = \fr{1}{2} M_{n-1} (\al_n + \al_n') + \fr{1}{2} M_{n-1}' (\al_n - \al_n'),
\eenr
where $M_n'$ are obtained from $M_n$ by interchanging all primed and
non-primed $\al$'s. The recursive relation (\ref{(3.11)}) gives, for all $ 1\leq k\leq n-1$ \cite{cgprs02} ,\cite{cabello652002}:
\benr
\label{(3.12)}
M_n = \fr{1}{2} M_{n-k} (M_k + M_k') + \fr{1}{2} M_{n-k}' (M_k - M_k').
\eenr
In particular, we have
\benr
%\label{(3.7)}
M_2 = BI  = \fr{1}{2} (\al_1 \al_2 + \al_1'\al_2 + \al_1\al_2' - \al_1' \al_2'),
\eenr
\benr
%\label{(3.8)}
 M_3 = MKI = \fr{1}{2}(\al_1\al_2 \al_3' + \al_1\al_2'\al_3 + \al_1' \al_2\al_3 -\al_1'\al_2'\al_3').
\eenr 
We now show that in HVT, 
\benr
\label{(3.13)}
 | \lan M_n \ran | \le s^{n}.
\eenr 

\noi First note that (\ref{(3.13)}) is true for $n = 2, 3$ (equations (\ref{(3.7)}), (\ref{(3.8)})). Suppose it is true for $n = k$ i.e. $Max |\lan M_k\ran |
=s^{k}$. Now
$$ | \lan M_{k+1} \ran | = \fr{1}{2} | \lan M_k \al_{k+1}\ran + \lan M_k \al_{k+1}' \ran + \lan M_k' \al_{k+1} \ran - \lan M_k' \al_{k+1}' \ran |.$$
Since HVT applies here we can use (\ref{(3.4)}) to get
$$ |\lan M_{k+1}\ran| = \fr{1}{2} |\lan M_k\ran (\lan \al_{k+1} \ran + \lan \al_{k+1}'\ran ) + \lan M_k'\ran (\lan \al_{k+1}\ran - \lan \al_{k+1}' \ran )|.$$
This implies, by induction hypothesis (and using the fact that ${\rm
max} |\langle M_2 \rangle| = s^2$), that
$$ \max | \lan M_{k+1} \ran | = s \max | \lan M_k\ran| = s^{k+1}.$$
This result is derived for $n$ spin-$s$ particles by Cabello \cite{cabello652002}.

We now define a quantity, denoted by ${\eta}_n$, which will be
required later on. $\eta_{n}$, is the ratio between maximum $| \lan
M_n\ran| $ given by quantum correlation between $n$ successive
measurement's outputs and the maximal classical one,
\benr
\label{(3.14)}
\eta_{n}=\frac{max| \lan M_n\ran_{QM}|}{s^{n}}.
\eenr

\section{Mixed Input State for Arbitrary Spin}
\subsection{Two Successive Measurements (BI) }

We first deal with the case when input state is a mixed state whose eigenstates coincide with those of $\vec{S} \cdot \h a_0$ for some $\h a_0$ whose eigenvalues we denote by $\al_0\in\{ -s, \cdots s\}$. For spin 1/2 this is the most general mixed state because given any density operator $\rho_0$ for spin 1/2 (corresponding to some point within the Bloch sphere), we can find an $\h a_0$ such that the eigenstates of $\vec{S} \cdot , \h a_0$ and $\rho_0$ coincide. However, for $s > 1/2$, our choice forms a restricted class of mixed states. We note that these are the only states accessible via SG experiments. Thus we have
\benr
\label{(4.1)}
 \rho_0 = \sum_{\al_0} p_{\al_0} | \vec{S} \cdot \h a_0 , \al_0 \ran \lan \vec{S} \cdot \h a_0, \al_0 | ;~~~~~~\left( \sum_{\al_0} p_{\al_0}=1 \right) 
\eenr
After the first measurement along $\h a_1$, the resulting state of the system is
\benr
\label{(4.2)}
\rho_1 = \sum_{\al_1} M^\dagger_{\al_1} \rho_0 M_{\al_1},
\eenr
where
$$ M^\dagger_{\al_1} = M_{\al_1} = | \vec{S} \cdot \h a_1, \al_1 \ran \lan \vec{S} \cdot \h a_1, \al_1 | .$$
Now $\lan \al_1 \al_2 \ran_{QM}$ is the expectation value (according
to QM) that given the initial state ${\rho}_0$ (given in equation
(\ref{(4.1)})), the 1st measurement along $\hat{a}_1$ will give rise to any
value ${\alpha}_1 \in \{-s, -s + 1, \ldots, s - 1, s\}$, and then,
on the after-measurement state ${\rho}_1$ (given in equation (\ref{(4.2)})),
if one performs measurement along $\hat{a}_2$, one of the values
${\alpha}_2 \in \{-s, -s + 1, \ldots, s - 1, s\}$ will arise. So
\benr
\label{(4.3)}
\lan \al_1 \al_2 \ran_{QM}&=&Tr(\rho_1 \vec{S} \cdot \h a_1
\vec{S} \cdot \h a_2)
 = \nonumber\\&& \sum_{\al_0 \al_1 \al_2} p_{\al_0} \al_1 \al_2 | \lan \vec{S} \cdot \h a_0, \al_0 | \vec{S} \cdot \h a_1, \al_1 \ran |^2 | \lan \vec{S} \cdot \h a_1, \al_1 | \vec{S} \cdot \h a_2, \al_2 \ran |^2.\nonumber\\
 \eenr
Note that, since $\vec{S} \cdot \h a_i$ are complete observables,
all of whose eigenvalues are non degenerate, the probabilities
factorize like those of a Markov chain \cite{beck92}. Every factor in (\ref{(4.3)})  corresponds to the transition amplitude between two successive measurements. By equation (A.12), we get
\benr
\label{(4.4)}
\lan \al_1 \al_2 \ran = \fr{1}{2} \cos \th_{12} [A \cos^2 \th_{1} + B], 
\eenr
where
$$A  =  3 \chi - s(s+1) ,~~~B  =  s(s+1) - \chi,~~~\chi  =  \sum^{+s}_{\al_0= - s} \al^2_0 p_{\al_0}.$$
It is to be noted that although $A$ can have positive and negative values, $B$
will always be positive. Moreover, for all $\theta \in [0, 2\pi]$, if $A\geq 0$, $A {\cos}^2 {\theta}_1 + B$ is always positive and  if $A<0$, $A {\cos}^2 {\theta}_1 + B \ge B + A = 2\chi \ge 0$. We now have
the following expression for the quantity $BI$, appeared in equation
(\ref{(3.7)}):
\benr
\label{(4.5)}
BI&=&\fr{1}{4}\{(A \cos^2 \th_{1} + B) (\cos \th_{12} + \cos \th_{12}^{\prime})
 + (A \cos^2 \th_{1}^{\prime} + B)(\cos \th_{12}^{\prime \prime} -  \cos \th_{12}^{\prime \prime \prime})\}\nonumber\\
&=& \frac{3\chi - s(s + 1)}{4}\left\{{\rm cos}^2 {\theta}_1 \left({\cos} {\theta}_{12} + {\cos} {\theta}_{12}^{\prime}\right) + {\cos}^2 {\theta}_1^{\prime} \left({\cos} {\theta}_{12}^{\prime \prime} - {\cos} {\theta}_{12}^{\prime \prime \prime}\right)\right\}\nonumber \\
&+& \frac{s(s + 1) - \chi}{4}\left\{\left({\cos} {\theta}_{12} + {\cos} {\theta}_{12}^{\prime}\right) + \left({\cos} {\theta}_{12}^{\prime \prime} - {\cos} {\theta}_{12}^{\prime \prime \prime}\right)\right\} ,
\eenr
where (according to Appendix A) ${\theta}_1$ is the angle between
$\hat{a}_0$ and $\hat{a}_1$, ${\theta}_1^{\prime}$ is the angle
between $\hat{a}_0$ and $\hat{a}_1^{\prime}$, ${\theta}_{12}$ is the
angle between $\hat{a}_1$ and $\hat{a}_2$, ${\theta}_{12}^{\prime}$
is the angle between $\hat{a}_1$ and $\hat{a}_2^{\prime}$,
${\theta}_{12}^{\prime \prime}$ is the angle between
$\hat{a}_1^{\prime}$ and $\hat{a}_2$, ${\theta}_{12}^{\prime \prime
\prime}$ is the angle between $\hat{a}_1^{\prime}$ and
$\hat{a}_2^{\prime \prime}$. We have used, in Eq (\ref{(4.5)}), the expressions for
$A$ and $B$ in terms of $\chi$ and $s$. Note that the second term in
equation (\ref{(4.5)}) ({\it i.e.}, the term with the factor $\frac{s(s + 1)
- \chi}{4}$) is similar to the expression for $\langle {\alpha}_1
{\alpha}_2 \rangle + \langle {\alpha}_1 {\alpha}_2^{\prime} \rangle
+ \langle {\alpha}_1^{\prime} {\alpha}_2 \rangle - \langle
{\alpha}_1^{\prime} {\alpha}_2^{\prime} \rangle$ corresponding to
the CHSH inequality \cite{chsh}. And hence, its maximum value will occur
when we choose all the four vectors $\hat{a}_1$,
$\hat{a}_1^{\prime}$, $\hat{a}_2$, $\hat{a}_2^{\prime}$ on the same
plane. But we also have to take care about maximization of the first
term in equation (\ref{(4.5)}) and that might require these four vectors to be on different planes. In order to resolve this issue, we now consider the spherical-polar co-ordinates $({\theta}_1, {\phi}_1)$,
$({\theta}_1^{\prime}, {\phi}_1^{\prime})$, $({\theta}_2,
{\phi}_2)$, $({\theta}_2^{\prime}, {\phi}_2^{\prime})$ of the
vectors $\hat{a}_1$, $\hat{a}_1^{\prime}$, $\hat{a}_2$,
$\hat{a}_2^{\prime}$ respectively, where ${\theta}_1,
{\theta}_1^{\prime}, {\theta}_2, {\theta}_2^{\prime} \in [0, \pi]$
and ${\phi}_1, {\phi}_1^{\prime}, {\phi}_2, {\phi}_2^{\prime} \in
[0, 2\pi]$. Then $BI$ has the form
\benr
\label{(4.5*)}
BI&=& \frac{1}{4}(A{ \cos}^2 \theta_1 + B)[
\cos \theta_1 ( \cos \theta_2 + \cos
\theta_2^{\prime}) + \sin \theta_1\sin\theta_2 \cos (\phi_1 - \phi_2)\nonumber\\& + &\sin \theta_1\sin \theta_2^{\prime}  \cos (\phi_1 -\phi_2^{\prime})]+ \frac{1}{4}(A \cos^2 \theta_1^{\prime} + B) [\cos \theta_1^{\prime} ( \cos \theta_2 - \cos\theta_2^{\prime})\nonumber\\&+& \sin\theta_1^{\prime}\sin \theta_2  \cos (\phi_1^{\prime} -\phi_2) - \sin\theta_1^{\prime}\sin\theta_2^{\prime}\cos
(\phi_1^{\prime} - \phi_2^{\prime})].
\eenr
Here also the maximum value of $|BI|$ will occur when all the
vectors $\hat{a}_1$, $\hat{a}_1^{\prime}$, $\hat{a}_2$,
$\hat{a}_2^{\prime}$ lie on the same plane. This is obtained by:
$$\frac{\partial BI}{\partial \phi_1}=\frac{\partial BI}{\partial \phi_2}=\frac{\partial BI}{\partial \phi_1^{\prime}}=\frac{\partial BI}{\partial \phi_2^{\prime}}=0\Rightarrow \phi_1=\phi_1^{\prime}=\phi_2=\phi_2^{\prime}.$$ 
In that case, the maximum value of $|BI|$ will occur when ${\theta}_1^{\prime} = \pi -
{\theta}_1$, ${\theta}_2 = \pi/2$, ${\theta}_2^{\prime} = 0$ and
(correspondingly) the quantity
\benr
\label{(4.6)}
 \eta_{2} = \fr{|BI|}{s^2} = \left( \fr{1}{2s^2}\right) |(\sin \th_1 +
\cos \th_1)| (A \cos^2 \th_1 + B)
\eenr
 is maximized over all possible values of ${\theta}_1$ ($A \cos^2 \th_1 + B\geq0$). If $\eta_{2}
> 1$, the correlations for two successive measurements violate the CHSH-type inequality (\ref{(3.7)}), and hence a contradiction with the above-mentioned HVT. In fact ${\ds \fr{\pa \eta_{2}}{\pa \th_1} = 0}$ implies that
\benr
\label{(4.7)}
 B \tan^3 \th_1 + (2A-B) \tan^2 \th_1+ (3A + B) \tan \th_1 - (A+B) = 0. 
\eenr 
Real roots (for ${\rm tan} {\theta}_1$) of this equation give values
of $\th_1$ for which $\eta_{2}$ is maximum. The maximum value of
$\eta_{2}$ is evaluated at these $\th_1$'s.

We find that for $s=\frac{1}{2} $, $ \chi = 1/4$ for all  $\rho_0$,
and so $A = 0$, $B = 1/4$. So, from equation (\ref{(4.6)}), we have $\eta_2
= {\rm sin} {\theta}_1 + {\rm cos} {\theta}_1$. Therefore equation
(\ref{(4.7)}) becomes ${\rm tan}^3 {\theta}_1 - {\rm tan}^2 {\theta}_1 +
{\rm tan} {\theta}_1 - 1 = 0$, whose only one real solution is ${\rm
tan} {\theta}_1 = 1$. So ${\theta}_1 = \pi/4$ or $5\pi/4$.
${\theta}_1 = \pi/4$ gives the maximum possible value $\eta_{2} = \sq
2 > 1$. Thus all possible spin-$1/2$ states break BI for (proper
choices of) two successive measurements. This can be compared with
the measurement correlations corresponding to measurement of spin
observables on space-like separated two particles scenario where
only the entangled pure states break BI while not all entangled
mixed states break it \cite{werner89}.

From now on, we will use the range of values of the quantity $\xi
\equiv \chi/s^2$ to identify the parametric region of the initial
density matrix ${\rho}_0$ where the inequality (\ref{(3.7)}) will be
violated. Thus we see that for all spin-$1/2$ input states
${\rho}_0$, $\xi = 1$.\\
For a spin-$1$ system, we first consider all input states ${\rho}_0$
none of which have a contribution of $S_z = 0$ eigenstate. In this
case $\chi = 1$, $A = B = 1$. So $\eta_2 = (1/2)({\rm sin}
{\theta}_1 + {\rm cos} {\theta}_1)({\rm cos}^2 {\theta}_1 + 1)$ and
equation (\ref{(4.7)}) takes the form ${\rm tan}^3 {\theta}_1 + {\rm tan}^2
{\theta}_1 + 4{\rm tan}^3 {\theta}_1 - 2 = 0$. The only real root of
this equation is ${\rm tan} {\theta}_1 \approx 0.433$. Thus the
maximum possible value of $\eta_2$ is (using equation (\ref{(4.6)}))
$1.2112$ ( approximately). Thus we see that all input spin-$1$ states
${\rho}_0$, none of which has a component along $|S_z = 0\rangle$,
break BI (equation (\ref{(3.7)})) for proper choice of the observables.

Next, for $s = 1$, we consider the state ${\rho}_0$, for which
$p_{{\alpha}_0 = 0} = 1$, {\it i.e.}, $ \rho_0 = | \vec{S} \cdot
\hat{a}_0, 0 \ran \lan  \vec{S} \cdot \hat{a}_0, 0 | $. In this
case, $\chi = 0$, and so, $A = -2$, $B = 2$, $\xi = 0$. Then
equation (\ref{(4.7)}) takes the form $2{\rm tan}^3 {\theta}_1 - 6{\rm
tan}^2 {\theta}_1 - 4{\rm tan} {\theta}_1 = 0$. It has three real
solutions, which corresponds to ${\theta}_1 = 0$ (or $\pi$),
$74.3165^o$ (approx.), $150.6836^o$ (approx.). The maximum possible
value of $\eta_2$ occurs at ${\theta}_1 = 74.3165^o$, and the
corresponding value is given by $\eta_2 \approx 1.1428$. Thus the
state $ \rho_0 = | \vec{S} \cdot \hat{a}_0, 0 \ran \lan  \vec{S}
\cdot \hat{a}_0, 0 | $ breaks the BI (equation (\ref{(3.7)})).

For $s = 1$, when $0 < p_{{\alpha}_0 = 0} \equiv p_0~ ({\rm say})~ <
1$, we have $\chi = 1 - p_0 = \xi$, $A = 1 - 3p_0 = 3\xi - 2$ and $B
= 1 + p_0 = 2 - \xi$. We then have $\eta_2 = (1/2)|{\rm sin}
{\theta}_1 + {\rm cos} {\theta}_1| \{(3\xi - 2){\rm cos}^2
{\theta}_1 + (2 - \xi)\}$ (by equation (\ref{(4.6)})), and equation (\ref{(4.7)})
becomes $(2 - \xi){\rm tan}^3 {\theta}_1 + (7\xi - 6){\rm tan}^2
{\theta}_1 + 4(2\xi - 1){\rm tan} {\theta}_1 - 2\xi = 0$. In this
case, one can show numerically that the BI will break ({\it i.e.},
$\eta_2 > 1$) if and only if either $0 < \xi < 0.33$ or $0.77 < \xi
< 1$ ( equivalently, either $0.67 < p_0 < 1$ or $0 < p_0 < 0.23$).

Thus we see that, when $s = 1$, only those input states ${\rho}_0$
( given in equation (\ref{(4.1)})) will break BI for each of which
$p_{{\alpha}_0 = 0} \in [0, 0.23) \cup (0.67, 1]$.

Next we consider the situations where $s > 1$. Note that,  by
definition (true for all $s$),
$$\xi = \left(p_s + p_{- s}\right) + \left(p_{s - 1} + p_{- s +
1}\right)\left(1 - \frac{1}{s}\right)^2 + \left(p_{s - 2} + p_{- s +
2}\right)\left(1 - \frac{2}{s}\right)^2 + \ldots,$$ where $0 \le
p_s, p_{- s}, p_{s - 1}, p_{- s + 1}, p_{s - 2}, p_{- s + 2}, \ldots
\le 1$ and $\sum_{{\alpha}_0 = - s}^{s} p_{{\alpha}_0} = 1$.
Therefore, we must have $0 \le \xi \le 1$. In this case, equations
(\ref{(4.6)}) and (\ref{(4.7)}) respectively take the forms
\benr
\label{(4.7.1)}
\eta_2 = \frac{1}{2s^2}\left|{\rm sin} {\theta}_1 + {\rm cos}
{\theta}_1\right|\left[\{3{\xi}s^2 - s(s + 1)\}{\rm cos}^2
{\theta}_1 + \{s(s + 1) - {\xi}s^2\}\right],
\eenr
\benr
\label{(4.7.2)}
&\{s(s + 1) - {\xi}s^2\}{\tan}^3 {\theta}_1 + \{7{\xi}s^2 - 3s(s
+ 1)\}{\tan}^2 {\theta}_1 +&\nonumber\\&\{8{\xi}s^2 - 2s(s + 1)\}{\tan}
{\theta}_1 - 2{\xi}s^2 = 0.&
\eenr
Let us first consider the input states of the form
\benr
\label{(4.8)}
 \rho_0 = p_s | \vec{S} \cdot \h a_0, s \ran \lan \vec{S} \cdot \h a_0,s| + p_{-s}|  \vec{S} \cdot \h a_0 , - s\ran \lan \vec{S} \cdot \h a_0, -s |,
 \eenr
 with $p_s + p_{-s} = 1$ and $p_{s - 1}$, $p_{- s + 1}$, $p_{s - 2}$,
$p_{- s + 2}$, $\ldots = 0$. Thus we see here that $\xi = 1$. Also
equations (\ref{(4.7.1)}) and (\ref{(4.7.2)}) have respectively been turned into the forms
\benr
\label{(4.7.1*)}
\eta_2 = (1/2s)|{\rm sin} {\theta}_1 + {\rm cos}
{\theta}_1|\{(2s - 1){\rm cos}^2 {\theta}_1 + 1\},
\eenr
\benr
\label{(4.7.1**)}
{\rm tan}^3 {\theta}_1 + (4s - 3){\rm tan}^2 {\theta}_1 + (6s -
2){\rm tan} {\theta}_1 - 2s = 0.
\eenr
 As here $s > 1$, therefore the
last equation will have only one positive root and the other two
roots will be complex. The positive root will correspond to an angle
${\theta}_1^{max} (s) \in (0, \pi/4)$ for which it can be shown that
$\eta_2^{max} (s) \equiv \eta_2({\theta}_1^{max} (s)) > 1$ for all
$s > 1$. Hence, in this case, BI is violated.

If ${\rho}_0$ has contribution from neither of the states
corresponding to ${\alpha}_0 = \pm s$ ({\it i.e.}, $p_s = p_{- s} =
0$), we have
$$\xi = \left(p_{s - 1} + p_{- s +
1}\right)\left(1 - \frac{1}{s}\right)^2 + \left(p_{s - 2} + p_{- s +
2}\right)\left(1 - \frac{2}{s}\right)^2 + \ldots \le \left(1 -
\frac{1}{s}\right)^2.$$ From equation (\ref{(4.6)}) it follows that
$$\eta_2 = \frac{1}{\sqrt{2} s}\left|{\rm sin} \left({\theta}_1 +
\pi/4\right)\right|\left[(s + 1 - s\xi) + (3s\xi - s - 1) {\rm
cos}^2 {\theta}_1\right]$$
$$< \frac{1}{\sqrt{2} s} \times 1 \times
[(s + 1 - s\xi) + (3s\xi - s - 1) \times 1] = \sqrt{2}\xi \le
\sqrt{2}\left(1 - \frac{1}{s}\right)^2.$$ But the quantity
$\sqrt{2}(1 - 1/s)^2$ is less than 1 for all $s = 1/2, 1, 3/2,
\ldots, 6$. Therefore, for $s > 1$, if the initial state ${\rho}_0$
has contribution from neither of the states corresponding to
${\alpha}_0 = \pm s$, BI will be satisfied for all $s \le 6$.

Thus we see that whenever $s \in \{3/2, 2, 5/2, \ldots, 6\}$, in
order that ${\rho}_0$ violates BI, the associated quantity $\xi$
must have values near $1$. In table 1, we have given the ranges of
values of $\xi$ (obtained numerically) for which BI is violated,
starting from $s = 1/2$. The case when $s \rightarrow \infty$ has
also been considered in table 1.

\newpage
\bc
{\bf Table 1} \\
\vspace{.1in}

\bt{||c|c||c|c||c|c||} \hline
$s$ & $\xi$ & $s$ & $\xi$ & $s$ & $\xi$ \\
\hline
$\fr{1}{2}$ & $\xi = 1$ & $\fr{5}{2}$ & $0.847 \le \xi \le 1$ & $\fr{9}{2}$ & $0.858 \le \xi \le 1$\\
\hline
1 & $ 0 \le \xi \le 0.33$ and  $0.77 \le \xi \le 1$ & 3 & $0.851 \le \xi \le 1$  & 5 & $0.859 \le \xi \le 1$\\
\hline
$\fr{3}{2} $ & $0.824 \le \xi \le 1 $  & $\fr{7}{2}$ & $0.854 \le \xi \le 1$ & $\fr{11}{2}$ & $0.860 \le \xi \le 1$  \\
\hline 2 & $0.84 \le \xi \le 1$  & 4 & $0.856 \le \xi \le 1$
& 6 & $0.862 \le \xi \le 1$ \\
\hline
& & & & $\iny$ & $0.87 \le \xi \le 1$  \\
\hline \et

\vspace{.2in}

{\small The ranges of $\xi$, for which BI is violated.}\\
\ec

The maximum violation of Bell inequality, characterized by
$\eta_{2}$, decreases monotonically with $s$. Table 2 summarizes
( obtained numerically) the maximum allowed value of $\eta_2$ for
each $s$. We see  from this table that for all spin values $s$,
BI is broken. Note that there is a sharp decrease in $\eta_{2}$ from
$s = \fr{1}{2}$ to $s = 1$, while $\eta_{2}$ decreases slowly as $s$
increases from 1. A possible reason is that, for $s = 1/2$, all
states break BI while for $s \ge 1$, only a fraction of spin states
break it.

\bc
{\bf Table 2} \\
\vspace{.1in}

\bt{||c|c||c|c||c|c||}
\hline
$s$ & $\eta_{2} $
 & $s$ & $\eta_{2}$  & $s$ & $\eta_{2}$   \\
\hline
$\fr{1}{2}$ & $\sq 2$ & $\fr{5}{2}$ & $1.1638$ &  $\fr{9}{2}$ & $1.1538$ \\
\hline
 1 & $1.2112$  & 3 & $1.1599$ & 5 & $1.1526$ \\
\hline
 $\fr{3}{2} $ & $1.1817 $ & $\fr{7}{2}$  & $1.1572$  & $\fr{11}{2}$ & $1.1517$ \\
\hline
2 & $1.17$ & 4 & $1.1553$ & 6 & $1.1509$ \\
\hline
& & &  & $\iny$ & $1.143$  \\
\hline
\et
\vspace{.1in}

{\small The maximum violation of BI for different spin values for
two successive measurements.}

\ec

We now consider a case where the initial state ${\rho}_0^{max}$
(given in equation (\ref{(4.8)})) is contaminated by the maximally noisy
state, resulting in the state
\benr
\label{(4.9)}
 \rho(f) = \left( 1-f\right)  \rho_0^{\max} + \fr{f}{2s +1} I,
 \eenr
where the positive parameter $f$ ($\le 1$) is the probability of the
noise contamination of the state $\rho_0^{\max}$. Proceeding as
before ( see equation (\ref{(4.4)})), we get
\benr
\label{(4.10)}
\lan \al_1 \al_2 \ran = \fr{1}{2} \cos \th_{12} [A' \cos^2 \th_{1}+B'] 
\eenr
where
$$ A' =(1-f)(2s-1)s; ~~~~ B' =(1-f) s + \fr{2}{3} f (s+1)s,$$
which leads to
\benr
\label{(4.11)}
\eta_{noise} = \left( \fr{1}{2s^2}\right) (\sin \th_1 + \cos \th_1) (A' \cos^2 \th_1 + B').
\eenr
Using the maximization procedure ({\it i.e.}, taking $\frac{\partial
\eta_{noise}}{\partial \theta_1} = 0$), ${\rm tan} \theta_1$ for
maximum $\eta_{noise}$ is given by a real root of
\benr
\label{(4.12)}
B'\tan^3\th_1 + (2A'-B') \tan^2\th_1 + (3A' + B') \tan \th_1 - (A'+B') = 0.
\eenr
The range of $f$ for which $\eta_{noise} > 1$ is tabulated in table
3. Note that for $s = \fr{1}{2}$ the state corresponding to $f=1$
( the random mixture) also breaks BI!  Of course we have already
shown that for $s = \fr{1}{2}$ , BI is broken for all states. This
indicates that the notion of ``classicality'', compatible with the
usual local HVT, is different in nature from the notion of
classicality that would arise from the non-violation of BI here.

\bc
{\bf Table 3} \\

\vspace{.1in}

\bt{||c|c||c|c||c|c||}
\hline
$s$ & $f$ & $s$ & $f$ & $s$ & $f$ \\
\hline
$\fr{1}{2}$ & $0 \le f \le 1$ & $\fr{5}{2}$ & $f < 0.287$ & $\fr{9}{2}$ & $f< 0.239$ \\
\hline
 1 & $f<0.696$ & 3 & $f<0.267$ & 5 & $f<0.234$ \\
\hline
$\fr{3}{2} $ & $f<0.395$ & $\fr{7}{2}$ & $f<0.254$ & $\fr{11}{2}$ & $f<0.230$\\
\hline
2 & $f<0.321$  & 4 & $f<0.245$ & 6 & $f<0.227$ \\
\hline
& & & & $\iny$ & $f<0.195$  \\
\hline
\et
\vspace{.1in}

{\small The range of the noise $f$ over which BI is violated.} \ec

Table 3 answers the question, ``what is the maximum fraction of
noise that can be added to $ \rho_0^{\max}$, which maximally breaks
BI, so that the state has stronger than ``classical
correlations\footnote{{\it i.e.}, correlations obeying ``realism''
and ``locality in time'', as described in section 3.}?'' We see that
the corresponding fraction of noise ({\it i.e.}, for which BI is
violated) decreases monotonically with $s$, or with the dimension of
the Hilbert space. This may be compared with the results of Collins
and Popescu \cite{collinpopescu01} who found that the nonlocal character of the correlations between the outcomes of measurements performed on entangled systems separated in space is robust in the presence of noise. They showed that, for any fraction of noise, by taking the Hilbert space of large enough dimension, one can find bipartite entangled states giving nonlocal correlations. These results have been obtained by considering two successive measurements on each
part of the system. On the other hand, in the present case of
successive measurements on the single spin state, we see that the
fraction of noise that can be added so that the quantum correlations
continue to break Bell inequality, falls off monotonically with $s$,
or the dimension of the Hilbert space. For $s=\frac{1}{2}$ all
fractions $f \leq 1$ are allowed, while for large s, $f < 0.195$.

%\vspace{.2in}

\subsection{Three Successive Measurements (MKI)}

We again assume the input state to be given by equation (\ref{(4.1)}). Using equation (A.19) :
\benr
\label{(4.13)}
\lan \al_1 \al_2 \al_3 \ran = \fr{1}{16} \cos \th_{23}\{ \cos \th_{1} [M \cos^2 \th_{12} + N]+R[3\cos^2 \th_{12}-1] \} 
\eenr
where
\begin{eqnarray*}
&&M=\sum^{+s}_{\al_0= - s}p_{\al_0} \al_0[9 \al^2_0 + s(s+1)-3],\\
&&N=\sum^{+s}_{\al_0= - s}p_{\al_0} \al_0[5s(s + 1)-3\al^2_0 +1],\\
&&R=\sum^{+s}_{\al_0= - s}p_{\al_0} \al_0[5 \al^2_0 -3s(s+1)+1],
\end{eqnarray*}
$\theta_1$ is the angle between $\hat{a}_0$ and $\hat{a}_1$
(measured with respect to the right-handed system $(\hat{a}_0,
\hat{a}_1, (\hat{a}_0 \times \hat{a}_1)/|\hat{a}_0 \times
\hat{a}_1|)$), $\theta_{12}$ is the angle between $\hat{a}_1$ ,
$\hat{a}_2$ (measured with respect to the right-handed system
$(\hat{a}_1, \hat{a}_2, (\hat{a}_1 \times \hat{a}_2)/|\hat{a}_1
\times \hat{a}_2|)$), etc.

We now consider the pure state $| \vec{S} \cdot \h a_0, s \ran \lan
\vec{S} \cdot \h a_0, s|$ instead of considering the most general
state ${\rho}_0$, given in equation (\ref{(4.1)}). So here $M = s(2s - 1)(5s + 3)$, $N = s(2s^2 + 5s + 1)$, and $R = s(2s - 1)(s - 1)$.
Substituting the correlations like that in equation (\ref{(4.13)}) in the MKI (given in equation (\ref{(3.8)})), using the above-mentioned values of $M$, $N$, $R$, and then finding out the conditions ( numerically) for
which $\eta_3 \equiv |MKI|/s^3$ is maximized, we get the maximum
possible $\eta_{3}$-values for different spins as summarized in
table 4.
%\newpage
\bc
{\bf Table 4} \\
\vspace{.1in}
\bt{||c|c||c|c||c|c||}
\hline
$s$ & $\eta_{3} $ & $s$ & $\eta_{3} $ & $s$ & $\eta_{3} $ \\
\hline
$\fr{1}{2}$ & $\sq 2$ &  $\fr{5}{2}$ & $1.1736$ & $\fr{9}{2}$ & $1.1634$ \\
\hline
1 & $1.2178$ & 3 & $1.1698$ & 5 & $1.1621$ \\
\hline
 $\fr{3}{2} $ & $1.1907 $  & $\fr{7}{2}$ & $1.1670$  & $\fr{11}{2}$ & $1.1610$\\
\hline
 2 & $1.1793$  & 4 & $1.1650$  & 6 & $1.1601 $ \\
\hline
& & & & $\iny$ & $1.1527 $  \\
\hline
\et

\vspace{.1in}

{\small The maximum violation of MKI for different spin values for
three successive measurements.}

\ec

We see that $\eta_{3}>1$ for all spins and $\eta_{3}>\eta_{2}$
except $s=\frac{1}{2}$, while $\eta_{3}=\eta_{2}=\sqrt{2}$ for $s =
1/2$. Also $\eta_{3}$, like $\eta_{2}$, decreases monotonically with
$s$. It is interesting, in the case of two and three successive
measurements of spin $s$ prepared in a pure state, the maximum
violation of BI and MKI tends to a constant  for arbitrary large $s$
.$$\eta_{3}(s\rightarrow\infty)=1.153~ ({\rm approx.}),$$
$$\eta_{2}(s\rightarrow\infty)=1.143~ ({\rm approx.}).$$
It is thus seen that large quantum numbers do not guarantee
``classical'' ( as defined in this chapter) behavior.

It is straightforward to check that, three successive measurements
satisfy  Svetlichny Inequality (SI) (equation (\ref{(3.9)})). The reason is that, for all $s$, the settings of the measurement directions which
maximize $MKI'$are obtained from those which maximize $MKI$ by
interchanging primes on the corresponding unit vectors. Thus these
two settings are incompatible so that we cannot get a single set of
measurement directions, which maximize both $MKI$ and $MKI'$.  In
fact, for all $s$, the measurement directions which maximize $MKI$
$(MKI')$ correspond to $MKI' = 0$ $(MKI = 0)$.\\ We now consider the situation of three consecutive observations but two-fold correlations for two measurements $\vec{S}.\hat{a}_1$ and $\vec{S}.\hat{a}_3$ performed, say, at time $t_1$ and $t_3$, but where an additional measurement ($\vec{S}.\hat{a}_2$) is performed at time $t_2$ lying between $t_1$ and $t_3$~$(t_1<t_2<t_3)$.\\
By substituting Eq (A.20) in Bell type inequality Eq(\ref{(3.7)}) and simplifying, we obtain:
\begin{eqnarray}
\label{(4.14)}
|BI|&=&\frac{1}{2}|[\cos\theta_{32}+\cos\theta_{3^{\prime}2}]\lan \al_1  \al_2 \ran + [\cos\theta_{32}-\cos\theta_{3^{\prime}2}]\lan \al_1^{\prime}\al_2 \ran| \nonumber \\ &\leq & \frac{1}{2}|[\cos\theta_{32}+\cos\theta_{3^{\prime}2}]| |\lan \al_1  \al_2 \ran |+
|[\cos\theta_{32}-\cos\theta_{3^{\prime}2}]| |\lan \al_1^{\prime}\al_2 \ran|\nonumber\\ &\leq& \cos\theta_{32}s^2\leq s^2. 
\end{eqnarray}
We have used $max|\lan \al_1  \al_2 \ran |=max |\lan \al_1^{\prime}\al_2 \ran|=s^2$.\\
So, the correlation function (\ref{(4.14)}) for a given measurement performed at $t_2$ cannot violate the Bell type inequality for measurements at $t_1$ and $t_3$. Therefore, any measurement performed at time $t_2$ ``disentangles" events at time $t_1$ and $t_3$ if $t_1<t_2<t_3$ \cite{vedral}.\\ 

\section{ n Successive Measurements for Spin-$\frac{1}{2}$}
\subsection{Violation Mermin-Klyshko Inequality (MKI) by n Successive Measurements }
We consider now $n$ successive measurements in direction $\vec{S}
\cdot \h a_i, (i = 1, 2, 3, \ldots,n)$ on a spin $s = \fr{1}{2}$
particle in  a mixed state. For simplicity we take the eigenvalues
to be $\al_k = \pm 1$, {\it i.e.}, the eigenvalues of $\si_z$ are
taken here as $\pm 1$ instead of $\pm (1/2)$. We also write $| \al_k
\ran$ for $|\vec{S} \cdot \h a_k, \al_k\ran$. The initial state is
taken as
\benr
\label{(5.1)}
 \rho_0 = p_{+ 1} | \al_0 = + 1 \ran \lan \al_0 = + 1| + p_{- 1} |\al_0 = - 1\ran \lan \al_0 = - 1|.
 \eenr
For a spin-$\fr{1}{2}$ system, we have
\benr
\label{(5.2)}
|\lan \al_{k-1} | \al_k \ran |^2 = \fr{1}{2} (1 + \al_{k-1} \al_k \cos \th_{k-1,k}) 
\eenr
$$ {\rm where}~ \cos \th_{k-1, k} = \h a_{k-1} \cdot \h a_k~ {\rm for}~ k = 1, 2, \ldots, n. $$
So, given the input state $|{\alpha}_0\rangle$, the (joint)
probability that the measurement outcomes will be ${\alpha}_1 \in
\{+ 1, - 1\}$ in the first measurement, ${\alpha}_2 \in \{+ 1, -
1\}$ in the second measurement, $\ldots$, ${\alpha}_n \in \{+ 1, -
1\}$ in the $n$-th measurement, will be given by
\benr
\label{(5.3)}
p(\al_1, \al_2, \cdots, \al_n) = \fr{1}{2^n} \prod^n_{i=1} (1 + \al_{i-1} \al_i \cos \th_{i-1, i}).
\eenr
Thus we see that given the input state ${\rho}_0 = \sum_{{\alpha}_0
= \pm 1} p_{{\alpha}_0} |{\alpha}_0 \rangle \langle {\alpha}_0|$,
the average output state after $n$ successive measurements will be
given by $${\rho}_n = \sum_{{\alpha}_0, {\alpha}_1, \ldots,
{\alpha}_n = \pm 1} p_{{\alpha}_0} p\left({\alpha}_1, {\alpha}_2,
\ldots, {\alpha}_n\right) \left| {\alpha}_n \rangle \langle
{\alpha}_n \right|.$$ Then, for $n$ successive measurements on
spin-$1/2$ system,
\begin{eqnarray}
\label{(5.4)}
\lan \al_{n-1} \al_n \ran_{QM} &=& \sum_{{\alpha}_{n - 1},
{\alpha}_n = \pm 1} {\alpha}_{n - 1} \{{\rm coeff.}~ {\rm of}~
| {\alpha}_{n - 1} \rangle \langle {\alpha}_{n - 1} |~
{\rm in}~ {\rho}_{n - 1}\} {\alpha}_n \left|\langle {\alpha}_{n
- 1} | {\alpha}_n \rangle\right|^2\nonumber\\
                               &=& \sum_{\al_0 = \pm 1} p_{\al_0} \sum_{\al_1, \al_2, \ldots, \al_n = \pm 1} \al_{n-1} \al_n p(\al_1, \al_2, \cdots, \al_n)\nonumber\\
& = & \sum_{\al_0 = \pm 1} p_{\al_0} 2^{-n} \sum_{{\alpha}_1,
{\alpha}_2, \ldots, {\alpha}_n = \pm 1} \prod^n_{i=1} \al_{n-1}
\al_n (1 + \al_{i-1} \al_i \cos \th_{i - 1, i})\nonumber\\ &=& \cos \th_{n-1, n}
\end{eqnarray}
by equation (\ref{(5.3)}). Further
\begin{eqnarray}
\label{(5.5)}
\lan \al_n \ran_{QM} & = & \sum_{{\alpha}_n = \pm
1} {\alpha}_n~ \{{\rm coeff.}~ {\rm of}~ \left| {\alpha}_n
\rangle \langle {\alpha}_n \right|~ {\rm in}~ {\rho}_n\} \nonumber\\
& = & \sum_{\al_0 = \pm 1} p_{\al_0} \sum_{\al_1, \al_2, \ldots, \al_n = \pm 1} \al_n p\left(\al_1, \al_2, \ldots, \al_n\right) \nonumber\\
& = & \sum_{\al_0 = \pm 1} p_{\al_0} 2^{-n} \sum_{\al_1, \al_2, \ldots, \al_n = \pm 1} \prod^n_{i=1} \al_n (1 + \al_{i-1} \al_i \cos \th_{i - 1, i}) \nonumber\\
& = & (p_{+ 1} - p_{- 1}) \cos \th_1 \cos \th_{12} \cdots \cos
\th_{n-1, n} 
\end{eqnarray}
where $\theta_1 \equiv \theta_{0, 1}$, $\theta_{12} \equiv
\theta_{1, 2}$, etc. Now equations (\ref{(5.4)}) and (\ref{(5.5)}) give,
\benr
\label{(5.6)}
 \lan \al_n \ran_{QM} = \lan \al_1 \ran_{QM} \lan \al_2 \al_3 \ran_{QM} \cdots \lan \al_{n-1} \al_n\ran_{QM}.
\eenr
Further,
\benr
\label{(5.7)}
&\lan \al_{n-k} \cdots \al_n \ran_{QM}=&\nonumber\\& \sum\limits_{\al_0} p_{\al_0} 2^{-n}  \sum\limits_{\al_1, \al_2, \ldots, \al_n = \pm 1} \prod^n_{i=1} (\al_{n-k} \cdots \al_n) (1 + \al_{i-1} \al_i \cos \th_{i-1,i})=&\nonumber
\\& \left\{ \ba{ll}  \lan \al_1 \ran_{QM} \lan \al_2 \al_3\ran_{QM} \cdots \lan \al_{n-1} \al_n\ran_{QM} & k ~\mbox{even} \\\\
  \lan \al_{n-k} \al_{n-k+1}\ran_{QM} \lan \al_{n-k+2}
 \al_{n-k+3} \ran_{QM} \cdots \lan \al_{n-1} \al_n \ran_{QM} & k ~ \mbox{odd} \ea \right.&\nonumber\\ 
 \eenr
All of the above results are inherently quantum and are not
compatible with HVT ( see the discussion in the next paragraph). The
first two results ((\ref{(5.5)}) and (\ref{(5.6)})) are the special cases of the last result (\ref{(5.7)}) for $k = 1$ and $k = 0$ (with $\al_0 = 1)$. If the number of variables ( which are averaged) is odd (i.e. $k$ is even)
the average depends on the measurements prior to $(n-k)$, while in
the other case the average does not depend on the measurements prior
to $(n-k)$. For example, for two successive measurements ( taking $n
= 2$ and $k = 1$), gives $\lan \al_1 \al_2 \ran = \cos \th_{12}$,
which is independent of the initial state. On the other hand, for
three successive measurements ( taking $n = 3$ and $k = 2$), we have
$\lan \al_1 \al_2 \al_3 \ran = \lan \al_1 \ran \lan \al_2 \al_3
\ran$ -- showing its dependence on the initial state (as $\langle
{\alpha}_1 \rangle = (p_{+ 1} - p_{- 1}) {\rm cos} {\theta}_1$
depends upon the initial state ${\rho}_0 = \sum_{{\alpha}_0 = \pm 1}
p_{{\alpha}_0} |{\alpha}_0 \rangle \langle {\alpha}_0|$). Moreover,
the correlation $\langle {\alpha}_1 {\alpha}_2 {\alpha}_3 {\alpha}_4
\rangle_{QM}$ for four successive measurements ( for example ) turns
out to be dependent only on the two `disjoint' correlations $\langle
{\alpha}_1 {\alpha}_2 \rangle_{QM}$ and $\langle {\alpha}_3
{\alpha}_4 \rangle_{QM}$ . In general, we have:
\begin{eqnarray}
\left\langle\alpha_1\alpha_2,\ldots,\alpha_{2p}\right\rangle=\langle\alpha_1\alpha_2\rangle\langle\alpha_3\alpha_4\rangle\ldots\langle\alpha_{2p-1}\alpha_{2p}\rangle
\end{eqnarray}
and
\begin{eqnarray}
\left\langle\alpha_1\alpha_2,\ldots,\alpha_{2p+1}\right\rangle=\langle\alpha_1\rangle\langle\alpha_2\alpha_3\rangle\ldots\langle\alpha_{2p}\alpha_{2p+1}\rangle.
\end{eqnarray}
Interestingly if $\h a_0 \perp \h a_1$ so that ${\rm cos} {\theta}_1 = 0$ ( and so,
$\lan \al_1 \ran_{QM} = 0$) or, if the initial state is the random
mixture $(1/2) \sum_{{\alpha}_0 = \pm 1} |{\alpha}_0 \rangle \langle
{\alpha}_0|$ ( and so $\langle {\alpha}_1 \rangle = 0$), then for all
even $k$, $$ \lan \al_{n-k} \cdots \al_n \ran_{QM} = 0$$ and so $$ \lan
\al_1 \al_2 \cdots \al_{n=2p+1} \ran_{QM} = 0.$$
We shall now show that for $n$ successive experiments ( with $n >
1$), QM violates the inequality $|\langle MKI \rangle| \le s^3$ ( see
equation (\ref{(3.8)})) up to $\sq 2$ for $ s = 1/2$ systems. We take the eigenvalues to be $\al_k = \pm 1$ so $ | \lan M_k \ran |_{HVT} \leq
1$). We have already shown that for $n = 2$ and $n = 3$, the
corresponding MKI's are violated (section 4).
Now, we know that temporal two-fold correlations $\lan \al_{k-1} \al_{k}'\ran$ , $\lan \al_{k-1}' \al_{k} \ran $, $ \lan \al_{k-1} \al_{k} \ran $ and $ \lan \al_{k-1}' \al_{k}' \ran $  are independent on the previous measurements. So by using equations (\ref{(5.7)}) and (\ref{(3.12)}) we find that
\benr
\label{(5.8)}
|\lan M_{k}\ran| = \fr{1}{2} | \lan M_{k-2}  \ran [\lan \al_{k-1} \al_{k}'\ran + \lan \al_{k-1}' \al_{k} \ran ] + \lan M_{k-2} '\ran [ \lan \al_{k-1} \al_{k} \ran - \lan \al_{k-1}' \al_{k}' \ran ] |.\nonumber\\ 
\eenr
We now consider the spherical-polar co-ordinates $({\theta}_{k-1}, {\phi}_{k-1})$,
$({\theta}_{k-1}^{\prime}, {\phi}_{k-1}^{\prime})$, $({\theta}_k,
{\phi}_k)$, $({\theta}_k^{\prime}, {\phi}_k^{\prime})$ of the
vectors $\hat{a}_{k-1}$, $\hat{a}_{k-1}^{\prime}$, $\hat{a}_k$,
$\hat{a}_k^{\prime}$ respectively, where all ${\theta}\in [0,\pi]$
and all ${\phi}\in[0, 2\pi]$. Then $| \lan M_{k}\ran|$ has the form
\begin{eqnarray}
\label{(5.9)}
|\lan M_{k}\ran|&=&\frac{1}{2}|\lan M_{k-2}\ran [\cos\theta_{k-1}\cos\theta_k^{\prime}+\sin\theta_{k-1}\sin\theta_k^{\prime}\cos(\phi_{k-1}-\phi_k^{\prime})\nonumber\\&+& \cos\theta_{k-1}^{\prime}\cos\theta_k+\sin\theta_{k-1}^{\prime}\sin\theta_k\cos(\phi_{k-1}^{\prime}-\phi_k)]\nonumber\\&+&\lan M_{k-2}^{\prime}\ran [\cos\theta_{k-1}\cos\theta_k+\sin\theta_{k-1}\sin\theta_k \cos(\phi_{k-1}-\phi_k)\nonumber\\&-& \cos\theta_{k-1}^{\prime}\cos\theta_k^{\prime}-\sin\theta_{k-1}^{\prime}\sin\theta_k^{\prime}\cos(\phi_{k-1}^{\prime}-\phi_k^{\prime})]|.
\end{eqnarray}
We know from
$$\frac{\partial |\lan M_{k}\ran|}{\partial \phi_{k-1}}=\frac{\partial | \lan M_{k}\ran|}{\partial \phi_k}=\frac{\partial | \lan M_{k}\ran|}{\partial \phi_{k-1}^{\prime}}=\frac{\partial | \lan M_{k}\ran|}{\partial \phi_k^{\prime}}=0\Rightarrow \phi_{k-1}=\phi_{k-1}^{\prime}=\phi_k=\phi_k^{\prime}.$$ 
The maximum value of $|\lan M_{k}\ran|$ will occur when all the
vectors $\hat{a}_{k-1}$, $\hat{a}_{k-1}^{\prime}$, $\hat{a}_k$,
$\hat{a}_k^{\prime}$ lie on the same plane. We obtain:
\begin{eqnarray}
\label{(5.10)}
|\lan M_{k}\ran|&\leq&\frac{1}{2}|\lan M_{k-2}\ran [\cos(\theta_{k-1}-\theta_k^{\prime})+\cos(\theta_{k-1}^{\prime}-\theta_k)]\nonumber\\&+&\lan M_{k-2}^{\prime}\ran [\cos(\theta_{k-1}-\theta_k)-(\cos\theta_{k-1}^{\prime}-\theta_k^{\prime})]|\nonumber\\&\leq&\frac{1}{2}|\lan M_{k-2}\ran| [\cos(\theta_{k-1}-\theta_k^{\prime})+\cos(\theta_{k-1}^{\prime}-\theta_k)]\nonumber\\&+&\frac{1}{2}|\lan M_{k-2}^{\prime}\ran| [\cos(\theta_{k-1}-\theta_k)-(\cos\theta_{k-1}^{\prime}-\theta_k^{\prime})]|.
\end{eqnarray}
By substituting $x=\theta_{k-1}-\theta_k^{\prime}$, $y=\theta_{k-1}^{\prime}-\theta_k$, $z=\theta_{k-1}-\theta_k$ and $\theta_{k-1}^{\prime}-\theta_k^{\prime}=x+y-z$ in above-equation and by using $$\frac{\partial |\lan M_{k}\ran|}{\partial x}=\frac{\partial |\lan M_{k}\ran|}{\partial y}=\frac{\partial |\lan M_{k}\ran|}{\partial z}=0,$$ we get $x=y=-z=\pi/4.$ Finally, by using  the fact  $| \lan M_{k}\ran |+| \lan M_{k}^{\prime}\ran |\leq 2,$ we obtain:
\benr
\label{(5.11)}
 | \lan M_{k}\ran | \leq \frac{\sqrt{2}}{2} \{ | \lan M_{k-2}\ran + \lan M_{k-2}^{\prime} \ran | \}\leq \sqrt{2}.
\eenr
 One can get this result by induction hypothesis. Thus, we conclude that QM violates the MKI ($|\langle M_n
\rangle| \le 1$) for $n$ successive measurements upto $ \sqrt{2}$,
{\it i.e.},
\benr
\label{(5.12)}
\eta_{n}=\sqrt{2}. 
\eenr
\subsection{Violation Scarani-Gisin Inequality (SCI) by Successive Measurements on Qubit}
Although in contrast to correlations in space there are no genuine multi-mode correlations in time, we will see that temporal correlations can be stronger than spatial ones in a certain sense. We denote by max$[B_{QM}^{space}(i,j)]$ the maximal value of the Bell expression for qubits $i$ and $j$ (Bell inequality is obtained by Eq(\ref{(3.7)})). Scarani and Gisin \cite{scaranigisin} found an interesting bound that holds for arbitrary state of three qubits:
\begin{equation}
\label{scarani}
\rm {max} [B_{QM}^{space}(1,2)]+\rm {max} [B_{QM}^{space}(2,3)]\leq 2.
\end{equation}
Physically, this means that no two pairs of qubits of a three-qubit system can violate the CHSH inequalities simultaneously. This is because if two systems are highly entangled, they can not be entangled highly to another systems. Let us denote by max$[B_{QM}^{time}(i,j)]$ the maximal value of the Bell expression for two consecutive observations of a single qubit at times $i$ and $j$. Since quantum correlations between two successive measurements do not depend on the initial state ( see Eq (\ref{(5.4)})), one can obtain:
\begin{eqnarray}
&&\left[B_{QM}^{time}(k-1,k)]+ [B_{QM}^{time}(k,k+1)\right]=\nonumber\\ &&\frac{1}{2}\left[\cos\theta_{k-1,k}+\cos\theta_{k-1,k^{\prime}}+\cos\theta_{k-1^{\prime},k}-\cos\theta_{k-1^{\prime},k^{\prime}}\right]+\nonumber\\&&\frac{1}{2}\left[\cos\theta_{k,k+1}+\cos\theta_{k,k+1^{\prime}}+\cos\theta_{k^{\prime},k+1}-\cos\theta_{k^{\prime},k+1^{\prime}}\right].
\end{eqnarray}

By selecting,$$\theta_{k-1,k}=\theta_{k-1,k^{\prime}}=\theta_{k-1^{\prime},k}=\theta_{k,k+1}=\theta_{k,k+1^{\prime}}=\theta_{k^{\prime},k+1}=\frac{\pi}{4}$$ and $$ \theta_{k-1^{\prime},k^{\prime}}=\theta_{k^{\prime},k+1^{\prime}}=\frac{3\pi}{4},$$ we obtain:
\begin{equation}
\label{scaranitime}
\rm{max}\left[B_{QM}^{time}(k-1,k)]+ \rm{max}[B_{QM}^{time}(k,k+1)\right]=\sqrt{2}+\sqrt{2}=2\sqrt{2}>2.
\end{equation}
Thus, although there are no genuine three-fold temporal correlations, a specific combination of two-fold correlations can have values that are not achievable with correlations in space for any three-qubit system. In fact, one would need two pairs of maximally entangled two-qubit states to achieve the bound in (\ref{scaranitime}). Also note that the local realistic bound is $2$, which is equal to the bound in (\ref{scarani}). Similar conclusion can be obtained for the sum of $n$ successive measurements.
 \begin{equation}
\label{nscaranitime}
\rm{max}\left[B_{QM}^{time}(1,2)\right]+ \rm{max}\left[B_{QM}^{time}(2,3)\right]+\ldots+\rm{max}\left[B_{QM}^{time}(n-1,n)\right]=n\sqrt{2}>n.
\end{equation}
\subsection{Violation Chained Bell  Inequalities (CHI) by Two Successive Measurement on Qubit}
Generalized CHSH inequalities may be obtained by providing more than two alternative experiments to each process. We consider two successive measurements on a spin-$\frac{1}{2}$ particle in a mixed state, such that the first experiment can measure spin component along one of the directions $\hat a_1, \hat a_3,\ldots,\hat a_{2n-1}$ and the second experiment along one of the directions $\hat b_2, \hat b_4, \ldots, \hat b_{2n}$. The results of these measurements are called $\alpha-r$ ($r=1,3,\ldots,2n-1$) and $\beta_s$ ($=2,4,\ldots,2n$), respectively, and their values are $\pm1$ (in unit if $\hbar/2$). We have a generalized CHSH inequality \cite{braunstein90},\cite{peres93}:
\begin{eqnarray}\rm{CBI}=\frac{1}{2}\left|\langle \alpha_1\beta_2\rangle+ \langle \beta_2\alpha_3\rangle + \langle\alpha_3\beta_4\rangle+\ldots+\langle\alpha_{2n-1}\beta_{2n}\rangle-\langle\beta_{2n}\alpha_1\rangle\right|\leq n-1\nonumber\\
\end{eqnarray}
This upper bound is violated by quantum correlations in two successive measurements, increasingly with larger $n$. In order to obtain the maximum value above-inequality, we consider the spherical-polar co-ordinates ($\theta_k,\phi_k$),($k=1,3,\ldots,2n-1$) of the vectors $\hat a_1, \hat a_3,\ldots,\hat a_{2n-1}$ and ($k=2,4,\ldots,2n$) for vectors $\hat b_2, \hat b_4, \ldots, \hat b_{2n}$. The maximum value $|CBI|$ will occur when all the vectors lie on the same plane. this is because of:$$\frac{\partial (\rm{CBI})}{\partial \phi_k}=0\Rightarrow \phi_1=\phi_2=\ldots=\phi_n.$$ After partial differential over all $\theta_k$, we get:\begin{equation*}\frac{\partial (\rm{CBI})}{\partial \theta_k}=0\Rightarrow \theta_{12}=\theta_{23}=\ldots=\theta_{2n-1,2n}=\theta.\end{equation*}Therefore, we obtain:\begin{equation}\rm{CBI}=\left(2n-1\right)\cos\theta-\cos(2n-1)\theta.\end{equation}So,\begin{equation}\frac{\partial (\rm{CBI})}{\partial \theta}=0\Rightarrow\theta=\pi/2n.
\end{equation}
By substituting , we obtain:
\begin{equation}
\rm{CBI}=(2n-1)\cos\frac{\pi}{2n}-\cos\frac{(2n-1)\pi}{2n}=2n\cos\frac{\pi}{2n}
\end{equation}
We know $\cos(\frac{\pi}{2n})$ tends to $(1-\frac{\pi^2}{8n^2})$ for $n\longrightarrow\infty$. Therefore the maximum CBI can be made arbitrarily close to $2n$.
\subsection{Violation Bell Inequalities Involving Tri and Bi-measurements Correlations by Three Successive Measurements on Qubit}
As a final remark, it would be interesting to consider Bell
inequalities involving both two and three successive measurement correlations. The simplest way of obtaining such an inequality would be by adding genuinely bipartite correlations to the tripartite correlations considered in Mermin's inequality. For instance,a straightforward calculation would allow us to prove
that any local realistic theory must satisfy the following inequality \cite{cabello02}:
\begin{eqnarray}
\label{tribi}
&-5\leq \lan \al_1 \al_2 \al_3'\ran - \lan \al_1 \al_2' \al_3' \ran - \lan \al_1' \al_2 \al_3' \ran - \lan \al_1' \al_2' \al_3\ran-&\nonumber\\&\lan \al_1 \al_2 '\ran -\lan \al_1 \al_3' \ran - \lan  \al_2 \al_3 \ran \leq 3 &
\end{eqnarray}
A numerical calculation shows that both the GHZ and W states give a same maximal violation of the inequality \ref{tribi}. However, if we assign a higher weight to the bipartite correlations appearing in the inequality, then we can reach a Bell inequality such as
\begin{eqnarray}
\label{bitri}
 &-8 \leq \lan \al_1 \al_2 \al_3'\ran - \lan \al_1 \al_2' \al_3' \ran - \lan \al_1' \al_2 \al_3' \ran - \lan \al_1' \al_2' \al_3\ran-&\nonumber\\&2\lan \al_1 \al_2 '\ran -2 \lan \al_1 \al_3' \ran - 2\lan  \al_2 \al_3 \ran \leq 4,& 
\end{eqnarray}
which is violated by the W state but not by GHZ state \cite{cabello02}.
It is not difficult to show that three successive measurements correlations for spin 1/2 break the hybrid Bell inequalities.
\begin{eqnarray}
\label{tribi*}
&-5.34\leq \lan \al_1 \al_2 \al_3'\ran - \lan \al_1 \al_2' \al_3' \ran - \lan \al_1' \al_2 \al_3' \ran - \lan \al_1' \al_2' \al_3\ran-&\nonumber\\&\lan \al_1 \al_2 '\ran - \lan \al_1 \al_3' \ran - \lan  \al_2 \al_3 \ran \leq 3.8& 
\end{eqnarray}
and
\begin{eqnarray}
\label{bitri*}
&-8.2 \leq \lan \al_1 \al_2 \al_3'\ran - \lan \al_1 \al_2' \al_3' \ran - \lan \al_1' \al_2 \al_3' \ran - \lan \al_1' \al_2' \al_3\ran-&\nonumber\\&2\lan \al_1 \al_2 '\ran -2 \lan \al_1 \al_3' \ran - 2\lan  \al_2 \al_3 \ran \leq 4.8. 
\end{eqnarray}
So two successive measurements correlations are relevant to those of three successive measurements. This behavior is analogous to three particle W state \cite{cabello02}.
\section{Classical Simulation of n Successive Measurements on a Spin-$\frac{1}{2}$ System}

We have seen that QM correlations between the outputs of $n$
successive measurements of incompatible observables $\vec{S} \cdot
\h a_k  (k = 1, 2, \cdots n)$ are stronger than their `classical'
({\it i.e.}, HVT) counterparts. An interesting question is whether
these quantum correlations can be simulated classically ({\it i.e.},
using classical resources, where `classical' means the conditions
obeyed by the HVT here in this chapter). Can we design a classical
protocol to produce $n$ sets of outputs which are correlated as if
these were the outputs of genuine successive measurements?  If this
is possible, what amount of classical information (cbits) has to be
shared between successive measurements \cite{toner03}?  Note that here sending cbits will be meaningful if the experimenter at later time does not have any information about the measurement direction ( as well as
measurement outcomes) of the previous experiment(s). We try and
answer some aspects of these questions in this section. Notice that,
there is no room for spatial non locality in this scenario, because
the events are time-like separated. When the particle is coming out
from $i$-th experiment there is no particle in any of the previous
experiments. The communication of information is carried by the
particle itself. We now describe our protocol for two successive
measurements.

We imagine that two {\it different} experimenters, Alice and Bob
perform two successive measurements of $\vec{S} \cdot \h a_1$ and
$\vec{S} \cdot \h a_2$ respectively at time $t_1$ and (at later
time) $t_2$. Directions $\h a_1$ and $\h a_2$ are chosen by each
experimenter randomly and independent of each other. Alice and Bob
do not know each others inputs $(\h a_1, \h a_2)$ and outputs
$(\al_1, \al_2)$. Alice knows the input state parameter $\h a_0$.
Bob does not know $\h a_0$. They share three random variables (unit
vectors) $\h \la_0, \h \la_1, \h \la_2$. They are chosen
independently and distributed uniformly over the unit sphere.\\ The
protocol proceeds as follows:\\ (i) Alice outputs $\al_1 = \rm sgn[\h a_1
\cdot (\h \la_0 + \h a_0)]$. \\(ii) Alice sends two cbits $c_1$ and
$c_2 \in \{ -1, 1\}$ to Bob where $c_1 = \rm sgn[\h a_1 \cdot (\h \la_0
+ \h a_0)]\rm sgn(\h a_1 \cdot \h \la_1) = \al_1 ~ \rm sgn(\h a_1 \cdot \h
\la_1),~ c_2 = \rm sgn[\h a_1 \cdot (\h \la_0 + \h a_0)] \rm sgn(\h a_1 \cdot
\h \la_2) = \al_1 ~\rm sgn(\h{a_1} \cdot \h{\la_2})$.\\  (iii) Bob
outputs $\al_2 =\rm sgn[\h a_2 \cdot (c_1 \h \la_1 + c_2 \h \la_2)]$,
where we have used the sgn function defined by $\rm sgn(x) = +1$ if $x
\ge 0$ and $\rm sgn(x) = -1$ if $x < 0$.\\
We note immediately that Bob cannot obtain any information about Alice's input and output from $c_1$ and $c_2$. We now show that the above protocol reproduces the statistics of two successive measurements of $\vec{S} \cdot \h a_1$ and $\vec{S} \cdot \h a_2$ on spin $1/2$ particle in initial state $|\vec{S} \cdot \h a_0, + 1\ran \lan \vec{S} \cdot \h a_0, + 1|$. As shown in Appendix B we have
$$ \lan \al_1\ran = \h a_0 \cdot \h a_1~,~  \lan \al_1 \al_2 \ran = \h a_1 \cdot \h a_2~,~
 \lan \al_2\ran = (\h a_0. \h a_1) (\h a_1. \h a_2) = \lan \al_1 \ran \lan \al_1 \al_2 \ran$$
which is consistent with the quantum case. We can generalize this
protocol to get the classical simulation of $n$ successive
experiments. Here, again, each experiment is performed by an
independent experimenter, who has no knowledge of the inputs and
outputs of the previous and the future experiments. All
experimenters share $(2n+1)$ random variables ( unit vectors ) $\h
\la_0, \h \la_1, \h \la_2, \cdots, \h \la_{2n}$. The $i$-th
experimenter $(i > 1)$ receives cbit $c_{2i-3}$ and $c_{2i-2}$ from
$(i-1)$-th experiment, defined by $c_{2i-3} = \al_{i-1} \rm sgn(\h
a_{i-1} \cdot \h \la_{2i-3}),~~ c_{2i-2} = \al_{i-1}\rm sgn(\h a_{i-1}
\cdot \h \la_{2i-2}).$ The $i$-th experimenter, then outputs $ \al_i
=\rm sgn[\h a_i \cdot (c_{2i-3} \h \la_{2i-3} + c_{2i-2} \h
\la_{2i-2})]$. For $i = 1$, the outputs $\al_1 =\rm sgn[\h a_1 \cdot
(\h \la_0 + \h a_2)].$

As shown in Appendix B, this protocol produces all quantum
correlations between $n$ successive measurements described in
equations (\ref{(5.4)}), (\ref{(5.5)}), (\ref{(5.6)}) and (\ref{(5.7)}).
Brukner et al. \cite{vedral} have described a communication complexity
( where space is replaced by time) protocol of computing the Boolean
function $f(y, x_1, x_2) = y.(-1)^{x_1.x_2}$ of three input bits $y,
x_1, x_2$. If the input bits are supplied one after another, the
computer ( where the computation has to be done) will have to store
all the three bits according to their arrival to compute the
function at the end. But if the random access memory (RAM) of the
computer is limited to one bit, then this process will not work, as
storing of three bits won't be possible in that case. But feeding
the first bit $y$ ( given at time $t_1$) into the RAM, then
performing measurement of one of the two observables $\hat{a}_1.{\bf
\sigma}$, $\hat{a}_2.{\bf \sigma}$ on the the input qubit according
to the value of the second bit $x_1$ (supplied at time $t_2 (>
t_1)$) and thereby multiplying $y$ by the output $s_1$ of the
measurement and then feeding the product $y.s_1$ into the RAM, then
repeating this process for the third input bit $x_2$ ( supplied at
time $t_3 (> t_2)$), by performing measurement of one of the two
observables $\hat{b}_1.{\bf \sigma}$, $\hat{b}_2.{\bf \sigma}$ on
the the output qubit after the first measurement qubit according to
the value of the third bit $x_2$, one can compute the function $f$
with probability of success higher than what one might get
classically ( {\it i.e.}, without using QM ).

Note that although the classical simulation, described above, seems
to be artificial ( as the same experimenter can, in principle,
perform the successive experiments and use the input and/or output
data), from the perspective of memory resource available, we can
always put some restriction on the capacity of the memory of the
experimenter, like limitation of RAM of the computer in the
above-mentioned description.

\section{Summary and Comments}

In this chapter we have considered a hidden variable theory of
successive measurements on a single spin-$s$ system. In all the
previous scenarios comparing HVT and QM the principal hypothesis
being tested was that, in a given state ( having spatial
correlation), HVT implies the existence of a joint probability
distribution for all observables even if some of them are not
compatible. QM is shown to contradict the consequence of this
requirement as it does not assign joint probabilities to the values
of incompatible observables. The particular implication that is
tested is whether the marginal of the observable $A$ in the joint
distribution of the compatible observables $A$ and $B$ is the same
as the marginal for $A$ in joint distribution for the observables
$A$ and $C$ even if $B$ and $C$ are not compatible. In other words,
HVT implies noncontextuality for which QM can be tested. The
celebrated theorem of Bell and Kochen-Specker showed that QM is
contextual \cite{bell66},\cite{kochenspecker}. In our scenario, the set of measured observables have a well defined joint probability distribution as each of them acts on a different state. Note that the Bell-type inequalities we have derived follow from equation (\ref{(3.4)}) which says that, for a given value of stochastic hidden variable $\lambda$, the joint probability for the outcomes of successive measurements must be statistically independent. In other words the hidden variable $\lambda$ completely decides the probabilities of individual measurement outcomes independent of other measurements. We show that QM is not consistent with this requirement of HVT. A Bell-type inequality (for single particle), testing contextuality of QM was proposed by Basu et al. \cite{basuban01} and was shown that it could be empirically tested. However, the approach given in the present paper furnishes a test for realistic nature of QM independent of contextuality. Our approach may be used to get a measure of the deviation of QM from HVT. One such measure is the amount of
information needed to be transferred between successive measurements
in order to classically simulate quantum correlations. As we have
shown in section 6, a pure spin 1/2 state can be classically
simulated by communicating two c-bits of information to get the
$k$-th  output from $(k-1)$-th output by using (2k+1) shared random
variables. Whether this is the minimum communication required is
still open. For our protocol, the amount of information needed is
twice as much in the case of bipartite nonlocal scenario \cite{toner03}.\\
In sections 4 and 5 we have compared QM with HVT for different values
of spin and for different number of successive measurements. The
dependence of the deviation of QM from HVT on the spin value and on
the number of successive measurements opens up new possibilities for
comparison of these models, and may lead to a sharper understanding
of QM.  We get many surprising results. First, for a spin $s$
particle, maximum deviation ($\eta$) is obtained for all convex
combinations (mixed states) of $\al_0 = \pm s$  states. This is
surprising as one would expect pure states to be more `quantum' than
the mixed ones thus breaking Bell inequalities by larger amount. In
particular, all spin $1/2$ states maximally break Bell inequality as
against only the entangled states break it in bipartite case. This
does not contradict Bell's explicit construction of HVT for spin
$1/2$ particle, as this construction does not apply for two or more successive
measurements. Further, the maximum deviation from Bell inequality
measured by $\eta_{max}$ falls off as the spin of the particle
increase. There is a large drop in $\eta_{max}$ value from $s=1/2$
to $s=1$, after which it drops monotonically with $s$, but very
weakly, asymptotically approaching $\eta=1.4$ . This can be compared
with the case of two spin $s$ operators in the singlet state where
the deviation from Bell's inequality is found to tend to a constant
\cite{peress92},\cite{gisinperes},\cite{cabello652002},\cite{mermin90}. All spins violate MKI and satisfy SI for three
successive measurements. The maximum violation is $\sqrt{2}$ for
spin $s=\frac{1}{2}$ . In the case of $n$ spin 1/2 particles in the
singlet state, Bell inequalities are broken by a factor which
increases exponentially with $n$ \cite{cabello652002}, \cite{mermin90}.
%\bc
\newpage

%{\large {\bf Appendix A}}
%\ec
\section{Appendix A}
We evaluate $ \lan \al_1 \ran $, $ \lan \al_1 \al_2 \ran $ and $ \lan \al_1 \al_2 \al_3 \ran $ in the state $ \rho_0 $ given in(4.1).\\$(|\vec{S} \cdot \h a_0, \al_0 \ran \equiv | \h a_0, \al_0 \ran )$
$$
 \lan \al_1 \ran = \sum^s_{\al_1 = -s} \al_1 p(\al_1) = \lan \h a_0, \al_0 |\vec{S} \cdot  \h a_1 | \h a_0, \al_0\ran  =   \lan \h a_1, \al_0 | e^{i\vec{S}\cdot \h n \th_{1}} (\vec{S} \cdot \h a_1) e^{-i\vec{S} \cdot \h n \th_{1}}| \h a_1, \al_0 \ran ~\eqno{(A.1)}$$
where $\th_{1}$ is the angle between $\h a_0$ and $\h a_1$ and $\h n$ is the unit vector along the direction defined by $\h n = \h a_0 \times \h a_1$.
By using Baker- Hausdorff Lemma
$$ e^{iG\la} Ae^{-iG\la} = A + i\la [G, A] + \left( \fr{i^2 \la^2}{2!}\right) [G, [G, A]] + \cdots ~\eqno{(A.2)}$$
we get,
\benrr
\lan \al_1\ran  &= &  \lan \h a_1, \al_0 | \vec{S} \cdot \h a_1 | \h a_1, \al_0 \ran + \fr{i\th_{1}}{1!} \lan \h a_1, \al_0| [\vec{S} \cdot \h n, \vec{S} \cdot \h a_1] | \h a_1, \al_0 \ran \\
& &  + \fr{i^2 \th^2_{1}}{2!} \lan \h a_1, \al_0 | [ \vec{S} \cdot \h n, [ \vec{S} \cdot \h n, \vec{S} \cdot \h a_1]] | \h a_1 \al_0 \ran + \cdots~~~~~~~~~~~~~~~~~~~~\mbox{(A.3)}
\eenrr
By using:
$$ \lan \h a_1, \al_0 | \vec{S} \cdot \h a_1 | \h a_1, \al_0 \ran = \al_0 \eqno{(A.4)},$$

$$ \lan \h a_1, \al_0 | [\vec{S} \cdot \h n, \vec{S} \cdot \h a_1] |\h a_1, \al_0 \ran = \lan \h a_1, \al_0 | (i\vec{S} \cdot (\h n \times \h a_1)) | \h a_1, \al_0 \ran = 0, \eqno{(A.5)}$$
and
$$ \lan \h a_1, \al_0 | [\vec{S} \cdot \h n, [\vec{S} \cdot \h n, \vec{S} \cdot \h a_1]] |\h a_1, \al_0 \ran = \lan \h a_1, \al_0 | \vec{S} \cdot \h a_1 |\h a_1, \al_0 \ran = \al_0.\eqno{(A.6)}$$
Terms with odd powers of $\th_{1}$ vanish
$$ \lan  \al_1 \ran = \al_0 - \fr{\th^2_{1}}{2!} \al_0 + \fr{\th^4_{1}}{4!} \al_0 - \cdots = \al_0 \cos \th_{1}. \eqno{(A.7)}$$
If the initial state is mixed state(4.1):
$$ \lan  \al_1 \ran =  \sum^{+s}_{\al_0= - s}p_{\al_0}\al_0 \cos \th_{1}. \eqno{(A.8)}$$
Further we compute
$$ \lan \al_1 \al_2 \ran = \sum_{\al_1} \al_1 | \lan \h a_0, \al_0 | \h a_1, \al_1 \ran |^2 \sum_{\al_2}\al_2 | \lan \h a_1, \al_1 | \h a_2, \al_2 \ran |^2.$$
By using (A.7)
\begin{eqnarray*}
\lan \al_1 \al_2 \ran & = & \cos \th_{12} \sum_{\al_1} \al^2_1 | \lan \h a_0, \al_0 |\h a_1, \al_1 \ran |^2
 =  \cos \th_{12} \lan \h a_0, \al_0 | (\vec{S} \cdot \h a_1)^2| \h a_0, \al_0 \ran \\
& =& \cos \th_{12}   \lan \h a_1, \al_0 | e^{i\vec{S}\cdot \hat{n} \th_{1}} (\vec{S} \cdot \h a_1)^2 e^{-i\vec{S} \cdot \hat{n}\th_{1}} | \h a_1, \al_0 \ran ~~~~~~~~~~~~~~~~~~~~(A.9)
\end{eqnarray*}
Using the Baker-Hausdorff Lemma, and using
$$\lan \hat a_{1} ,\al_{0} | [\vec{S} \cdot \h n, [ \vec{S} \cdot \h n, [ \vec{S} \cdot \h n, \cdots [\vec{S} \cdot \hat{n}, (\vec{S} \cdot \hat{a_1})^2]] \cdots]]| \hat a_{1} ,\al_{0}\ran \eqno{(A.10)}$$
$$ = \left\{ \ba{l} 0~~~~~~~~~~~~~~~~~~~~~\mbox{if $\vec{S} \cdot \h n$ occurs odd number of times } \\ \\ 3\al^2_0 - s^2 - s~~~~~~~\mbox{if $\vec{S} \cdot \h n$ occurs $2p$ times} \ea \right. $$
we get,
$$\lan  \al_1 \al_2  \ran = \fr{1}{2} \cos \th_{12} [(s^2 + s - \al^2_0) + (3\al^2_0 - s^2 - s) \cos^2 \th_{1}]. \eqno{(A.11)}$$
If the initial state is mixed state (4.1),
$$\lan  \al_1 \al_2  \ran = \fr{1}{2} \cos \th_{12}  \sum^{+s}_{\al_0= - s}p_{\al_0}[(s^2 + s - \al^2_0) + (3\al^2_0 - s^2 - s) \cos^2 \th_{1}]. \eqno{(A.12)}$$
Next we calculate,
$$ \lan \al_1 \al_2 \al_3 \ran = \sum_{\al_1} \al_1 | \lan \h a_0,  \al_0 | \h a_1, \al_1 \ran |^2 \sum_{\al_2} \al_2 | \lan \hat{a_1}, \al_1 | \h a_2, \al_2 |^2 \sum_{\al_3} \al_3 | \lan \h a_2, \al_2 | \h a_3, \al_3 \ran |^2.\eqno{(A.13)}$$
By using (A.7) and (A.11) we get,

$$ \lan \al_1 \al_2 \al_3 \ran = \fr{1}{2} \al_0 \cos \th_{1} \cos \th_{23} \sin^2 \th_{12} s(s+1) + \fr{1}{2} \cos \th_{23} (3 \cos^2 \th_{12} - 1) A, \eqno{(A.14)}$$
where,
$$ A = \sum_{\al_1} \al^3_1 | \lan \h a_0,  \al_0 |\h a_1, \al_1 \ran |^2 = \lan \h a_1, \al_0| e^{i\vec{S}\cdot \hat{n}\th_{1}} (\vec{S} \cdot \h a_1)^3 e^{-i\vec{S}\cdot \hat{n}\th_{1}}|\h a_1, \al_0\ran. \eqno{(A.15)}$$
Using Baker-Hausdorff lemma and
$$\lan \hat a_{1} ,\al_{0} | [\vec{S} \cdot \h n, [ \vec{S} \cdot \h n, [ \vec{S} \cdot \h n, \cdots [\vec{S} \cdot \hat{n}, (\vec{S} \cdot \hat{a_1})^3]] \cdots]]| \hat a_{1} ,\al_{0}\ran $$
$$ = \left\{ \ba{l} 0~~~~~~~~~~~~~~~~~~~~~\mbox{if $\vec{S} \cdot \h n$ occurs odd number of times } \\ \\ Y (X - a^3_0) + X ~~ \mbox{if $\vec{S} \cdot \h n$ occurs $2p$ times} \ea \right.\eqno{(A.16)} $$
 where,
\begin{eqnarray*}
&&X = 6\al_0^3 + \al_0 (1 - 3s(s+1))\\
&&Y=3^{2p-2} + 3^{2p-4} + \cdots + 3^2=(\fr{9}{8}) [9^{2p-2}-1] ~~~~~~~~~~~~~~~~~~~~~~~~~~(A.17)
\end{eqnarray*}
we get,
$$ A=\fr{1}{8}\sum^{\infty}_{j= 0}(-1)^{j}[(9^{j}-1)X-(9^{j}-9)\alpha_{0}^{3}]\;(\frac{\theta_{01}^{2j}}{2j!}).$$
This gives,
$$A=\frac{1}{8}\al_0\{[3 \al^2_0 +3s(s+1)-1]\cos \th_{01}+[5 \al^2_0 -3s(s+1)+1]\cos 3\th_{01}\}\eqno{(A.18)}$$

After substituting (A.18) in (A.14) and simplifying,
$$ \lan \al_1 \al_2 \al_3 \ran = \fr{1}{16} \cos \th_{23}\{ \cos \th_{1} [M \cos^2 \th_{12} + N]+R[3\cos^2 \th_{12}-1] \}\eqno{(A.19)}$$
where,
\begin{eqnarray*}
&&M= \al_0[9 \al^2_0 + s(s+1)-3]\\
&&N= \al_0[5s(s + 1)-3\al^2_0 +1]\\
&&R= \al_0[5 \al^2_0 -3s(s+1)+1].
\end{eqnarray*}
If the initial state is a mixed state (4.1),
\begin{eqnarray*}
&&M=\sum^{+s}_{\al_0= - s}p_{\al_0} \al_0[9 \al^2_0 + s(s+1)-3]\\
&&N=\sum^{+s}_{\al_0= - s}p_{\al_0} \al_0[5s(s + 1)-3\al^2_0 +1]\\
&&R=\sum^{+s}_{\al_0= - s}p_{\al_0} \al_0[5 \al^2_0 -3s(s+1)+1].
\end{eqnarray*}
Also, by using equations (A.12) and (A.13), one obtains:
\begin{eqnarray*}
 \lan \al_1  \al_3 \ran &=& \sum_{\al_1} \al_1 | \lan \h a_0,  \al_0 | \h a_1, \al_1 \ran |^2 \sum_{\al_2}  | \lan \hat{a_1}, \al_1 | \h a_2, \al_2 |^2 \sum_{\al_3} \al_3 | \lan \h a_2, \al_2 | \h a_3, \al_3 \ran |^2\\ &=&\cos\theta_{32}\lan \al_1  \al_2 \ran\\&=& \frac{1}{2}\cos\theta_{32}\cos\theta_{21}[(s^2+s-\al_0^2)+(3\al_0^2-s^2-s)\cos^2\theta_1],~~~~\mbox{(A.20)} 
\end{eqnarray*}
and we can obtain:
\begin{eqnarray*}
 \lan \al_2 \al_3 \ran &=& \sum_{\al_1} | \lan \h a_0,  \al_0 | \h a_1, \al_1 \ran |^2 \sum_{\al_2} \al_2| \lan \hat{a_1}, \al_1 | \h a_2, \al_2 |^2 \sum_{\al_3} \al_3 | \lan \h a_2, \al_2 | \h a_3, \al_3 \ran |^2\\&=&\cos\theta_{32}\lan  \al_2^2 \ran \\ &=& \frac{1}{2}s\cos\theta_{32}\{(s+1)\sin^2\theta_{12}+\frac{1}{2}(3\cos^2\theta_{12}-1)[1+(2s-1)\cos^2\theta_1]\}.\\
\end{eqnarray*}
\begin{flushright}
\mbox{(A.21)}
\end{flushright}
%\bc
\newpage
%{\large {\bf Appendix B}} \\
%\ec
\section{Appendix B}
To evaluate $\lan \al_1 \ran = \h a_1 \cdot \h a_0$ we integrate over $\h \la_0$, taking $\h a_1$ to point along the positive $\h z$ axis.
\benrr
\lan \al_1 \ran & = & \fr{1}{4\pi} \int d\la_0 sgn[\h a_1 \cdot (\h \la_0 + \h a_0)] \\
& = & \fr{1}{4\pi} \int^{2\pi}_0 d\b_0 \int^\pi_0 \sin~ \al_0 d \al_0 ~ sgn(cos \al_0 + \cos \th_1) = \cos \th_1 = \h a_1 \cdot \h a_0 
\eenrr
\begin{flushright}
\mbox{(B-1)}
\end{flushright}
where $\cos \al_0 = \h a_1 \cdot \h \la_0$ and $\h \la_0 = (\sin \al_0  \cos \b_0, \sin \al_0 \sin \b_0, \cos \al_0)$.

To evaluate $\lan \al_2 \ran = (\h a_0 \cdot \h a_1) (\h a_1 \cdot \h a_2)$
\benrr
\lan \al_2 \ran & = & \fr{1}{(4\pi)^3} \int d\la_0 d\la_1 d\la_2 ~ sgn[\h{a_2}\cdot (c_1\h\la_1 + c_2\h\la_2)] \\
& = & \fr{1}{(4\pi)^3} \int d\la_0d\la_1 d\la_2 \fr{1}{4} \sum_{d_1 =\pm1} \sum_{d_2=\pm 1} (1+c_1d_1) (1+c_2d_2) sgn[\h a_2 \cdot (d_1\h \la_1 + d_2\h \la_2)]\\
& = & \fr{1}{(4\pi)^3} \fr{1}{2} \int d\la_0 d \la_1 d\la_2 \{c_1(sgn[\h a_2 \cdot (\h \la_1 + \h \la_2)] + sgn[\h a_2 \cdot (\h \la_1 - \h \la_2)]) \\
& & + c_2(sgn[\h a_2 \cdot (\h \la_1 + \h \la_2)] - sgn[\h a_2 \cdot (\h \la_1 - \h \la_2)])\} \\
& = & \fr{1}{(4\pi)^2} \fr{1}{2} \int d\la_0 sgn[\h a_1 \cdot (\h \la_0 + \h a_0)]
 \{\int d\la_1 sgn(\h a_1 \cdot \h \la_1) 2(\h a_2 \cdot \h \la_1)\\
& &  + \int d\la_2 sgn(\h a_1 \cdot \h \la_2) 2(\h a_2 \cdot \h \la_2)\}
 =  (\h a_0 \cdot \h a_1) (\h a_1 \cdot \h a_2) 
\eenrr
\begin{flushright}
\mbox{(B-2)}
\end{flushright}
The same way we can prove
$$ \lan \al_1 \al_2 \ran = \fr{1}{(4\pi)^3} \int d \la_0 d \la_1 d\la_2 \al_1 \al_2 = (\h a_1 \cdot \h a_2) $$
\begin{flushright}
\mbox{(B-3)}
\end{flushright}
By using induction, we shall show that for $n(n > 2)$ successive measurements is simulated by this protocol. We suppose for $n = k-1$, it is true i.e.
$$ \lan \al_{k-1}\ran = \fr{1}{(4\pi)^{2k-4}} \int d\la_0 d\la_1 \cdots d\la_{2k-4} \al_{k-1} = \lan \al_1\ran \lan \al_2\al_3 \ran \cdots \lan \al_{k-2} \al_{k-1}\ran \eqno{(B-4)} $$
$$ \lan \al_{k-2} \al_{k-1} \ran = \fr{1}{(4\pi)^{2k-4}} \int d\la_0 d\la_1 \cdots d\la_{2k-4} \al_{k-2} \al_{k-1}  = \h a_{k-2} \cdot \h a_{k-1} \eqno{(B-5)}$$

$$ \lan \al_{k-1-m} \cdots \al_{k-1}\ran = \left\{ \ba{ll} \lan \al_1 \ran \lan \al_2 \al_3 \ran \cdots \lan \al_{k-2} \al_{k-1} \ran & m~~ \mbox{even} \\ \\
\lan \al_{k-1-m} \al_{k-m} \ran \cdots \lan \al_{k-2} \al_{k-1} \ran & m~~ \mbox{odd} \ea \right. \eqno{(B-6)}$$

So, for $n = k$, first we show that,
\begin{eqnarray*}
& & \int d\la_{2k-3} d\la_{2k-2} \al_k = \int d\la_{2k-3} d\la_{2k-2} sgn[\h a_k \cdot (c_{2k-3} \h \la_{2k-3} + c_{2k-2} \h \la_{2k-2})]=\\
&&\fr{1}{2} \int d\la_{2k-3} c_{2k-3} \int d\la_{2k-2} [sgn(\h a_k \cdot (\h \la_{2k-3} + \h \la_{2k-2})) + sgn(\h a_k \cdot (\h \la_{2k-3} - \h \la_{2k-2}))]+\\
&& \fr{1}{2} \int d\la_{2k-2} c_{2k-2} \int d\la_{2k-3} [sgn(\h a_k \cdot (\h \la_{2k-3} + \h \la_{2k-2})) - sgn(\h a_k \cdot (\h \la_{2k-3} - \h \la_{2k-2}))]\\
& & = \int d\la_{2k-3} c_{2k-3} (4\pi) (\h a_k \cdot \h \la_{2k-3}) + \int d\la_{2k-2} c_{2k-2} (4\pi) (\h a_k \cdot \h \la_{2k-2}) \\
& & = (4\pi) \al_{k-1} \{ \int d\la_{2k-3} sgn(\h a_{k-1} \cdot \h \la_{2k-3})(\h a_k \cdot \h \la_{2k-3})\\
& & + \int d\la_{2k-2} sgn(\h a_{k-1} \cdot \h \la_{2k-2}) (\h a_k \cdot \h \la_{2k-2})\}\\
& & = (4\pi)^2 \al_{k-1} \{ \fr{1}{2} (\h a_{k-1} \cdot \h a_k)+ \fr{1}{2} (\h a_{k-1} \cdot \h a_k)\} = (4\pi)^2 \al_{k-1} (\h a_{k-1} \cdot \h a_k) 
\end{eqnarray*}
\begin{flushright}
\mbox{(B-7)}
\end{flushright}
By using (B-7), (B-4), (B-5) and (B-6) all quantum correlations are obtained by this protocol.
\newpage
\chapter{Non Locality Without Inequality}
%\end{center}
\section{Introduction}
An early thought-provoking analysis of quantum composite system explicitly pointing out the surprising nature of entangled quantum states that also introduced consideration of locality and realism in regard to microscopic physical systems was made by Einstein, Podolsky and Rosen (EPR)\cite{epr}, who provided a specific argument for the incompleteness-though importantly, not the incorrectness- of the quantum mechanical description of the microscopic world.\\
To arrive at this conclusion, EPR assumed locality and a criterion for the result of a measurement of a physical quantity to be considered an element of physical reality prior to the measurement, and established a necessary condition for a physical theory to be considered complete. This led to the search of a ``complete theory" by adding ``hidden" variables to the wave function in order to implement realism, the most celebrated of this kind of theories being the de Broglie-Bohm theory \cite{bohmtheory}.\\
Bohm, just prior to developing his HV interpretation, introduced a simplified scenario involving two spin-half particles with correlated spins, rather than two particles with correlated positions and momenta as used by EPR \cite{bohm57}. The EPR-Bohm scenario has the advantage of being experimentally accessible.\\
In 1964 John Bell \cite{bell64} derived an inequality ( which is a statistical result, and is called Bell's inequality BI) using locality and reality assumptions of EPR-Bohm, and showed that the singlet state of two spin-$1/2$ particles violates this inequality, and hence the contradiction with quantum mechanics.\\
Bell inequalities are statistical predictions about measurements made on two particles, typically photons or particles with spin $\frac{1}{2}$. So some people were trying to show a direct contradiction ( which is not a statistical one) of quantum mechanics with local realism. Greenberger, Horne and Zeilinger (GHZ)\cite{ghz} found a way to show more immediately, without inequalities, that results of quantum mechanics are inconsistent with the assumptions of EPR. A proof of non locality without inequalities for two spin-$1$ particles had been given earlier by Heywood and Redhead \cite{heywoodredhead} which was much simplified by Brown and Svetlichny \cite{brsv}. This employed a Kochen-Specker \cite{kochenspecker} type argument to demonstrate that elements of reality corresponding to space separated measurements must be contextual. The GHZ proof used three spin half particles and the Heywood and Redhead proof used two spin \textit{one } particles. Thus both proofs required a minimum total of eight dimensions in Hilbert space rather than the four required by Bell in his proof.\\
Hardy \cite{hardy92}gave a proof of non locality for two particles with spin $\frac{1}{2}$ that only requires a total of \textit{four} dimensions in Hilbert space like Bell's proof but does not require inequalities.\\
This was accomplished by considering a particular experimental setup consisting of two over-lapping Mach-Zehnder interferometers, one for positrons one for electrons, arranged so that if the electron and positron each take a particular path then they will meet and annihilate one another with probability equal to 1. This arrangement is required to produce assymetric entangled state which only exhibits non locality without any use of inequality. The argument has been generalized to two spin s particles by Clifton and Niemann \cite{clinie} and to N spin half particles by Pagonis and Clinton \cite{pagcli}.\\
Later, Hardy showed that this kind of non locality argument can be made for almost all entangled states of two spin-$\frac{1}{2}$ particles except for maximally entangled one \cite{hardy93}. This proof was again simplified by Goldstein \cite{goldstein} who extended it to the case of bipartite systems whose constituents belong to Hilbert spaces of arbitrary dimensions.\\
Hardy's proof works only for entangled states but non-maximally entangled states ( also known as ``Hardy states" \cite{clinie}). This curious feature has led to several attempts and suggestions to develop a Hardy-like argument for maximally entangled states of two qubits. However, so far none of these proposals has worked\cite{cabello2000}. In addition, Hardy's argument works only for $9\%$ of the runs of a certain experiment. On the other hand, nonlocality in GHZ works for 100\% of the runs, but requires three observers ( instead of two, as in Hardy's proof).\\
The converse of the Hardy's result \textit{i.e.} for any choice of two measurement possibilities for each particle, there is a state which exhibits Hardy's non locality, was first presented by Jordan \cite{jordan94} and later on by Cereceda \cite{cereceda}.\\
For almost all entangled states of three spin-$\frac{1}{2}$ particles, Wu and Xie \cite{wuxie} have recently proved Hardy's non locality theorem. They have used a particular type of relation among the coefficients of the given entangled state; but this type of relation does not arise in the case of Hardy's non locality proof for two spin-$\frac{1}{2}$ particles, so, the proof of Wu and Xie lacks generality.\\
Recently, Cabello has introduced a logical structure to prove Bell's theorem without inequality for three particles GHZ and W state \cite{cabello02}. Logical structure presented by Cabello is as follows: consider four events D, E, F and G where D and F may happen in one system and E and G happen in another system which is far apart from the first. The probability of joint occurrence of D and E is nonzero, E always implies F, D always implies G, but F and G happen with lower probability than D and E. These four statements are not compatible with local realism. The difference between these two probabilities is the measure of violation of local realism. Cabello's logical structure was originally proposed for showing non locality for three particle states, but Liang and Li \cite{liangli05}exploited it in establishing non locality without inequality for a class of two qubit mixed entangled states. In this sense, Hardy's logical structure is a special case of Cabello's structure as the logical structure of Hardy for establishing non locality is as follows: D and E sometimes happen, E always implies F, D always implies G, but F and G never happen. Based on Cabello's logical structure Kunkri and Choudhary \cite{kunkchou}, recently, have shown that there may be many classes of two-qubit mixed states which exhibit non locality without inequality. It is noteworthy here that in contrast there is no two-qubit mixed state which shows Hardy type non locality \cite{gkar97}. So, it seems interesting to study that whether maximally entangled states follow this more general ( than Hardy's), Cabello's non locality argument or not because Hardy's non locality argument is not followed by a maximally entangled state.\\
In section 2,  we review Hardy's non locality.
In section 3, we describe Cabello's argument for two qubits. In section 4, we show that maximally entangled states do not respond even to Cabello's logic. Next in section 5, we show for all other pure entangled states, Cabello's argument runs. In section 6, we obtain the highest value of difference between the two probabilities which appear in Cabello's argument. Surprisingly, for almost all two qubit pure entangled states, this value turns out to be larger than the highest value of probability that appears in Hardy's argument. In section 7, we extend previous result to three probabilities which appear in Cabello's argument and we show that this value is larger than the previous result. Finally we conclude with summary and comments in section 8. Mathematical details are relegated to appendix.
\section{Non Locality Without Inequalities for Almost All Entangled States for Two Particles ( Hardy Non Locality)}
Here we give the Hardy's non locality proof ( as presented by Goldstein \cite{goldstein}) for given states of two spin-$1/2$ particles. Consider a system of two spin-$1/2$ particles in a pure non maximal entangled state $|\psi\rangle$. We see that $|\psi\rangle$ can be written as 
\begin{equation}
\label{hardy}
|\psi\rangle=a|v_1,v_2\rangle+b|u_1,v_2\rangle+c|v_1,u_2\rangle ~~~(abc\neq0),
\end{equation}
for a proper choice of orthonormal basis $\left\{ \left|u_i\right\rangle,\left|v_i\right\rangle\right\}$ for $i$-th particle, $i=1,2$. Now we define the following projectors 
\begin{eqnarray}
U_i&=&\left|u_i\right\rangle\left\langle u_i\right|,\\
W_i&=&\left|w_i\right\rangle\left\langle w_i\right|,
\end{eqnarray}
where 
\begin{equation}
\left|w_i\right\rangle=\frac{a\left|v_i\right\rangle+b\left|u_i\right\rangle}{\sqrt{\left|a\right|^2+\left|b\right|^2}}.
\end{equation}
Now from the above state in (\ref{hardy}), we have the following conclusions:
If we measure $U_1$ and $U_2$ then 
\begin{equation}
\label{i}
U_1 U_2=0,
\end{equation}
since there is no $\left|u_1,u_2\right\rangle$ term.\\If we measure $U_1$ on particle $1$ and $W_2$ on particle $2$ then
\begin{equation}
\label{ii}
 U_1=0 ~\Rightarrow ~W_2=1.
\end{equation}
If we measure $W_1$ on particle $1$ and $U_2$ on particle $2$ then
\begin{equation}
\label{iii}
 U_2=0 ~\Rightarrow ~W_1=1.
\end{equation}
Finally, if we measure $W_1$ and $W_2$, with non vanishing probability
\begin{equation}
\label{iv}
W_1=W_2=0,
\end{equation}
since $abc\neq0$.
But  if we assume local hidden variable theory for these observables, it follows from (\ref{i}), (\ref{ii}) and (\ref{iii}) that the result of measurements of both $W_1$ and $W_2$ can not be simultaneously $0$, which contradicts (\ref{iv}).
We can rearrange the above logic,as follows:
\begin{equation}
\label{1}
p(U_1=1, U_2=1)=0.
\end{equation}
\begin{equation}
\label{2}
p(U_1=0, W_2=1)=1.
\end{equation}
\begin{equation}
\label{3}
p(W_1=1, U_2=0)=1.
\end{equation}
\begin{equation}
\label{4}
p(W_1=0, W_2=0)\neq 0. 
\end{equation}
There is contradiction between result obtained from (\ref{1}), (\ref{2}),(\ref{3}) and result (\ref{4}).\\
Thus a normalized state $\left|\phi\right\rangle_{AB} $ of two spin-$1/2$ particles $A$ and $B$ is said to satisfy Hardy type non locality, if we can find two pairs of one dimensional projectors ($P\left[\left|u\right\rangle_A\right],P\left[\left|w\right\rangle_A\right]$) and ($P\left[\left|u\right\rangle_B\right],P\left[\left|w\right\rangle_B\right]$)
for the systems $A$ and $B$ respectively, such that\\
(I)~~$_{AB}\langle\phi|P[|u \rangle_A]\otimes P [|u\rangle_B]|\phi\rangle_{AB}=0,$\\
(II) if A is found in the state $|u^{\bot}\rangle_A$ ( after doing the projective measurement in the orthonormal basis ($|u\rangle_A$,$|u^{\bot}\rangle_A$)on $A$), the state of $B$ will be projected on to the state $|w_B\rangle$,\\ 
(III) if B is found in the state $|u^{\bot}\rangle_B$ ( after doing the projective measurement in the orthonormal basis ($|u\rangle_B$,$|u^{\bot}\rangle_B$)on $B$), the state of $A$ will be projected on to the state $|w_A\rangle$,\\
 (IV) $_{AB}\langle\phi|P[|w^{\bot} \rangle_A]\otimes P [|w^{\bot}\rangle_B]|\phi\rangle_{AB}>0.$\\
 Now, we show that no maximally entangled state of two spin-$1/2$ particles will satisfy Hardy's non locality.$$ |\phi\rangle_{AB}=\frac{1}{\sqrt{2}}(|u\rangle_{A}\otimes|v\rangle_{B}+|u^{\bot}\rangle_{A}\otimes|v^{\bot}\rangle_{B})$$
In order that this maximally entangled state will satisfy Hardy's non locality, we must have, in the above-mentioned conditions (\ref{1})- (\ref{4}):
\begin{eqnarray}
\label{aa}
|w^{\bot}\rangle_{B}&\equiv& |v \rangle_B,\\
|w^{\bot}\rangle_{A}&\equiv& |u^{\bot} \rangle_A,
\end{eqnarray}
which follow from (\ref{1})-(\ref{3}), above. From (\ref{4}), we have: 
\begin{equation}
(_{A}\langle u^{\bot}|\otimes _{B}\langle v|)|\phi\rangle_{AB}\neq 0.
\end{equation}
But this last condition does not hold by quantum mechanics, as can be easily seen from the maximally entangled state $|\phi\rangle$.
So any non-maximally entangled state of two spin-$1/2$ particles A and B will satisfy Hardy's non locality with a maximum probability $p_{max}(|\phi\rangle)$, obtained by taking maximum of the probabilities $_{AB}\langle\phi|P[|w^{\bot} \rangle_A]\otimes P [|w^{\bot}\rangle_B]|\phi\rangle_{AB}$ over the choice af all two pairs of one dimensional projectors ($P\left[\left|u\right\rangle_A\right],P\left[\left|w\right\rangle_A\right]$) and ($P\left[\left|u\right\rangle_B\right],P\left[\left|w\right\rangle_B\right]$) for the systems A and B respectively. Now every non-maximally entangled state $|\phi\rangle_{AB}$ of two spin-$1/2$ particles A and B has its Schmidt decomposition:$$|\phi\rangle_{AB}=\alpha|e\rangle_A\otimes |f\rangle_B +\beta|e^{\bot}\rangle_A\otimes |f^{\bot}\rangle_B,$$ where \{$ |e\rangle_A,|e^{\bot}\rangle_A$\} is an orthonormal basis for system A, \{$ |f\rangle_B,|f^{\bot}\rangle_B$\} is an orthonormal basis for system B, and $0< \alpha,\beta <1$. It can be shown that the maximum possible value  of $p_{max}(|\phi\rangle)$'s over all possible choice of non-maximally entangled states of two spin-$1/2$ particles A and B, is approximately equal to $0.09$ , and this maximum value will occur when the ratio $\frac{\alpha}{\beta}$ is equal to $0.46$ \cite{hardy93}.

\section{Cabello's Argument for Two Qubits}

Let us consider two spin-1/2 particles A and B. Let F, D, G and E represent the spin observables along

$$n_{F}(\sin\theta_{F}\cos\phi_{F}, \sin\theta_{F}\sin\phi_{F}, \cos\theta_{F}), $$
$$n_{D}(\sin\theta_{D}\cos\phi_{D}, \sin\theta_{D}\sin\phi_{D}, \cos\theta_{D}), $$
$$n_{G}(\sin\theta_{G}\cos\phi_{G}, \sin\theta_{G}\sin\phi_{G}, \cos\theta_{G}), $$
and
$$n_{E}(\sin\theta_{E}\cos\phi_{E}, \sin\theta_{E}\sin\phi_{E}, \cos\theta_{E}) ,$$
respectively. Every observable has the eigenvalue $\pm1$. Let F and D be measured on particle A whereas G and E are measured on particle B. Now we consider the following equations:
\begin{eqnarray}
\label{fg}
	p(F=+1,G=+1)=q_{1}\neq 0,
\end{eqnarray}
\begin{eqnarray}
\label{dg}
	p(D=+1,G=-1)=q_{2}=0,
\end{eqnarray}
\begin{eqnarray}
\label{fe}
	p(F=-1,E=+1)=q_{3}=0,
\end{eqnarray}
\begin{eqnarray}
\label{de}
	p(D=+1,E=+1)=q_{4}\neq 0.
\end{eqnarray}
Equation (\ref{fg}) tells that if F is measured on particle A and G is measured on particle B, then the probability that both will get $+1$ eigenvalue is $q_{1}$. Other equations can be analyzed in a similar fashion. These equations form the basis of Cabello's non locality argument. It can easily be seen that these equations contradict local realism if $q_{1}<q_{4}$. To show this, let us consider those hidden variable states $\lambda$ for which $D=+1$ and $E=+1$. For these states, Eqs. (\ref{dg}) and (\ref{fe}) tell that the values of G and F must be equal to $+1$. Thus according to local realism $p(F=+1,G=+1)$ should be at least equal to $q_{4}$. This contradicts Eq. (\ref{fg}) as $q_{1}<q_{4}$. It should be noted here that $q_{1}=0$ reduces this argument to Hardy's one. So by Cabello's argument, we specifically mean that the above argument runs, even with nonzero $q_{1}$.\\
Now we will show that for almost all two-qubit pure entangled states other than maximally entangled one, this kind of non locality argument runs. Following Schmidt decomposition procedure; any entangled state of two particles A and B can be written as
\begin{eqnarray}\label{5}
	\vert \psi \rangle=(\cos\beta)\vert 0\rangle_{A}\vert0\rangle_{B}+(\sin\beta)e^{i\gamma}\vert 1\rangle_{A}\vert1\rangle_{B}
\end{eqnarray}
If either $\cos\beta $ or $\sin \beta$ is zero, we have a product state not an entangled state. Then it is not possible to satisfy Eqs. (\ref{fg})-(\ref{de}). Hence, we assume that neither  $\cos\beta $ or $\sin \beta$ is zero; both are positive.The density matrix for the above state is
\begin{eqnarray}\label{6}
	\rho&=&\frac{1}{4}[I^{A}\otimes I^{B}+(\cos^{2}\beta-\sin^{2}\beta)I^{A}\otimes \sigma_{z}^{B}
	+(\cos^{2}\beta-\sin^{2}\beta)\sigma_{z}^{A}\otimes I^{B}\nonumber \\ &&+
	(2\cos\beta \sin\beta \cos\gamma ) \sigma_{x}^{A}\otimes \sigma_{x}^{B}
+(2\cos\beta \sin\beta \sin\gamma ) \sigma_{x}^{A}\otimes \sigma_{y}^{B}\nonumber \\
&&+(2\cos\beta \sin\beta \sin\gamma ) \sigma_{y}^{A}\otimes \sigma_{x}^{B}
-(2\cos\beta \sin\beta \cos\gamma ) \sigma_{y}^{A}\otimes \sigma_{y}^{B}
\nonumber\\&&+ \sigma_{z}^{A}\otimes \sigma_{z}^{B}],	
\end{eqnarray}
where $\sigma_{x}$, $\sigma_{y}$ and $\sigma_{z}$ are pauli operators. For this state if F is measured on particle A and G is measured on particle B, then the probability that both will get $+1$ eigenvalue is given by
\begin{eqnarray}\label{7}
	P(F=+1,G=+1)&=&(\frac{1}{4})[1+(\cos^{2}\beta-\sin^{2}\beta)(\cos\theta_{F}+ \cos\theta_{G})\nonumber\\&&\cos\theta_{F}\cos\theta_{G}  
	+ 2 \cos\beta  \sin\beta \sin\theta_{F} \sin\theta_{G}\nonumber\\&&\cos(\phi_{F}+\phi_{G}-\gamma)]\nonumber\\.
\end{eqnarray}
Rearranging the above expression we get 
\begin{eqnarray}\label{8}
P(F=+1,G=+1)&=&\cos^{2}\beta \cos^{2}\frac{\theta_{F}}{2}\cos^{2}\frac{\theta_{G}}{2}+\sin^{2}\beta
\sin^{2}\frac{\theta_{F}}{2}\sin^{2}\frac{\theta_{G}}{2}\nonumber \\
 &&+2 \cos \beta \sin\beta \cos\frac{\theta_{F}}{2}\sin\frac{\theta_{F}}{2}
\cos\frac{\theta_{G}}{2}\sin\frac{\theta_{G}}{2}\nonumber\\&&\cos(\phi_{F}+\phi_{G}-\gamma)=q_{1}(say).
\end{eqnarray}
Similar calculations for other probabilities give us
\begin{eqnarray}\label{9}
P(D=+1,G=-1)&=&\cos^{2}\beta \cos^{2}\frac{\theta_{D}}{2}\sin^{2}\frac{\theta_{G}}{2}+\sin^{2}\beta
\sin^{2}\frac{\theta_{D}}{2}\cos^{2}\frac{\theta_{G}}{2}\nonumber \\
&&+2 \cos \beta \sin\beta \cos\frac{\theta_{D}}{2}\sin\frac{\theta_{D}}{2}
\cos\frac{\theta_{G}}{2}\sin\frac{\theta_{G}}{2}\nonumber \\ &&\cos(\phi_{D}+\phi_{G}+\pi-\gamma)=q_{2}(say),
\end{eqnarray}

\begin{eqnarray}\label{10}
P(F=-1,E=+1)&=&\cos^{2}\beta \cos^{2}\frac{\theta_{E}}{2}\sin^{2}\frac{\theta_{F}}{2}+\sin^{2}\beta
\sin^{2}\frac{\theta_{E}}{2}\cos^{2}\frac{\theta_{F}}{2}\nonumber \\
&&+2 \cos \beta \sin\beta \cos\frac{\theta_{F}}{2}\sin\frac{\theta_{F}}{2}
\cos\frac{\theta_{E}}{2}\sin\frac{\theta_{E}}{2}\nonumber \\ &&\cos(\phi_{F}+\phi_{E}+\pi-\gamma)=q_{3}(say),
\end{eqnarray}

\begin{eqnarray}\label{11}
P(D=+1,E=+1)&=&\cos^{2}\beta \cos^{2}\frac{\theta_{D}}{2}\cos^{2}\frac{\theta_{E}}{2}+\sin^{2}\beta
\sin^{2}\frac{\theta_{D}}{2}\sin^{2}\frac{\theta_{E}}{2}\nonumber \\
&&+2 \cos \beta \sin\beta \cos\frac{\theta_{D}}{2}\sin\frac{\theta_{D}}{2}
\cos\frac{\theta_{E}}{2}\sin\frac{\theta_{E}}{2}\nonumber \\ &&\cos(\phi_{D}+\phi_{E}-\gamma)=q_{4}(say).
\end{eqnarray}
For running Cabello's non locality argument, the following conditions should be satisfied:
\begin{eqnarray}\label{12}
	q_{2}=0,~~~~q_{3}=0,~~~~ (q_{4}-q_{1})>0, ~~~~q_{1}>0.
\end{eqnarray}
since $q_{2}$ represents probability, it cannot be negative. If it is zero, it is at its minimum value. Then its derivative must be zero. From its derivative with respect to $\phi_{D}$ we see that $\sin(\phi_{D}+\phi_{G}+\pi -\gamma)$ must be zero. Evidently 
\begin{eqnarray}\label{13}
\cos(\phi_{D}+\phi_{G}+\pi -\gamma)=-1.
\end{eqnarray}
We conclude that if $q_{2}$ is zero, then
\begin{eqnarray}
\label{14}
	\cos\beta \cos\frac{\theta_{D}}{2} \sin\frac{\theta_{G}}{2}=
	\sin\beta \sin\frac{\theta_{D}}{2} \cos\frac{\theta_{G}}{2}.
\end{eqnarray}
A similar sort of argument for $q_{3}$ to be zero will give
\begin{eqnarray}
\label{15}
\cos(\phi_{F}+\phi_{E}+\pi -\gamma)=-1.
\end{eqnarray}
and
\begin{eqnarray}
\label{16}
	\cos\beta \cos\frac{\theta_{E}}{2} \sin\frac{\theta_{F}}{2}=
	\sin\beta \sin\frac{\theta_{E}}{2} \cos\frac{\theta_{F}}{2}.
\end{eqnarray}

\section{Maximally Entangled States of Two Spin-1/2 Particles Do Not Exhibit Cabello Type Non Locality}
For maximally entangled state $\tan\beta=1$, then from Eqs. (\ref{14})and (\ref{16}) we get 
\begin{eqnarray}
\label{17}
	\frac{\theta_{G}}{2}=\frac{\theta_{D}}{2}+n\pi,
\end{eqnarray}
\begin{eqnarray}
\label{18}
	\frac{\theta_{F}}{2}=\frac{\theta_{E}}{2}+m\pi.
\end{eqnarray}

Using Eqs. (\ref{17}) and (\ref{18}), first in Eq. (\ref{8}) and in Eq. (\ref{11}), we get $q_{1}$ and $q_{4}$ for maximally entangled state as
\begin{eqnarray}\label{19}	q_{1}&=&\frac{1}{2}\cos^{2}\frac{\theta_{D}}{2}\cos^{2}\frac{\theta_{E}}{2}+
	\frac{1}{2}\sin^{2}\frac{\theta_{D}}{2}\sin^{2}\frac{\theta_{E}}{2}\nonumber \\ &&+	\cos\frac{\theta_{D}}{2}\sin\frac{\theta_{D}}{2}\cos\frac{\theta_{E}}{2}\sin\frac{\theta_{E}}{2}\cos(\phi_{F}+\phi_{G}-\gamma),
\end{eqnarray}
\begin{eqnarray}\label{20}	q_{4}&=&\frac{1}{2}\cos^{2}\frac{\theta_{D}}{2}\cos^{2}\frac{\theta_{E}}{2}+
	\frac{1}{2}\sin^{2}\frac{\theta_{D}}{2}\sin^{2}\frac{\theta_{E}}{2}\nonumber \\ &&+	\cos\frac{\theta_{D}}{2}\sin\frac{\theta_{D}}{2}\cos\frac{\theta_{E}}{2}\sin\frac{\theta_{E}}{2}\cos(\phi_{D}+\phi_{E}-\gamma).
\end{eqnarray}

From Eqs. ( \ref{19}) and (\ref{20}) it is clear that $q_{4}$ will be greater than $q_{1}$ for a maximally entangled state only when $\cos(\phi_{D}+\phi_{E}-\gamma)> \cos(\phi_{F}+\phi_{G}-\gamma)$. But Eq. ( \ref{13}) together with    Eq.(\ref{15}) says that $\cos(\phi_{D}+\phi_{E}-\gamma)= \cos(\phi_{F}+\phi_{G}-\gamma)$, i.e., $q_{4}=q_{1}$. So one can conclude that there is no choice of observable which can make maximally entangled state to show Cabello type of non locality.\\

\section{Cabello's Argument Runs for Other Two Particle Pure Entangled States}
To show that for every pure entangled state other than maximally entangled state of two spin-$1/2$ particles, a Cabello-like argument runs; it will be sufficient to show that one can always choose a set of observables, for which set of conditions given by Eq.(\ref{12}) is satisfied. This is equivalent of saying that for $0<\beta<\pi/2$ ( except for $\beta=\pi/4$) there is at least one value for each of $\theta_{D}$, $\theta_{E}$, $\theta_{G}$, $\theta_{F}$ $\phi_{D}$, $\phi_{E}$, $\phi_{G}$, $\phi_{F}$, for which conditions mentioned in (\ref{12}) are satisfied.\\
Let us choose our $\phi$'s in such a manner that
\begin{eqnarray}
	\cos(\phi_{D}+\phi_{E}-\gamma)= \cos(\phi_{F}+\phi_{G}-\gamma)=-1.\nonumber
\end{eqnarray}
For these $\phi$'s Eqs.(\ref{8}) and (\ref{11}), respectively, will read as
\begin{eqnarray}
\label{21}
	q_{1}=(\cos\beta \cos\frac{\theta_{F}}{2}\cos\frac{\theta_{G}}{2}-\sin\beta \sin\frac{\theta_{F}}{2}\sin\frac{\theta_{G}}{2})^{2},
\end{eqnarray}
\begin{eqnarray}
\label{22}
	q_{4}=(\cos\beta \cos\frac{\theta_{D}}{2}\cos\frac{\theta_{E}}{2}-\sin\beta \sin\frac{\theta_{D}}{2}\sin\frac{\theta_{E}}{2})^{2}.
\end{eqnarray}
So
\begin{eqnarray}
\label{23}
(q_{4}-q_{1})&=&\cos^{2}\beta(\cos^{2}\frac{\theta_{D}}{2}\cos^{2}\frac{\theta_{E}}{2}-\cos^{2}\frac{\theta_{F}}{2}\cos^{2}\frac{\theta_{G}}{2})\nonumber\\ &&\sin^{2}\beta(\sin^{2}\frac{\theta_{D}}{2}\sin^{2}\frac{\theta_{E}}{2}-\sin^{2}\frac{\theta_{F}}{2}\sin^{2}\frac{\theta_{G}}{2})\nonumber \\ && 2 \sin\beta \cos\beta (\cos\frac{\theta_{F}}{2}\cos\frac{\theta_{G}}{2}\sin\frac{\theta_{F}}{2}\sin\frac{\theta_{G}}{2}-\nonumber \\ &&\cos\frac{\theta_{D}}{2}\cos\frac{\theta_{E}}{2}\sin\frac{\theta_{D}}{2}\sin\frac{\theta_{E}}{2}).	
\end{eqnarray}
Now we will have to choose at least one set of values of $\theta$'s in such a way that $(q_{4}-q_{1})$ and $q_{1}$ are nonzero and positive. Moreover, these values of $\theta$'s should also not violate conditions given in Eqs. (\ref{14}) and(\ref{16}). Let us try with $\theta_{D}/2=0$, i.e., with
$$ \sin\frac{\theta_{D}}{2}=0,~~~~~~~~~~~~ \cos\frac{\theta_{D}}{2}=1.$$
This makes Eq. (\ref{14}) to read as 
$$\sin\frac{\theta_{G}}{2}=0,\Rightarrow \frac{\theta_{G}}{2}=0.$$
Then from Eq. (\ref{23}) we get
$$ (q_{4}-q_{1})=\cos^{2}\beta
(\cos^{2}\frac{\theta_{E}}{2}-\cos^{2}\frac{\theta_{F}}{2}).$$
Thus $(q_{4}-q_{1})>0$ if
\begin{eqnarray}
\label{24}
	\cos\frac{\theta_{E}}{2}>\cos\frac{\theta_{F}}{2}
\end{eqnarray}
rewriting Eq.(\ref{16}) as
\begin{eqnarray}
\label{25}
	\tan\frac{\theta_{F}}{2}=\tan\beta \tan\frac{\theta_{E}}{2}.
\end{eqnarray}
Values of $\theta$'s satisfying inequality (\ref{24}) will not violate Eq. (\ref{25}) provided $\tan\beta>1$. Now for these values of $\theta$'s, from Eq. (\ref{21}), we get: $q_{1}=(\cos\beta \cos\frac{\theta_{F}}{2})^{2}$ which is greater than zero.\\
 So for the above values of $\theta$'s, i.e., for $\frac{\theta_{D}}{2}=\frac{\theta_{G}}{2}=0$ and $ \cos\frac{\theta_{E}}{2}>\cos\frac{\theta_{F}}{2}$, all the states for which $\tan\beta>1$; Cabello's non locality argument runs.\\
 For other states, i.e., for the states for which $\tan\beta<1$, let us choose $\frac{\theta_{D}}{2}=\frac{\theta_{G}}{2}=\pi/2$. Then from Eq. (\ref{23}) we get 
 $$ (q_{4}-q_{1})=\sin^{2}\beta
(\sin^{2}\frac{\theta_{E}}{2}-\sin^{2}\frac{\theta_{F}}{2}).$$
Thus $(q_{4}-q_{1})>0$ if
\begin{eqnarray}
\label{26}
	\sin\frac{\theta_{E}}{2}>\sin\frac{\theta_{F}}{2}.
\end{eqnarray}
One can easily check that for above mentioned values of $\theta$'s; $q_{1}$ is also positive and Eq. (\ref{25}) is satisfied too.\\
Thus if we choose $\frac{\theta_{D}}{2}=\frac{\theta_{G}}{2}=\pi/2$ and 
$\sin\frac{\theta_{E}}{2}>\sin\frac{\theta_{F}}{2}$, then all the states for which $\tan\beta$, satisfy Cabello's non locality argument. So for every $\beta$ ( except for $\beta=\pi/4$), we can choose $\theta$'s and  $\phi$'s and hence the observables in such a way that Cabello's argument runs.\\

\section{Maximum Probability of Success}
For getting maximum probability of success of Cabello's argument in contradicting local realism, we will have to maximize the quantity $q_{4}-q_{1}$, for a given $\beta$, over all observable parameters $\theta$'s and $\phi$'s and under the restrictions given by Eqs. (\ref{13})-(\ref{16}), we get 
\begin{eqnarray}\label{27}
(q_{4}-q_{1})&=&\cos^{2}\beta[(k_{2}-k_{1})+ \tan^{2}\beta ~\tan^{2}\frac{\theta_{D}}{2}~\tan^{2}\frac{\theta_{E}}{2}(k_{2}-k_{1}\tan^{4}\beta)\nonumber \\ &&+2\tan\beta ~\tan\frac{\theta_{D}}{2}~\tan\frac{\theta_{E}}{2}(k_{2}-k_{1}\tan^{2}\beta)\cos(\phi_{D}+\phi_{E}-\gamma)]\nonumber\\
\end{eqnarray}
where
\begin{eqnarray}
	k_{1}&=&(\tan^{2}\beta ~ \tan^{2}\frac{\theta_{D}}{2}+1)^{-1}(\tan^{2}\beta ~ \tan^{2}\frac{\theta_{E}}{2}+1)^{-1},\nonumber\\
%\end{eqnarray}
%\begin{eqnarray}
k_{2}&=&(\tan^{2}\frac{\theta_{D}}{2}+1)^{-1}(\tan^{2}\frac{\theta_{E}}{2}+1)^{-1}.\nonumber
\end{eqnarray}
It is clear from Eq. (\ref{27}) that one can obtain maximum value of $(q_{4}-q_{1})$, when $\cos(\phi_{D}+\phi_{E}-\gamma)=\pm 1 $. Let us first consider $\cos(\phi_{D}+\phi_{E}-\gamma)=-1$, then from Eq. (\ref{27}) we get
\begin{eqnarray}\label{28}
&&(q_{4}-q_{1})=\cos^{2}\beta \nonumber \\&&\left[\frac{(1-\tan\beta ~\tan\frac{\theta_{D}}{2}~\tan\frac{\theta_{E}}{2})^{2}}{(1+\tan^{2}\frac{\theta_{D}}{2})(1+\tan^{2}\frac{\theta_{E}}{2})}
 -\frac{(1-\tan^{3}\beta ~\tan\frac{\theta_{D}}{2}~\tan\frac{\theta_{E}}{2})^{2}}{(1+\tan^{2}\beta ~\tan^{2}\frac{\theta_{D}}{2})(1+ \tan^{2}\beta ~\tan^{2}\frac{\theta_{E}}{2})}\right]\nonumber\\.	
\end{eqnarray}
From the above equation one can show that $(q_{4}-q_{1})$ will be maximum when $\theta_{D}=\theta_{E}$ ( see the Appendix ) which, in turn, implies $\theta_{G}=\theta_{F}$, i.e., $(q_{4}-q_{1})$ becomes maximum when measurement settings in both the sides are same as was the situation in Hardy's argument. Now for the optimal case, i.e., for $\theta_{G}=\theta_{F}$ and $\theta_{D}=\theta_{E}$, $(q_{4}-q_{1})$ becomes
\begin{eqnarray}\label{29}
&&(q_{4}-q_{1})=\cos^{2}\beta \nonumber\\&&\left[\frac{(1-\tan\beta ~\tan^{2}\frac{\theta_{D}}{2})^{2}}{(1+\tan^{2}\frac{\theta_{D}}{2})^{2}}
 -\frac{(1-\tan^{3}\beta ~\tan^{2}\frac{\theta_{D}}{2})^{2}}{(1+\tan^{2}\beta ~\tan^{2}\frac{\theta_{D}}{2})^{2}}\right].	
\end{eqnarray}
Numerically we have checked that $(q_{4}-q_{1})$ has a maximum value of 0.1078 when $\cos\beta=0.485$ with $\theta_{D}=\theta_{E}=0.59987$. This is interesting as maximum probability of success of Hardy's argument is only $9\%$ whereas in case of Cabello's argument it is approximately $11\%$.\\ Here we are comparing the maximum probability of success of Hardy's argument with that of Cabello's argument for all states ( see Fig.3 ).\\Figure 3 shows that for $\cos\beta\approx 0.7$, i.e., for $\beta=\pi/4$ and $\cos\beta=1$, i.e., for $\beta=0$ ; maximum of $(q_{4}-q_{1})$ vanishes. This is as expected because these values of $\beta$ represent, respectively, the maximally entangled and product states for which Cabello's argument does not run. For most of the other values of $\beta$, the probability for success of Cabello's argument inestablishing their nonlocal feature, is more than the maximum probability of success of Hardy's argument in doing the same.\\
\begin{figure}
\begin{center}
\includegraphics[width=14cm,height=12cm]{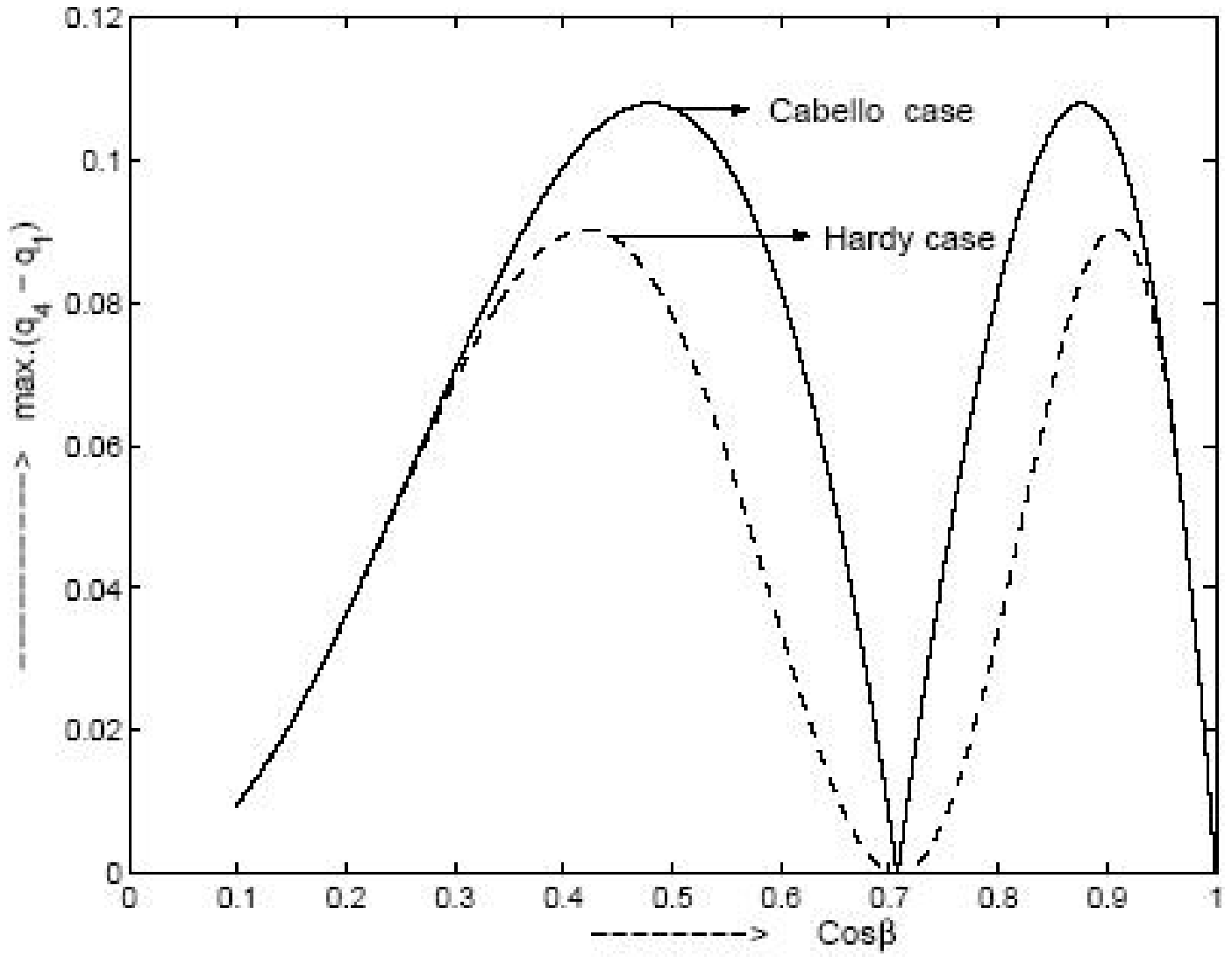}
\vspace{0.4cm}
\underline{Figure 3 :} \small{ Figure3: Comparison of the Maximum probability of success between Hardy's and Cabello's case.}
\end{center}
\end{figure}
\vspace{0.8cm}
\newpage
 As we have mentioned earlier [just before Eq. (\ref{28})] that $ \cos(\phi_{D}+\phi_{E}-\gamma)=-1$ also optimizes $(q_{4}-q_{1})$. This also gives the same maximum value for $(q_{4}-q_{1})$ as given by $\cos(\phi_{D}+\phi_{E}-\gamma)=-1$, but for $\theta_{D}=-\theta_{E}$.\\
\section{Extension Cabello's Argument to Three Probabilities Are Nonzero}
We consider only $ q_{3}=0$, for running Cabello's non locality argument, the following conditions should be satisfied:
\begin{eqnarray}
\label{30}
	q_{3}=0,~~~~ q_{4}-(q_{1}+q_{2})>0.
\end{eqnarray}
We show that in this case, maximum value $q_{4}-(q_{1}+q_{2})$ is higher than $q_{4}-q_{1}$ with $q_{2}=q_{3}=0$.
since $q_{3}$ represents probability, it cannot be negative. If it is zero, it is at its minimum value. Then its derivative must be zero. From its derivative with respect to $\phi_{F}$ we see that $\sin(\phi_{D}+\phi_{G}+\pi -\gamma)$ must be zero. Evidently 
\begin{eqnarray}\label{31}
\cos(\phi_{F}+\phi_{E}+\pi -\gamma)=-1.
\end{eqnarray}
We conclude that if $q_{3}$ is zero, then
\begin{eqnarray}\label{32}
	\cos\beta \cos\frac{\theta_{E}}{2} \sin\frac{\theta_{F}}{2}=
	\sin\beta \sin\frac{\theta_{E}}{2} \cos\frac{\theta_{F}}{2}.
\end{eqnarray}
The optimization $q_{4}-(q_{1}+q_{2})$, for a given $\beta$, over all observable parameters $\theta$'s and $\phi$'s and under restrictions given by Eqs. (\ref{31}) and (\ref{32}), occurs when
\begin{eqnarray}\label{33}
	\cos(\phi_{F}+\phi_{G}-\gamma)=\cos(\phi_{D}+\phi_{E}-\gamma)=\cos(\phi_{D}+\phi_{G}+\pi-\gamma)=\pm 1\nonumber\\.
\end{eqnarray}
The maximum value $q_{4}-(q_{1}+q_{2})$ is obtained by  selection:
\begin{displaymath}
\left\{\begin {array}{ll}
 \phi_{F}+\phi_{G}-\gamma=\pi \\
 \phi_{D}+\phi_{G}-\gamma=\pi \\
 \phi_{F}+\phi_{E}-\gamma=2\pi.
\end{array} \right.	
\end{displaymath}
So
\begin{eqnarray}
	q_{4}-(q_{1}+q_{2})&=&[\cos \beta \cos \frac{\theta_{D}}{2}\cos \frac{\theta_{E}}{2}+\sin \beta \sin \frac{\theta_{D}}{2}\sin \frac{\theta_{E}}{2}]^{2}-\nonumber \\ &&[\cos \beta \cos \frac{\theta_{D}}{2}\sin \frac{\theta_{G}}{2}+\sin \beta \sin \frac{\theta_{D}}{2}\cos \frac{\theta_{G}}{2}]^{2}-\nonumber \\ && \frac{\cos^{2}\beta \cos^{2}\frac{\theta_{G}}{2}}{1+\tan ^{2}\beta \tan ^{2} \frac{\theta_{E}}{2}}(1-\tan^{2}\beta \tan \frac{\theta_{E}}{2}\tan \frac{\theta_{G}}{2})^{2}
\end{eqnarray}
Numerically we have checked that $q_{4}-(q_{1}+q_{2})$ has a maximum value of $0.1588$ when $\cos \beta =0.95$ with $\theta_{D}=0.45$, $\theta_{E}=1.4$ and $\theta_{E}=0.46$. So, the maximum probability of success of Hardy's argument is only $9\%$ whereas in case of Cabello's argument with two probability nonzero it is approximately $11\%$ and for Cabello's argument with three probability nonzero it is approximately $16\%$.
\section{Conclusion}
In conclusion, here we have shown that maximally entangled states do not respond even to Cabello's argument which is a relaxed one and is more general than Hardy's argument. All other pure entangled states response to Cabello's argument. These states also exhibit Hardy's type non locality. But, interestingly, for most of these non maximally entangled states, fraction of runs in which Cabello's argument succeeds in demonstrating their nonlocal feature, can be made more than the fraction of runs in which Hardy's argument succeeds in doing the same. So it seems that in some sense, for demonstrating the nonlocal features of most of the entangled states, Cabello's argument is a better candidate.
\newpage
\section{Appendix}

We want to optimize $(q_{4}-q_{1})$ given in Eq. (\ref{28}) with respect to $\theta_{D}$ and $\theta_{E}$ for a given $\beta$. Differentiating Eq. (\ref{28}) with respect to $\theta_{D}$ and equating it to zero, we get the following two equations:
\begin{eqnarray}
	(\tan\beta ~ \tan\frac{\theta_{E}}{2}+ \tan\frac{\theta_{D}}{2})=0
\end{eqnarray}
and
\begin{eqnarray}
	&&(\tan\beta ~ \tan\frac{\theta_{E}}{2}~\tan\frac{\theta_{D}}{2}-1)(\tan^{2}\beta ~ \tan^{2}\frac{\theta_{E}}{2}+1)(\tan^{2}\beta ~ \tan^{2}\frac{\theta_{D}}{2}+1)^{2}=\nonumber \\ &&
	(\tan^{2}\beta ~ \sec^{2}\frac{\theta_{D}}{2})(\tan^{3}\beta ~\tan\frac{\theta_{E}}{2}~ \tan\frac{\theta_{D}}{2}-1)(\sec^{2}\frac{\theta_{D}}{2}~\sec^{2}\frac{\theta_{E}}{2}).
\end{eqnarray}
Similarly differentiating Eq. (\ref{28}) with respect to $\theta_{E}$ and equating it to zero, we have 
\begin{eqnarray}
	(\tan\beta ~ \tan\frac{\theta_{D}}{2}+ \tan\frac{\theta_{E}}{2})=0
\end{eqnarray}
and
\begin{eqnarray}
	&&(\tan\beta ~ \tan\frac{\theta_{D}}{2}~\tan\frac{\theta_{E}}{2}-1)(\tan^{2}\beta ~ \tan^{2}\frac{\theta_{D}}{2}+1)(\tan^{2}\beta ~ \tan^{2}\frac{\theta_{E}}{2}+1)^{2}=\nonumber \\ &&
	(\tan^{2}\beta ~ \sec^{2}\frac{\theta_{E}}{2})(\tan^{3}\beta ~\tan\frac{\theta_{D}}{2}~ \tan\frac{\theta_{E}}{2}-1)(\sec^{2}\frac{\theta_{E}}{2}~\sec^{2}\frac{\theta_{D}}{2}).
\end{eqnarray}
Analyzing the above four conditions we find that $\theta_{D}=\theta_{E}$ will give the optimal solution. Similarly for $ \cos(\phi_{D}+\phi_{E}
-\gamma)=+1$, we will get same kind of results.
\newpage

\chapter{Summary and Future Directions}

\begin{center}
\scriptsize \textsc{This is not the end. Nor is this a beginning of the end.\\ This may at most be the end of a beginning.\\
 -Sir Winston Churchill\footnote{From his last address to the British parliament as the prime minister of U.K. ( 3rd tune 1946)}.}
\end{center}

In this thesis I have tried to understand different aspects of quantum non-locality, particularly:\\
I-  Measure of non locality in quantum correlations between two spin-$s$ particles in the singlet state by the number of  classical bits required to simulate the corresponding quantum correlations.\\
II- Investigation of non locality in time for a spin-$s$ particle in successive measurements scenario and  deviation of quantum mechanics from realistic hidden variable theory.\\
III- Investigation of non locality in quantum mechanics without using inequalities in Hardy's argument and Cabello's argument.\\
\emph{Entanglement in space} displays one of the most interesting features of quantum mechanics, often called quantum non locality. Locality in space and realism impose constraints -Bell's inequalities- on certain combinations of correlations for measurements of spatially separated systems, which are violated by quantum mechanics. Non locality is one of the strangest properties of quantum mechanics, and understanding this notion remains an important problem. \\ \emph{Entanglement in time} is not introduced in quantum mechanics because of different roles time and space play in quantum theory. The meaning of locality in time is that the results of measurement at time $t_2$ are independent of any measurement performed at some earlier time  $t_1$ or later time $t_3$. The temporal Bell's inequalities are derived from the realistic hidden variable theory.\\The non-local nature of quantum mechanics is, in general, a consequence of entangled states ( non locality in space), although they are independent properties of quantum systems, or consequence of successive measurements (non locality in time). A possible approach to measure non locality ( in space or in time) is the number of  classical bits required for simulation of non local quantum correlations. The goal is to quantify how powerful quantum mechanics is by comparing its achievements to those of more familiar resources. For instance, one may naturally ask how much information should be sent from one party ( or one experiment) to the other party (or other consecutive measurement) in order to reproduce the correlations distributed by entangled pairs ( or two successive measurements). Information is something that we are able to quantify, thus the number of  communicated cbits measure  the non-locality in space ( or in time). In the following, I bring some of the key results obtained in this thesis.\\
\textbf{Chapter II:}\\1- We give a classical protocol to exactly simulate quantum correlations implied by a spin-$s$ singlet state for all spins satisfying $2s+1=P^n$, in the worst-case scenario, $P$ and $n$ are a positive integer.  We can find out a unique positive integer $m$ such that $2^m-1< P= (2s+1)^{-n}\leq 2^{m+1}-1$, so we can write $P$ according to the binary representation, $P=a_02^m+a_12^{m-1}+\ldots +a_m 2^0\equiv\underline{a_0a_1\ldots a_m}$ ( where all $a_i$$\in\{0,1\}$). We must have $a_0\neq0$. The required amount of communication ($n_c$)is found to be $$n_c=nm=\left[\lg_2P\right]^{-1}\left[\lg_2(2s+1)\right]\left[\left\lceil \lg_2(P+1) \right\rceil -1\right],$$
and the number of independent and uniformly distributed shared random variables ($n_{srv}$) is found to be $$n_{srv}=n(2m+\sum_{i=1}^m a_i)=\left[\lg_2P\right]^{-1}\left[\lg_2(2s+1)\right]\left[2\left[\left\lceil \lg_2(P+1) \right\rceil -1\right]+\sum_{i=1}^m
a_i\right].$$
For example, Case $P=2$: The number of cbits is $n_c=\lg_2(2s+1)$ and the number of share randomness is found $n_{srv}=2n=2\lg_2(2s+1)$.\\
2- We also introduce another classical model to exactly simulate quantum correlations corresponding to the spin $s$ singlet state for all spins. The required amount of communication is found to be $n_c=\left\lceil \lg_2(s+1)\right\rceil$ and the number of shared randomness variables is found to be $n_{srv}=2n+\sum_{i=1}^n a_i,$ where $n$ satisfy $2^{n-1}<s+1\leq 2^n$ ($d=2s+1=a_02^n+a_12^{n-1}+\ldots +a_n 2^0\equiv\underline{a_0a_1\ldots a_n}$). It is interesting to see that for all $s$ values satisfying $2^n\leq s+1< 2^{n+1}$ the required communication does not vary and equals $n$ cbits.\\
3- The correlations of a pure non-maximally entangled state can not be simulated using one cbit of communication \cite{brunnergisin}. We give a classical protocol for simulation of quantum correlations between two parties who share a non maximally entangled two qubit state which  requires two cbits of communication.\\
\textbf{Chapter III}\\1- We obtain temporal Mermin-Klyshko inequality (MKI) for $n$ successive measurements by using realism and non locality in time. We showed quantum  correlations violate temporal Mermin-Klyshko inequality.\\ 2- It was interesting that, for a spin-$s$ particle, maximum deviation of quantum mechanics from realism was obtained for all convex combinations  of $\alpha_0=\pm1$ states (the case when input state is a mixed state whose eigenstates coincide with those of $\vec{S}\cdot\hat a_0$ for some $\hat a_0$ whose eigenvalues we denote by $\alpha_0\in\{-s,\ldots,s\}$). This is surprising as one would expect pure states to be more `quantum' than the mixed ones thus breaking Bell inequalities by larger amount.\\
3- All spin 1/2 states maximally break  Mermin-Klyshko inequalities for $n$ successive measurements ($\eta_n=\sqrt{2}$) as against only the entangled states break it in  multipartite case. Interestingly that for $s=\frac{1}{2}$ the random mixture ( maximum noisy state) also breaks BI. This indicates that the notion of ``classicality", compatible with the usual local HVT, is different in nature from the notion of classicality that would arise from the non-violation of BI here.\\ 4- We see that  for all spins, BI and MKI is violated in two and three successive measurements ($\eta_2>1$,$\eta_3>1$) and the value of violation MKI in three successive measurements is a little more than the value of violation BI in two successive measurements $\eta_3>\eta_2$ except $s=\frac{1}{2}$, while $\eta_3=\eta_2=\sqrt{2}$ for spin $s=\frac{1}{2}$. Also $\eta_3$ and $\eta_2$ decrease monotonically with increase in the value of $s$. It is interesting, in the case of two and three successive measurements of spin $s$ prepared in a pure state, that the maximum violation of BI and MKI  falls off as the spin of the particle increases, but tends to a constant for arbitrary large $s$, $\eta_2(s\longrightarrow\infty)=1.143 $ and $\eta_3(s\longrightarrow\infty)=1.153 $. It is thus seen that large quantum numbers do not guarantee classical behavior.\\ 5- We show that for $s=\frac{1}{2}$, the correlation between the outputs of measurements from last $k$ out of $n$ successive measurements ($k<n$) depend on the measurement prior to ($n-k$), when $k$ is even, while for odd $k$, these correlations are independent of the outputs of measurements prior to $n-k$.\\
6- Interestingly if the initial state is the random mixture or first Stern-Gerloch measurements for the qubit component along the directions $a_1$ perpendicular to initial state $\hat a_0 \perp \hat a_1$ so that, $\langle \alpha_1 \rangle_{QM} = 0$, then always quantum averages for all odd number of successive measurements are zero.\\
7- We proved that the correlation function between first and third measurement ($t_1 $ and $t_3$) on spin-$s$ particle for a given measurement performed at $t_2$ can not violate the temporal Bell inequality. Therefore, any measurement performed at time $t_2$ disentangles events at time $t_1$ and $t_3$ if $t_1<t_2<t_3$.\\
8- Three successive measurements on spin-$s$ particles do not break Svetlinchi Inequality. But it is proved that three successive measurements on qubit violate Scarani-Gisin inequality. Thus, although there are no genuine three-fold temporal correlations, a specific combination of two-fold correlations can have values that are not achievable with correlations in space for any three-qubit system.\\ 9- Also we showed that three successive measurements  violate two types of  Bell inequalities involving two and three successive measurements. So two successive measurement correlations are relevant to those of three successive measurements. This behavior is analogous to three particle W-state.\\
10- Quantum correlations between two successive measurements on a qubit violates chained Bell inequality which is obtained by providing more than two alternative experiments in every step.\\
11- By using just two cbits communications between  successive measurements on qubit it is possible to simulate all quantum correlations. The amount of information needed to be transferred between successive measurements in order to classically simulate quantum correlations, measure the deviation of quantum mechanics from HVT.\\
\textbf{Chapter IV} \\ 1- In this chapter we showed that for most of non maximally entangled states, fraction of runs in which Cabello's argument succeeds in demonstrating their nonlocal feature, can be made more than the fraction of runs in which Hardy's argument succeeds in doing the same. The maximum probability of success of Hardy's argument is only $9\%$ whereas in case of Cabello's argument with two probability nonzero it is approximately 11\% and for Cabello's argument with three probability nonzero it is approximately 16\%.\\
Finally, there are deeper question that remain unanswered:\\
Until now, we had investigated on the simulation $\langle \alpha\rangle$, $\langle \beta\rangle$ and $\langle \alpha\beta\rangle$ for singlet states on spin-$s$ only for spin observables ( Stern-Gerlach measurement ), namely the observables of the form $\vec{S}\cdot\hat{a}$ on each individual spin-$s$ system where $\hat{a}$ is an arbitrary unit vector in $\mathbb{R}^3$ and $\vec{S}=(S_x,S_y,S_z)$. For $s=\frac{1}{2}$, this choice contains all observables but for $s>\frac{1}{2}$ this choice corresponds to a restricted class of observables. There are some directions in which this work can be generalized.\\
 1- In classical model that introduced in chapter II, if the spin value crosses $2^{n+1}$, the required number of cbits takes a ``quantum jump" from $n$ to $n+1$.  It will be interesting to find an explanation of this behavior and is one of the problems we want to tackle in future.\\
2- Toner and Bacon's protocol \cite{toner03} to simulate quantum correlations for $s=\frac{1}{2}$ singlet state, simulates $\langle\alpha\rangle=0$, $\langle\beta\rangle=0$ and $\langle\alpha\beta\rangle=-\hat{a}\cdot\hat{b}$ where $\alpha\in\{-1,+1\}$ and $\beta\in\{-1,+1\}$ are the outputs of Alice and Bob respectively. This specifies all the joint probabilities $p(\alpha,\beta)$, through the system of equations
\begin{eqnarray*}
\langle\alpha\rangle &=&p(1,1)+p(1,-1)-\left[p(-1,1)+p(-1,-1)\right],\\
\langle\beta\rangle &=&p(1,1)+p(-1,1)-\left[p(1,-1)+p(-1,-1)\right],\\
\langle\alpha\beta\rangle &=&p(1,1)+p(-1,-1)-\left[p(-1,1)+p(1,-1)\right],\\
1 &=&p(1,1)+p(1,-1)+p(-1,1)+p(-1,-1).
\end{eqnarray*}
The question is, can we have a classical simulation protocol which can simulate all the joint probabilities implied by $s>\frac{1}{2}$ singlet state?\\ To do this for spin $s=1$, classical protocol must simulate eight quantum averages for the spin $1$ singlet state $\langle \alpha\rangle=0 , \langle \beta\rangle=0 , \langle \alpha\beta\rangle=\frac{2}{3}\hat{a}\cdot\hat{b}, \langle \alpha^2\rangle=\frac{2}{3}, \langle \beta^2\rangle=\frac{2}{3}, \langle \alpha^2\beta\rangle=0 , \langle \alpha\beta^2\rangle=0$ and $\langle \alpha^2\beta^2\rangle=\frac{1}{3}[1+(\hat{a}\cdot\hat{b})^2]$.\\ This can then be used to obtain all nine joint probabilities through the system of nine equations, eight of them are obtained using the definitions of the corresponding average in terms of the joint probabilities and the ninth equation is given by normalization condition, ( as described above for the $s=\frac{1}{2}$ case).\\For high spin ($s>1$) singlet state, the classical protocol  must simulate all quantum correlations averages $\langle \alpha^{q}\rangle$, $\langle\beta^{r}\rangle$ and $\langle\alpha^{q}\beta^{r}\rangle$ ( for all $q,r=1,2,\ldots$). By comparison, our protocols presented in chapter II simulate only $\langle\alpha\rangle ,\langle\beta\rangle ~\mbox{and}~ \langle\alpha\beta\rangle$ for all spin-$s$ singlet state.\\ 
3- If we consider most general measurements on both the sides of a maximally entangled state of two qudits, How about exactly simulating arbitrary entangled state- even mixed states- under arbitrary measurements, including POVMs?\\
4- How can we simulate all quantum correlations- even spin observables- between more than two parties (like GHZ-state and W-state)?\\
5- Until now, there are no any classical models to simulate  all quantum correlations  implied by two qudit  non maximally entangled states. This open problem can be generalized to  more than two parties.\\
6- How much $n$ ($n>3$) successive measurements on spin-$s$ ($s>\frac{1}{2}$) violate MKI?\\
7- How much is the maximum violation  of CGLMP inequality \cite{cglmp} ( in terms of probabilities)  by $n$ successive measurements?\\
8- We may consider $n$-successive measurements on a spin $s$-system and each step, experimenter select $m$ ($v=2s+1$)-valued measurements. This gives a total of $(mv)^n$ experimentally accessible probabilities.  Investigation on the set of all Bell's inequalities is  a open problem.\\
9- How can we simulate all quantum correlations in successive measurements for high spin-$s$?\\
10- Exploring the  possibility to find a non locality in time argument without using inequality (like Hardy's argument or Cabello's argument) for successive measurements scenario.\\

\end{document}